\documentclass[review]{elsarticle}
\usepackage{amsmath}
\usepackage[cmintegrals]{newtxmath}
\usepackage{supertabular}
\usepackage{multirow}
\usepackage{textcase}
\usepackage[tablename=TABLE]{caption}
\DeclareCaptionTextFormat{up}{\MakeTextUppercase{#1}}
\usepackage{placeins}
\usepackage{graphicx}
\usepackage{amsmath}
\usepackage{amssymb,amsfonts}
\usepackage{algorithm,algorithmicx}
\usepackage{algpseudocode}
\usepackage{url}
\usepackage{longtable}
\usepackage{amssymb}
\usepackage{array}
\usepackage{pgfplots}
\usepackage{tikz}
\usetikzlibrary{arrows,shapes,positioning,shadows,trees}
\usepackage{forest}
\usepackage{longtable}
\usetikzlibrary{arrows.meta}
\tikzstyle{arrow} = [thick,->,>=stealth]

\pgfplotsset{compat=1.14}
\usepackage{lineno,hyperref}
\hypersetup{pdfauthor={Name}}
\modulolinenumbers[5]

\journal{Journal of Network and Computer Applications}

\begin{document}

\begin{frontmatter}

\title{A Review on Impact of Bloom Filter on Named Data Networking: The Future Internet Architecture}

\author{Sabuzima Nayak}
\author{Ripon Patgiri\corref{mycorrespondingauthor}}
\ead{ripon@cse.nits.ac.in}
\ead[url]{http://cs.nits.ac.in}
\author{Angana Borah}
\address{National Institute of Technology Silchar}


\cortext[mycorrespondingauthor]{Corresponding author: Ripon Patgiri, Department of Computer Science \& Engineering, National Institute of Technology Silchar, Cachar-788010, India}


\begin{abstract}
Today is the era of smart devices. Through the smart devices, people remain connected with systems across the globe even in mobile state. Hence, the current Internet is facing scalability issue. Therefore, leaving IP based Internet behind due to scalability, the world is moving to the Future Internet Architecture, called Named Data Networking (NDN). Currently, the number of nodes connected to the Internet is in billions. And, the number of requests sent is in millions per second. NDN handles such huge numbers by modifying the IP architecture to meet the current requirements. NDN is scalable, produces less traffic and congestion, provides high level security, saves bandwidth, efficiently utilizes multiple network interfaces and have many more functionalities. Similarly, Bloom Filter is the only good choice to deploy in various modules of NDN to handle the huge number of packets. Bloom Filter is a simple probabilistic data structure for the membership query. This article presents a detailed discussion on the role of Bloom Filter in implementing NDN. The article includes a precise discussion on Bloom Filter and the main components of the NDN architecture, namely, packet, content store, forward information base and pending interest table are also discussed briefly. 
\end{abstract}


\end{frontmatter}


\section{Introduction}
Named Data Networking (NDN) is the most recent emerging research area which is known as the future Internet. It is introduced in 2010 and has become one of the most emerging research fields. Since then, the NDN is able to influence academician, industrialist, scientist and practitioner. IPv4 has serious scalability issue and therefore, the NDN overtaking the IPv6 and creates a new future possibilities. The Internet using IP address permits point-to-point connectivity between two nodes. With the advancement of packet switching, not only text, audio and video are also sent through the Internet. However, IP is unable to satisfy the customer requirements due to increasing demands of the Internet, changing nature of applications, user requirements, and huge increment of the Internet users. Network layer of the OSI communication model has only IP. This does not permit IP to modify existing functionality nor add new functionality to satisfy new requirements of the users. The network today have multiple interfaces and very mobile. IP only forward within a spanning tree. Therefore, IP is not suitable for today's network \cite{Jacobson}.  The new user requirement is \textit{``what data''} without knowing \textit{``where data''}. These requirements lead to the ingress of the Internet to a new era of content-centric communication  \cite{Fang}. 

Internet with content-centric communication provide ubiquitous inter-connectivity and a wide range of services \cite{NDN1}. It enables information-intensive business (eg. banking, travelling and financial services) to use the Internet to expand their business. Internet helps in storing data in different geolocations, staying connected to their data around the clock, providing their services easily to the users, and interacting with the user without physical interaction. In addition, hardware advancement also contributed largely to increase the inter-connectivity between the systems. With the growth of mobile devices, the Internet is also becoming mobile. Mobile Internet helps in inter-connectivity between every smart device with our mobile phones. Not just text, large sized audio, video and images are also transmitted through the Internet. Several projects are proposed for designing content-based future Internet paradigms \cite{NSF, IRTF, GreenICN}. To achieve Information-Centric Network \cite{ICN1, ICN2}, the Named Data Networking (NDN) \cite{NDN, NDN2, NDN3} is proposed which has an efficient design and simple communication model. NDN is capable of handling variable-length, location-independent names, searching and retrieving content upon content request. 

Bloom Filter is a simple set membership filtering data structure. Simplicity of Bloom Filter makes it suitable to implement along with an application with a negligible overhead. Its advantages far exceed its issues. Bloom Filter is implemented in diverse network applications such as Network Traffic \cite{Sasaki}, Network Security \cite{GERAVAND1,Xiao2,DDoS}, Routing Algorithms \cite{isBF}, Web Cache \cite{countingBF}, Wireless Sensor Networking \cite{Talpur}, wired/wireless networking \cite{Tarkoma}. Packet filtering in networking had been always an issue. However, with the onset of the Big Data era, the importance of Bloom Filter escalated. Big Data being a problem in every domain, all domain felt a need for a simple deduplicating data structure. Therefore, Bloom Filter emerged as a solution for Big data. Some emerging topics implementing Bloom Filter are Internet of Things \cite{Al-Fuqaha,Singh}, Cloud Computing \cite{Xiong,Mukherjee}, and 5G \cite{B-MaFIB}. Moreover, interdisciplinary domains such as Biometrics \cite{Rath,Biomet} and Bioinformatics \cite{Abyss,Nayakbio} are preferring Bloom Filter due to the above mentioned advantages along with tiny memory consumption. Bloom Filter is able to attract the research community; because, it provides a faster lookup or a filter using a tiny on-chip memory. Bloom Filter is highly adaptable and is able to enhance a system performance significantly. For instance, BigTable claims that its performance is increased dramatically after using Bloom Filter. 


NDN consists of many components such as Content store, Forward Information Base and Pending Interest Table. Each component has to deal with Big Data. Therefore, NDN requires Bloom Filter to deal with its problem. Currently, Bloom Filter is playing a vital role in NDN. Bloom Filter is an approximation data structure to check membership with some error tolerance. It uses a tiny amount of memory to check whether a given element is a member of a set or not. Also, it can answer any query in $O(1)$ time complexity. Therefore, NDN is able to use Bloom Filter in its all components to check whether a given item is a member of a set or not. NDN is enhanced drastically by engaging Bloom Filter. Therefore, we carried out a rigorous review to expose the impact of Bloom Filter in NDN.

\subsection{Motivation}
Numerous surveys are available on both Bloom Filter and NDN individually. A list of Bloom Filter survey papers are presented in Table \ref{bf_survey} and Table \ref{ndn_survey} representing a list of NDN survey articles. Observing those papers it is concluded that no survey paper is published which focuses on the role of Bloom Filter in NDN. Bloom Filter is a crucial part of NDN, and hence, this survey is carried out to explore the state-of-the-art NDN, which fully depends on the Bloom Filter data structure. The literature search exposes the similar survey published by reputed publishers, for instance, IEEE. However, this survey focuses on the significance of Bloom Filter in implementing NDN. Bloom Filter is used to implement the Pending Interest Table (PIT), Content Store (CS) and Foreward Information Base (FIB). Bloom Filter is able to enhance the performance of these modules of NDN using a tiny amount of main memory.

\begin{table}
\centering
    \caption{\textbf{List of survey papers available on Bloom Filter}}
    \begin{tabular}{|p{1.7cm}|p{1.6cm}|p{4.2cm}|p{4.5cm}|}
    \hline
    \textbf{Authors} & \textbf{Year} & \textbf{Title of the paper} & \textbf{Domain}\\ \hline
     
Broder \textit{et al.} \cite{Broder} & 2004, Taylor \& Francis  & \small{ Network Applications of Bloom Filters: A Survey} & \small{ Bloom Filter, Distributed Caching, P2P/Overlay Networks, Resource Routing, Packet Routing} \\ \hline

Tarkoma \textit{et al.} \cite{Tarkoma} & 2012, IEEE & \small{Theory and practice of bloom filters for distributed systems} & \small{Variants of Bloom Filter, Distributed Computing} \\ \hline
 
Geravand \textit{et al.} \cite{GERAVAND1} & 2013, Elsevier & \small{Bloom filter applications in network security: A state-of-the-art survey} &  \small{Bloom Filter, Network (Wireless and Wired), Network security} \\ \hline

Sangeetha \textit{et al.} \cite{Sangeetha} & 2015, IEEE & \small{A survey of hardware signature implementations in multi-core systems} &  \small{Hardware signature} \\ \hline

Gupta and Batra \cite{Gupta} & 2017, IEEE & \small{A short survey on bloom filter and its variants} &  \small{Variants of Bloom Filter} \\ \hline

Patgiri \textit{et al.} \cite{Shed} & 2018, \small{CSREA Press} & \small{Shed More Light on Bloom Filter's Variants} &  \small{Variants of Bloom Filter} \\ \hline

Patgiri \textit{et al.} \cite{DDoS} & 2018, EAI & \small{Preventing DDoS using Bloom Filter: {A} Survey} &  \small{DDoS} \\ \hline

Luo \textit{et al.} \cite{surv1} & 2019, IEEE & \small{Optimizing Bloom Filter: Challenges, Solutions, and Comparisons} &  \small{Mathematical analysis of various aspects of Bloom Filter}  \\ \hline

Patgiri \textit{et al.} \cite{patgiri1} & 2019, Hindawi  & \vspace{-\baselineskip}\small{Hunting the pertinency of bloom filter in computer networking and beyond: A survey} &  \small{Computer Network (Wireless Sensor, P2P), Security (Network, Big Data, Biometric), IoT Environment, Metadata Server} \\ \hline

Nayak \textit{et al.} \cite{access} & 2019, IEEE & \small{A Review on Role of Bloom Filter on DNA Assembly} &  \small{Variants of Bloom Filter, DNA Assembly} \\ \hline
 
Patgiri \textit{et al.} \cite{BD} & 2019, \small{TheSAI.org}  & \small{Role of Bloom Filter in Big Data Research: A Survey} &  \small{Data Storage} \\ \hline
    
    \hline
    \end{tabular}
    \label{bf_survey}
\end{table}

\begin{longtable}{|p{1.7cm}|p{1cm}|p{4.1cm}|p{4.1cm}|}
    \caption{\textbf{List of survey papers available on Named Data Network}}
    \label{ndn_survey}\\ 
     \hline
     \textbf{Authors} & \textbf{Year} & \textbf{Title of the paper} & \textbf{Domain}\\ \hline
     \endfirsthead
\hline
\centering 
\textbf{Authors} & \textbf{Year} & \textbf{Title of the paper} & \textbf{Domain}\\ \hline
\hline
\endhead
Ariefianto \textit{et al.} \cite{Syambas} & 2017, IEEE & Routing in NDN network: A survey and future perspectives &  Routing \\ \hline

Aboodi \textit{et al.} \cite{Aboodi1} & 2019, IEEE  & Survey on the Incorporation of NDN/CCN in IoT &  \small IoT, Data naming, In-network caching, Access control and policies, Routing \\ \hline

Soniya and Kumar \cite{Soniya} & 2015, IEEE & A survey on named data networking &  Forwarding plane, Adaptive forwarding \\ \hline

Zhang \textit{et al.} \cite{Zhang1} & 2016, IEEE & A survey of mobility support in Named Data Networking &  Mobility support, Producer mobility support \\ \hline

Chen and Mizero \cite{Chen2} & 2015, CoRR & A Survey on Security in Named Data Networking  &  Security and Privacy \\ \hline

Bakkouchi \textit{et al.} \cite{Bakkouchi} & 2019, IEEE & A hop-by-hop Congestion Control Mechanisms in NDN Networks -- A Survey  &  Congestion control, Hop-by-hop Congestion control \\ \hline

Li \textit{et al.} \cite{Li} & 2019, IEEE  & Packet Forwarding in Named Data Networking Requirements and Survey of Solutions &  Data structure used in FIB, content store and PIT \\ \hline

Aloulou \textit{et al.} \cite{NDN-FS} & 2017, IEEE  & Taxonomy and comparative study of NDN forwarding strategies &  Taxonomy of NDN Forwarding strategies \\ \hline

Tariq \textit{et al.} \cite{NDN-FS1} & 2019, IEEE  & Forwarding Strategies in NDN based Wireless Networks: A Survey &  Forwarding In Ad Hoc (wireless, mobile, vehicular), Forwarding Based on Wireless Sensor and Wireless Mesh \\ \hline

Feng \textit{et al.} \cite{Feng1} & 2016, Springer  & Mobility support in Named Data Networking: a survey  & Mobility support schemes \\ \hline

Khelifi \textit{et al.} \cite{Khelifi} & 2019, IEEE & Named Data Networking in Vehicular Ad hoc Networks: State-of-the-Art and Challenges &  Vanet based (Naming Schemes, Routing, Data Discovery, Forwarding And Content Dissemination, Caching Schemes, Mobility, Security And privacy solutions, Issues) \\ \hline

Chen \textit{et al.} \cite{Chen3} & 2016, IEEE & Transport Control Strategies in Named Data Networking: A Survey &  Comparisons of transport control between NDN and IP-based Internet, Recent proposals for NDN Transport control, Challenges \\ \hline

Rahel \textit{et al.} \cite{Rahel} & 2017, IEEE & Energy-efficient on caching in named data networking: A survey  &  System Services, Energy Efficiency  \\ \hline

Chatterjee \textit{et al.} \cite{NDN-sec} & 2018, IEEE & Security Issues in Named Data Networks &  Types of security attack \\ \hline

Kumar \textit{et al.} \cite{kumar1} & 2019, Springer & Security Attacks in Named Data Networking: A Review and Research Directions & 
Types of security attack \\ \hline

Saxena \textit{et al.} \cite{NDN} & 2016, Elsevier  & Named data networking: a survey &  System services, Applications, Issues \\ \hline

Ren \textit{et al.} \cite{ren2016} & 2016, Elsevier & Congestion control in named data networking - a survey &  Congestion control issues, Mechanism of congestion control, Issues \\ \hline

Ahed \textit{et al.} \cite{Ahed} & 2019, IEEE & Content Delivery in Named Data Networking based Internet of Things &  IoT, IoT application, Content Forwarding In NDN-based IoT \\ \hline

Hussaini \textit{et al.} \cite{hussaini} & 2018, IAES & Producer mobility support schemes for named data networking: A survey &  Producer Mobility Support \\ \hline

Yovita \textit{et al.} \cite{yovita} & 2018, IAES & Caching on Named Data Network: a Survey and Future Research &  Caching \\ \hline
    
\end{longtable} 

\subsection{Organization of the article}
The article is organized as follows- initially, Section \ref{BF} provides a brief tutorial on Bloom Filter to understand Section \ref{VBF} which exposes various variants of Bloom Filter developed for NDN. Again, Section \ref{NDN} presents a brief tutorial on NDN and its architecture. Also, Section \ref{PKT} establishes the relation between packet and Bloom Filter to understand the rest section. The rest sections present a rigorous survey on NDN using Bloom Filter. Thus, the survey begins from Section \ref{CS} which illustrates the impact of Bloom Filter on Content Store. Furthermore, Section \ref{PIT} exploits Bloom Filter in the capability of enhancement of Pending Interest Table. Similarly, Section \ref{FIB} exposes the necessity of Bloom Filter in implementing Forward Information base. Moreover, Section \ref{OTH} emphasizes on implementation of Bloom Filter in other components of NDN. Section \ref{SEC} reviews the security requirements of NDN. Section \ref{trend} discuss some future trends of NDN along  with some techniques implementing Bloom Filter. Section \ref{dis} discusses on effects of Bloom Filter in NDN implementation, and also, discusses lesson learnt from the survey. Finally, the article draws a suitable conclusion in Section \ref{con}.


\section{Review plan}
The research papers/articles are collected from leading indexing engines, namely, IEEE Xplore, ACM Digital Library, Springer Link and ScienceDirect. We have also searched research papers/articles from SCOPUS and Web of Science. The query terms are \{"Named Data Network"\}, \{"Bloom Filter"\}, and \{"Named Data Network" AND "Bloom Filter"\}. We have excluded Google Scholar for this review. This paper presents the review of query \{"Named Data Network" AND "Bloom Filter"\} performed on the indexing engines.

\subsection{Study of NDN Trend}

\begin{figure}[!ht]
    \centering
    \includegraphics[width=0.8\textwidth]{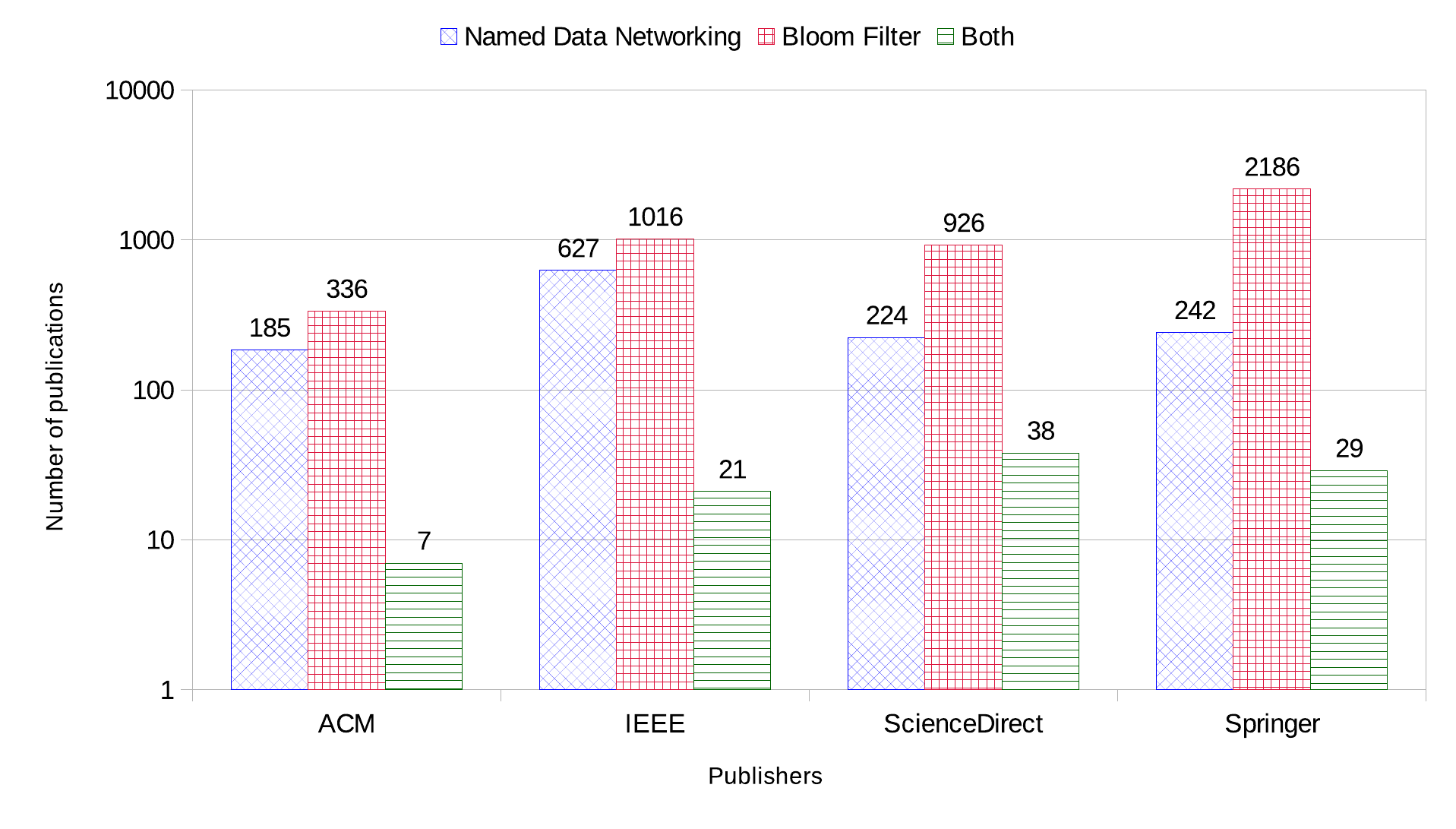}
    \caption{\textbf{Number of publications in various publishers.}}
    \label{pub}
\end{figure}

Figure \ref{pub} illustrates the number of publications published by a prominent publisher like IEEE (ieeexplore.ieee.org), ACM (dl.acm.org), Springer (link.springer.com), and Elsevier (sciencedirect.com). The publication of NDN starts from 2010 while the publication of Bloom Filter starts from 1970. In our search, we have concluded that a few papers have been published for both the term "named data networking" and "Bloom filter". However, IEEE publisher has published 627 papers of "named data networking", and Springer has published 2186 papers of "Bloom Filters" which is the highest. Interestingly, ACM publishes 7, IEEE publishes 21, ScienceDirect publishes 38 and Springer publishes 29 papers for both the terms "Named Data Networking" and "Bloom filter".

\begin{figure}[!ht]
    \centering
    \includegraphics[width=0.8\textwidth]{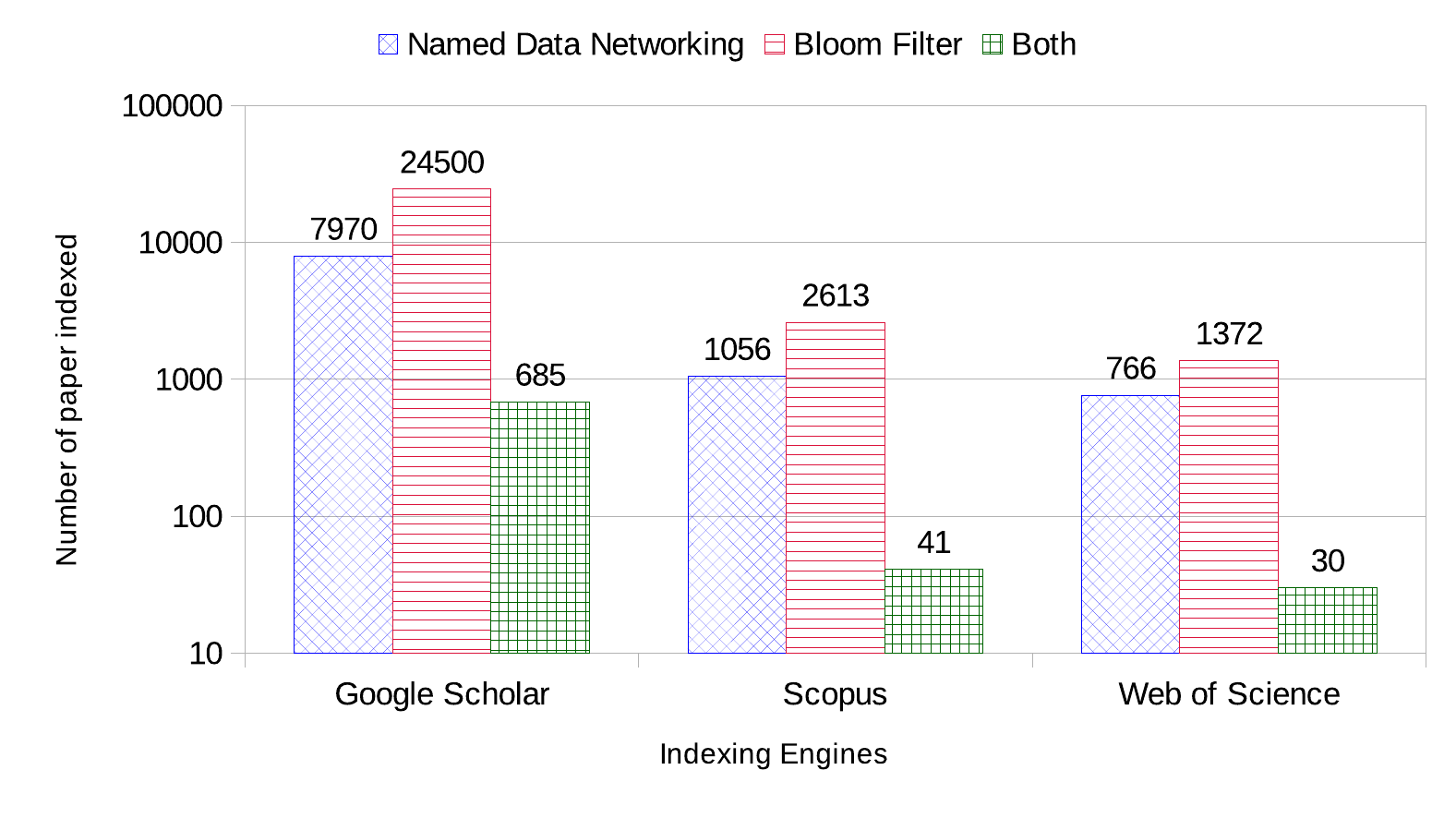}
    \caption{\textbf{Number of paper indexed by various indexing engines.}}
    \label{index}
\end{figure}

Nowadays, Google Scholar is the most popular search engine to discover research papers. However, quality paper indexing is not assured in Google Scholar. On the contrary, Scopus and Web of Science, both are popular search and indexing engine for high quality papers. Surprisingly, Scopus and Web of Science have indexed 41 papers and 30 papers that contain both terms "Named Data Networking" and "Bloom filter". Google Scholar indexed 685 papers in the said terms.
 
\begin{figure}[!ht]
    \centering
    \includegraphics[width=0.45\textwidth]{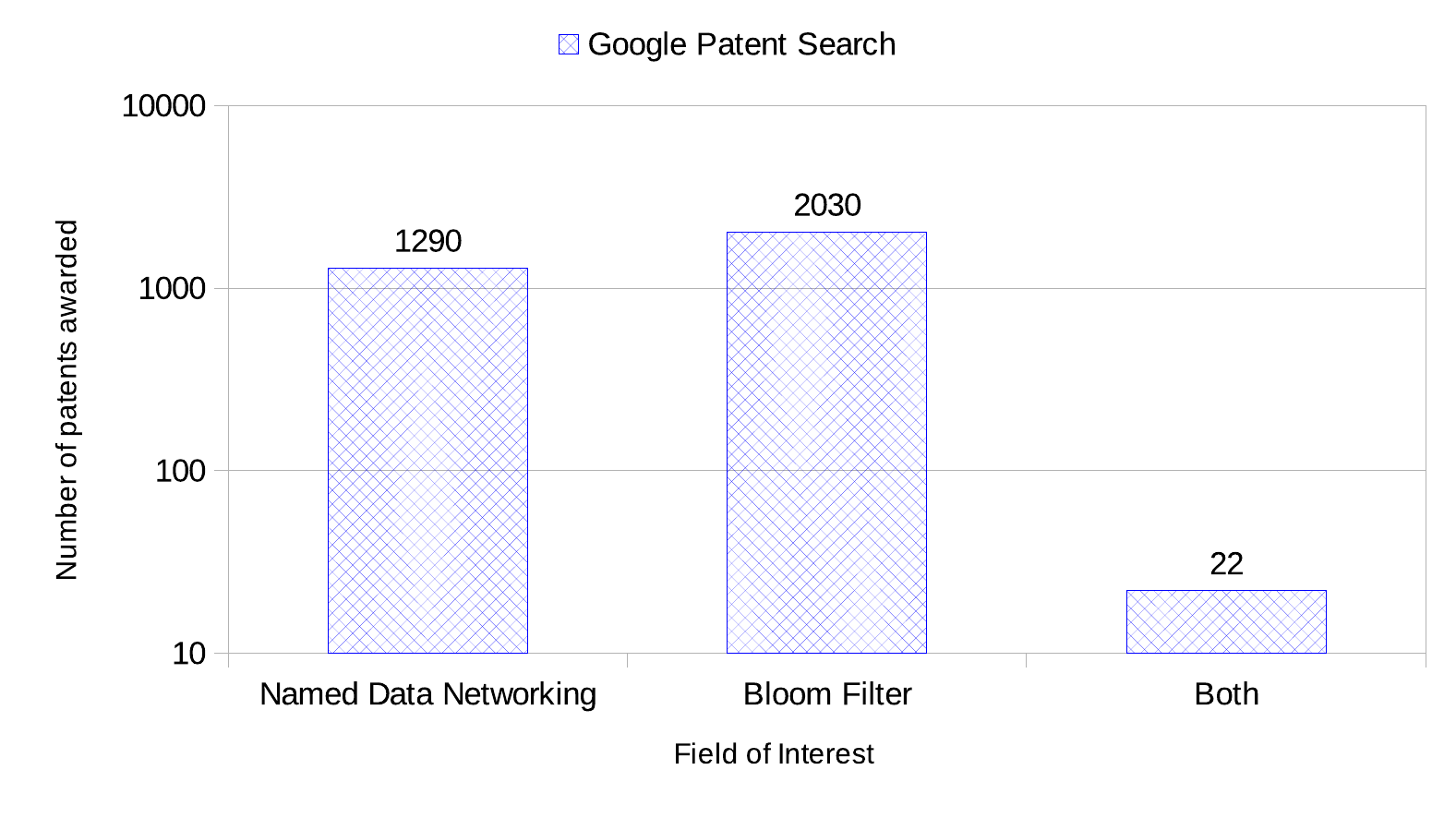}
    \caption{\textbf{Number of patent awarded by various organizations.}}
    \label{patent}
\end{figure}

The patent is an intellectual property right that preserves the copyright of any intellectual creations. This is also important to discuss to understand the current scenario of Bloom Filter and NDN. NDN already has 1290 patents awarded since its inception (2010) while Bloom Filter has 2030 patents awarded from 2010. With the increasing popularity of NDN, Bloom Filter is also intensively experimented to deploy.

\begin{figure}[!ht]
    \centering
    \includegraphics[width=0.9\textwidth]{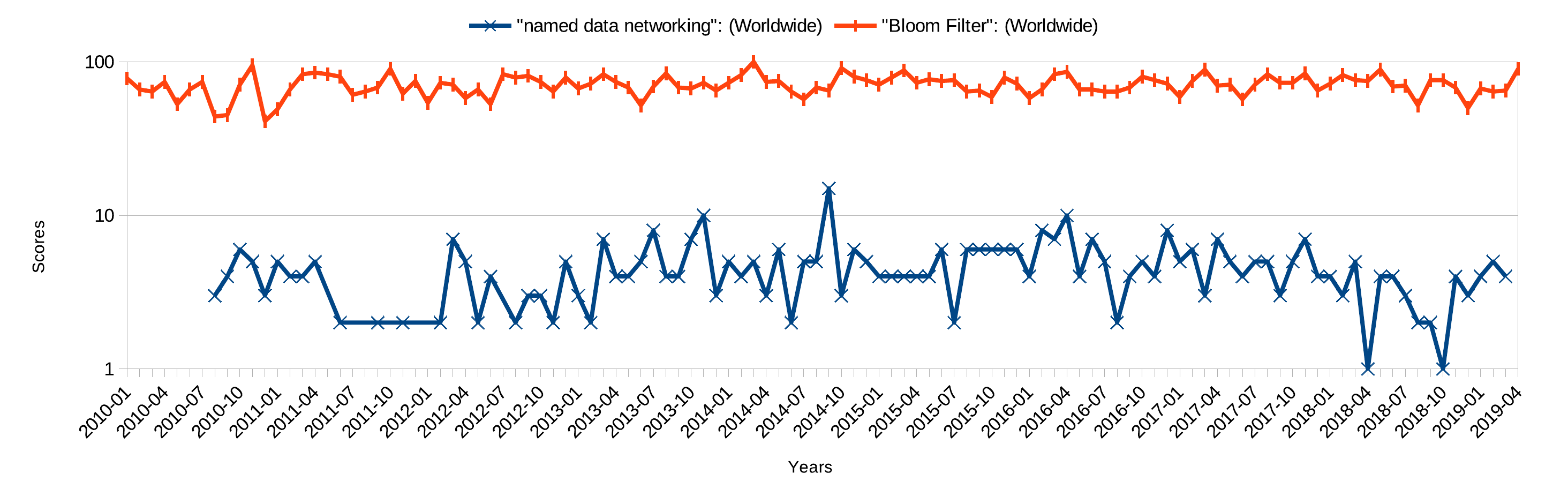}
    \caption{\textbf{Popularity of Bloom Filter and Named Data Networking in Google trends.}}
    \label{trends}
\end{figure}

Current trends of Bloom Filter are higher than NDN, because, NDN has been introduced in 2010 while Bloom Filter is introduced in 1970. The trend gap clearly depicts that a few research community is engaged with NDN and it has more miles to go.

\section{Bloom Filter}
\label{BF}
Bloom Filter is a probabilistic data structure for set membership checking \cite{Bloom}. Bloom Filter has a few issues and challenges \cite{Luo}. The key issues of Bloom Filter are false positive and false negative. Most of the modern Bloom Filter does not have a false negative. However, some variants of Bloom Filter suffer from false negative. For instance, counting Bloom Filter (CBF) \cite{Li_Fan}. On the other hand, the Bloom Filter has many challenges to achieve, namely, high scalability, high accuracy, high lookup performance and low memory consumption.

\subsection{Architecture}

\begin{figure}[!ht]
    \centering
    \includegraphics[width=0.8\textwidth]{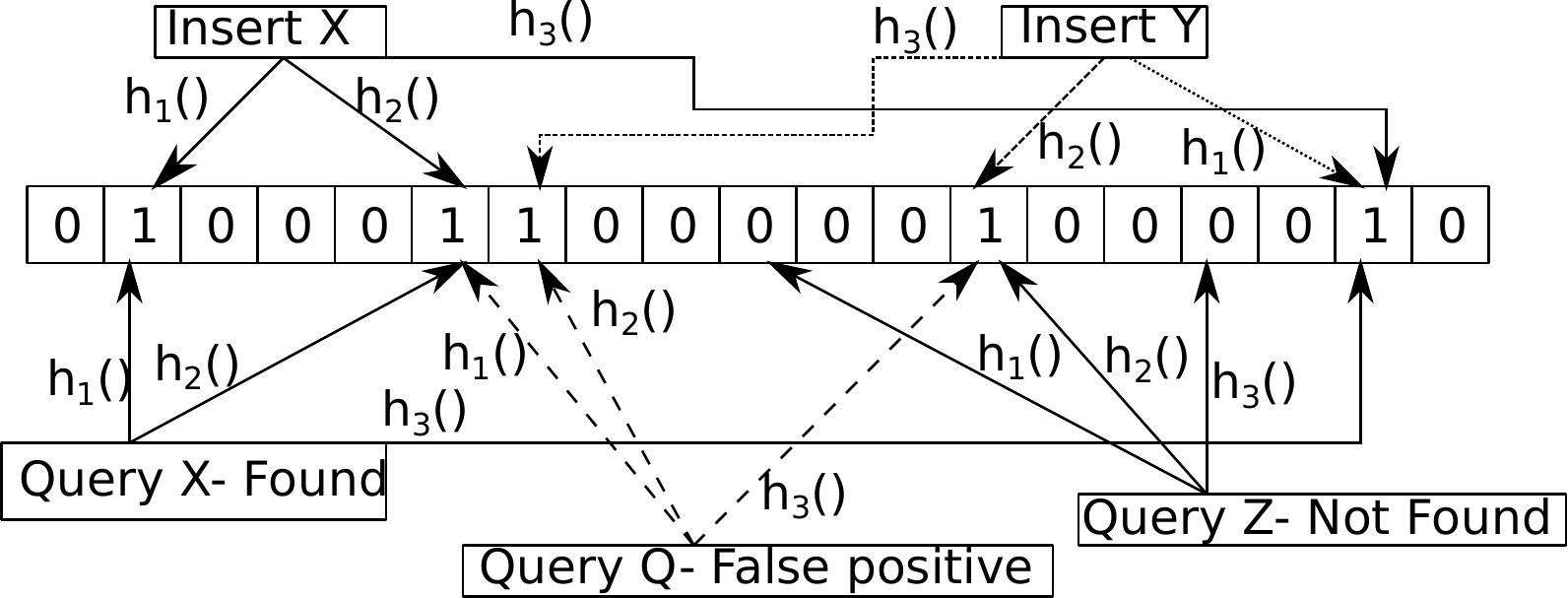}
    \caption{\textbf{Architecture of Bloom Filter. Figure depicts insertion and query when $k=3$. Source \cite{access}}}
    \label{bf_arch}
\end{figure}

In Bloom Filter \cite{Bloom}, the input items are hashed into bitmap array \cite{bitmap1}. Unlike a conventional hash data structure, Bloom Filter uses a bitmap array instead of object array (for instance, int, char, float etc.) and it permits errors in lookups. The bitmap array consists of bits \cite{bitmap1} and the input items are hashed into these bits. Initially, bitmap array is set to zero. $m$ is the size of the Bloom array. $m$ is a prime number, otherwise, collision probability increases. The Bloom Filter implements multiple hash functions to map the input item to bit locations. $k$ is the number of hash functions. The Less number of hash functions cause more collision probability. On the contrary, the number of hash functions trigger invoking many operations which affect the performance adversely. This trade-off is removed by calculating the optimal number of hash functions.

Bloom Filter returns response in $True$ or $False$. $True$ indicates the item is present. $False$ indicates the item is absent. The $True$ response is further classified into $True~positive$ and $False~positive$. Similarly, $False$ response is further classified into $True~Negative$ and $False~Negative$.
Let, \texorpdfstring{$\mathbb{B}$} be a Bloom Filter, $\mathbb{S}$ be a set whose elements are inserted into \texorpdfstring{$\mathbb{B}$}, $\kappa$ be the input item, then 
\begin{itemize}
    \item \textbf{True Positive:} If $\kappa\in \mathbb{S}$, and $\kappa\in \mathbb{B}$, then the result of \texorpdfstring{$\mathbb{B}$} is true positive.
    \item \textbf{False Positive:} If $\kappa\not\in \mathbb{S}$, but $\kappa\in \mathbb{B}$, then the result of \texorpdfstring{$\mathbb{B}$} is false positive. 
    \item \textbf{True Negative:} If $\kappa\not\in \mathbb{S}$, and $\kappa\not\in \mathbb{B}$, then the result of \texorpdfstring{$\mathbb{B}$} is true negative. 
    \item \textbf{False Negative:} If $\kappa\in \mathbb{S}$, but $\kappa\not\in \mathbb{B}$, then the result of \texorpdfstring{$\mathbb{B}$} is false negative.

\end{itemize}
\subsection{Operations}
\begin{figure}[!ht]
\centering 
\newcolumntype{C}[1]{>{\centering}p{#1}}
\begin{forest}
  	for tree={
  		if level=0{align=center}{
    		align={@{}C{20mm}@{}},
  		},
  		grow=east,
  		draw,
  		font=\sffamily\bfseries,
  		edge path={
    		\noexpand\path [draw, \forestoption{edge}] (!u.parent anchor) -- +(4mm,0) |- (.child  anchor)\forestoption{edge label};
  		},
  		parent anchor=east,
  		child anchor=west,
  		l sep=10mm
  }
  [Operations
    [Deletion]
    [Lookup]
    [Insertion]
  ]
\end{forest}
\caption{\textbf{Operations of Bloom Filter}}
\label{taxo1}
\end{figure}
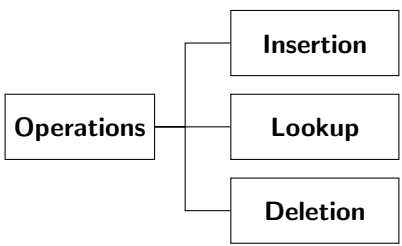

Bloom Filter supports three key operations, namely, insert, lookup, and delete.
\subsubsection{Insertion}
Bloom Filter stores bit information about an input item in bitmap array. Any input item is hashed into a particular slot of the bitmap array and the slot is set to `1' in insertion operation. As shown in Figure \ref{bf_arch}, input item $X$ is hashed and three hash functions (i.e $h_1()$, $h_2()$, and $h_3()$) generates three bit locations. These locations are set to 1. Similarly, the input item $Y$ is inserted. 

Any hash function can be used to insert an item in Bloom Filter, namely, CRC32, MD5, FNV, DJB and other hash functions. The Murmur hash function is the fastest hash algorithm among the string hash functions including FNV1, FNV1a, CRC32, DJB, DJB2a, SuperFastHash and xxHash. The Murmur hash function has three version Murmur, Murmur2 and Murmur3. Algorithm \ref{Algo1} uses Murmur hash function to hash the input items in the Bloom Filter. The input item $Key$, string length $length$ and seeds $S_1,~S_2,~and~S_3$ are passed to Murmur hash function. The seed values are prime numbers. The Murmur hash function returns a ten digit number. The ten digit number is hashed using $m$ where $m$ is the size of Bloom array. The performance of Bloom Filter depends on the Murmur hash function and the number of modulus operations in Algorithm \ref{Algo1}. 

The insertion operation performance depends on the number of hash functions. Therefore, the time complexity of insertion operation is $O(k)$ where $k$ = number of hash functions. $k$ is constant, therefore the time complexity of insertion operation is $O(1)$.

\begin{algorithm}
\caption{Insertion of an input item $Key$ into Bloom Filter $\mathbb{B}$ using three hash functions. The seed values $Seed~S_1,~S_2,~S_3$ are used to create different hash functions.}
\begin{algorithmic}[1]
\Procedure{Insert}{$\mathbb{B}[],~Key,~S_1,~S_2,~S_3$}
    \State $v_1=\Call{Murmur}{Key,~length,~S_1}\%m$ \Comment{$m$ is the size of $\mathbb{B}[]$ which is a prime number.}
    \State $v_2=\Call{Murmur}{Key,~length,~S_2}\%m$ \Comment{\% is modulus operator}
    \State $v_3=\Call{Murmur}{Key,~length,~S_3}\%m$
    \State $\mathbb{B}[v_1]\leftarrow 1$ \Comment{Key is hashed into Bloom Filter array $\mathbb{B}[]$ to insert the input item $Key$}
    \State $\mathbb{B}[v_2]\leftarrow 1$
    \State $\mathbb{B}[v_3]\leftarrow 1$
\EndProcedure
\end{algorithmic}
\label{Algo1}
\end{algorithm}

\subsubsection{Lookup}
Bloom Filter is most popular for its deduplication. Both in insertion and deletion operations first the item is searched in the Bloom Filter. Therefore, lookup operation is the most invoked operation in Bloom Filter. As shown in Figure \ref{bf_arch}, lookup operation is invoked for item $X$. Similar to insertion operation, $X$ is hashed by $k$ hash functions. The bit value of the location generated by the hash function is checked. In case all bits are 1, then $True$ is returned. In case at least one bit is 0, then $False$ is returned.

Algorithm \ref{Algo2} is Lookup operation invoking three Murmur hash functions. The AND-ing operations are performed to examine whether the item $Key$ is a member of the Bloom Filter or not, i.e., $\mathbb{B}[v_1]~AND~\mathbb{B}[v_2]~AND~\mathbb{B}[v_3]$. If any slot contains $0$ values, then the item is not a member of the Bloom Filter and the Bloom Filter returns negative.

Similar to insertion operation, the time complexity of lookup operation is $O(k)\approx O(1)$.

\begin{algorithm}
\caption{Lookup an item $Key$ in Bloom Filter using three hash functions.}
\begin{algorithmic}[1]
\Procedure{Lookup}{$\mathbb{B}[],~Key,~S_1,~S_2,~S_3$}
    \State $v_1=\Call{Murmur}{Key,~length,~S_1}\%m$ 
    \State $v_2=\Call{Murmur}{Key,~length,~S_2}\%m$ 
    \State $v_3=\Call{Murmur}{Key,~length,~S_3}\%m$
    
    \If{$\mathbb{B}[v_1]~AND~\mathbb{B}[v_2]~AND~\mathbb{B}[v_3]$} \Comment{AND-ing operation is perform to check the bit values of $\mathbb{B}[]$.}
        \State $\Return~True$
    \Else
        \State $\Return~False$
    \EndIf
\EndProcedure
\end{algorithmic}
\label{Algo2}
\end{algorithm}

\subsubsection{Deletion}
Algorithms \ref{Algo3} performs deletion operation on non-counting Bloom Filter. Before deleting an item, lookup operation is performed. If the item exists in the Bloom Filter, then reset the bit values to $0$. The bit locations are obtained as discussed in insertion and lookup operations. Bloom Filter may reset the bit location to `0' in case of an absent key, if the lookup operation is not performed before the deletion operation. To elaborate, let us assume $\mathbb{B}[v_1]=1$, $\mathbb{B}[v_2]=0$ and $\mathbb{B}[v_3]=1$. In this case, the $\mathbb{B}[v_1]$ and $\mathbb{B}[v_3]$ are reset to zero unnecessarily, albeit, the item was not actually present in the Bloom Filer. This causes a false negative issue. Thus, the lookup operation is performed before removal of an item from Bloom Filter. However, this process does not ensure free from false negatives due to collision probability in deletion operation. For instance, let us assume $\mathbb{B}[v_1]=1$, $\mathbb{B}[v_2]=1$ and $\mathbb{B}[v_3]=1$, and the item was not actually inserted into Bloom Filter. The slots are exhibiting value `1' due to the false positives. In deletion operation, the slots are reset to `0' assuming that the item was inserted into the Bloom Filter while it is not the case. Thus, conventional Bloom Filter exhibits false negative if and only if it supports the deletion operation. Therefore, deletion operation is not permitted in the conventional Bloom Filter. To permit the deletion operation, counting Bloom Filter was introduced. The time complexity of deletion operation is similar to insertion and lookup operations, i.e., $O(1)$.

\begin{algorithm}
\caption{Deletion of an item $Key$ from Bloom Filter using three hash functions.}
\begin{algorithmic}[1]
\Procedure{Delete}{$\mathbb{B}[],~Key,~S_1,~S_2,~S_3$}
    \State $v_1=\Call{Murmur}{Key,~length,~S_1}\%m$ 
    \State $v_2=\Call{Murmur}{Key,~length,~S_2}\%m$ 
    \State $v_3=\Call{Murmur}{Key,~length,~S_3}\%m$
    \If{$\mathbb{B}[v_1]~AND~\mathbb{B}[v_2]~AND~\mathbb{B}[v_3]$}
        \State $\mathbb{B}[v_1]\leftarrow 0$ \Comment{Key is hashed into Bloom Filter array $\mathbb{B}[]$ to reset to $0$ for the input item $Key$}
        \State $\mathbb{B}[v_2]\leftarrow 0$
        \State $\mathbb{B}[v_3]\leftarrow 0$
    \Else
        \State False 
    \EndIf
\EndProcedure
\end{algorithmic}
\label{Algo3}
\end{algorithm}

\subsection{Taxonomy of Bloom Filter}
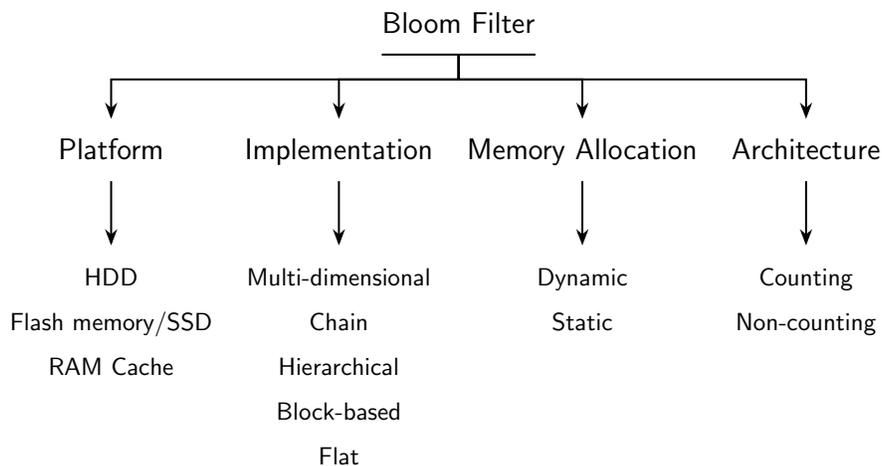
\begin{figure}[!ht]
\centering
\resizebox{\linewidth}{!}{
\newcolumntype{C}[1]{>{\centering}p{#1}}
\begin{forest}
 for tree={
    align=center,
    parent anchor=south,
    child anchor=north,
    font=\sffamily,
    edge={thick, -{Stealth[]}},
    l sep+=10pt,
    edge path={
      \noexpand\path [draw, \forestoption{edge}] (!u.parent anchor) -- +(0,-10pt) -| (.child anchor)\forestoption{edge label};
    },
    if level=0{
      inner xsep=0pt,
      tikz={\draw [thick] (.south east) -- (.south west);}
    }{}
  }
  [\large Bloom Filter
  	[\large Platform [	HDD \\ Flash memory/SSD \\RAM Cache ]]
    [\large Implementation [Multi-dimensional \\ Chain \\ Hierarchical \\ Block-based \\ Flat] ]
  	[\large Memory Allocation [Dynamic \\ Static] ]
    [\large Architecture [Counting \\ Non-counting] ]
  ]
\end{forest}
}
\caption{\textbf{Taxonomy of Bloom Filters}}
\label{taxo2}
\end{figure}

Figure \ref{taxo2} classifies the variants of Bloom Filters depending on the characteristics. Bloom Filter poses different characteristics depending on the architecture, memory allocation, implementation, and platform. For instance, a Bloom Filter can be counting or non-counting type. The next subsequent subsection illustrates about taxonomy depicted in Figure~\ref{taxo2}. 

\subsubsection{Architecture}
Bloom Filter is classified into two key categories, particularly, counting Bloom Filter, and non-counting Bloom Filter depending on the architecture. Counting Bloom Filter (CBF) \cite{Li_Fan} typically count the number of input and increment the counter upon insertion of an item. The counter is decremented upon deletion of an item. Let, $X=3$ be the value of counter for certain states. The $X$ value is decremented upon removal of an item, and thus, $X=2$. Now, if the deleted item is looked up again, then Bloom Filter returns true, since, $X>0$. Thus, the false positive become a challenge in CBF, but CBF has a low false negative probability. Furthermore, CBF offers high scalability. On the other hand, non-counting Bloom Filter set the bit to 1 during insertion of an item, and set the bit to 0 during deletion of an item. Thus, there is a false negative issue in non-counting Bloom Filter.

\subsubsection{Memory allocation}
We classify the Bloom Filter into two types depending on the memory allocation strategy, specifically, static Bloom Filter and dynamic Bloom Filter. Static Bloom Filter allocates memory once, and does not permit to increase the allocated memory size. Also, static Bloom Filter does not allow to alter the parameters of the Bloom Filter. However, another instance of static Bloom Filter can be created on demand. Thus, static Bloom Filter meets the requirement of scalability. On the contrary, dynamic Bloom Filter allocates memory as per the requirement of the current situation. Dynamic Bloom Filter grows the size over a time period and adjust automatically. However, the dynamic Bloom Filter requires adjustment of the Bloom Filter to keep the same rule of insertion, lookup, and deletion. These requirements make the designing of dynamic Bloom Filter more complex compared to static Bloom Filter.

\subsubsection{Implementation}
Based on implementation, Bloom Filter is classified into five categories, particularly, flat Bloom Filter, block-based Bloom Filter, hierarchical Bloom Filter, chained Bloom Filter, and multidimensional Bloom Filter. Flat Bloom Filter uses an array of bits. The bit array contains information about membership of a large dataset. Block-based Bloom Filter creates multiple blocks on the given bit array to reduce the collision probability. However, this is similar to flat Bloom Filter. Notably, hierarchical Bloom Filters create multiple Bloom Filters and forms treelike structure. Each node of the hierarchical Bloom Filter is embedded with conventional Bloom Filter. The hierarchical Bloom Filters greatly helpful in very large scale membership filtering. Similarly, the chained Bloom Filter creates a chain of multiple Bloom Filters to increase scalability, i.e., the chained Bloom Filter creates a chain of Bloom Filters. A new Bloom Filter is linked to chained Bloom Filter to increase the scalability. On the contrary, multidimensional Bloom Filter is implemented using multidimensional Bloom Filter array \cite{rDBF}. For instance, 3D array. Multidimensional Bloom Filter is similar to flat Bloom Filter except it has more dimension. 

\subsubsection{Platform}
Finally, we categorize the Bloom Filter into four key categories based on platform, namely, cache, RAM, Flash/SSD, and HDD. Cache-based Bloom Filter is designed to improve the caching performance. However, caching can be implemented in cache memory, and main memory. The block size of Bloom Filter is ought to be small such that the block size should fit in cache memory. On the other hand, the RAM-based Bloom Filters are most popular. Bloom Filters are stored in RAM. Flash/SSD is used to increase the scalability and fault tolerance. This type of Bloom Filter adapts lazy write policy, whereas any update does not immediately cause write into Flash/SSD. Immediate update cause performance issue because the update operation writes 3-8 bits per update into Flash/SSD. Thus, Flash/SSD based Bloom Filter adapts lazy write-back policy. Similar to Flash/SSD based Bloom Filter, the HDD is used to backup the Bloom Filter to implement persistence Bloom Filter. If the system shuts down, then Bloom Filter can be reconstructed from the backup which is stored in HDD.

\subsection{Key objective}
\begin{figure}[!ht]
\centering
\scalebox{0.8}{
\newcolumntype{C}[1]{>{\centering}p{#1}}
\begin{forest}
  	for tree={
  		if level=0{align=center}{
    		align={@{}C{40mm}@{}},
  		},
  		grow=east,
  		draw,
  		font=\sffamily\bfseries,
  		edge path={
    		\noexpand\path [draw, \forestoption{edge}] (!u.parent anchor) -- +(4mm,0) |- (.child anchor)\forestoption{edge label};
  		},
  		parent anchor=east,
  		child anchor=west,
  		l sep=10mm
  }
  [Objective
    [Maximize Performance]
    [Increase Scalability]
    [Reduce False negative]
    [Reduce False positive]
  ]
\end{forest}
}
\caption{\textbf{Key objectives of Bloom Filters}}
\label{taxo3}
\end{figure}
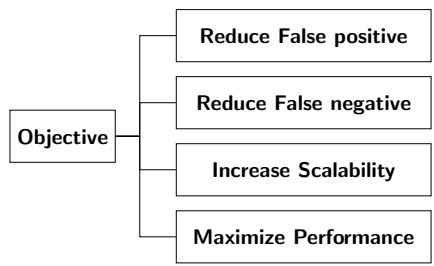
The prime objectives of developing new variants of Bloom Filter are \textit{maximize performance}, \textit{increase scalability}, \textit{reduce false negative} and \textit{reduce false positive}. False positive and false negative are the key issues of Bloom Filter. Therefore, there are numerous variants of Bloom Filter to reduce both issues. However, performance of Bloom Filter is effected while reducing those issues.  Moreover, increasing scalability is also a challenge. Because, the  applications of Bloom Filter demand large scale Bloom Filter, for instance, deduplication. There is a trade-off between false positive and performance. A few Bloom Filters sacrifice performance to reduce false positive. For instance, Cuckoo Filter \cite{Cuckoo}. Also, sometime the space consumption increases to reduce false positive.

\subsection{Issues and Challenges}
 \label{issue}
The simplicity and performance of Bloom Filter positively impact the research community and developers. However, there are still some issues and challenges in the Bloom Filter.

\subsubsection{False Positive and false negative}
False positive and false negative are a key issue in Bloom Filter data structure. False positive negatively impacts the Bloom Filter. A false positive is an overhead to Bloom Filter. It increments the procedural steps by unnecessary searching in the application. Hence, the key challenge is to reduce the false positive probability. The counting Bloom Filter exhibits high false positive probability while it lowers the false negatives. Similarly, the non-counting Bloom Filter lowers the false positive probability while it exhibits high false negative probability provided if it allows the deletion operation. Therefore, non-counting Bloom Filter does not support deletion operation to prevent the false negative. On the other hand, false positive introduces errors in duplicate key filtering. False positive filter out the wrong elements due to collision. On the other hand, false negative also introduces error in some applications. Therefore, it's a grand challenge to reduce both the false positive and false negative probability of Bloom Filter.

The false positive of Bloom Filter can be derived from $m$, $n$ and $k$ where $m$ is the size of memory, $n$ is  the number of input items and $k$ is  the number of hash functions, the probability of a particular slot is not set to $1$ by a particular hash function is
\[\left(1-\frac{1}{m}\right)\]
Let us assume, $k$ hash functions are used by the Bloom Filter and the above probability is 
\[\left(1-\frac{1}{m}\right)^k\]
If $n$ elements are inserted into the Bloom Filter and the above probability is \[\left(1-\frac{1}{m}\right)^{nk}\]
The probability of that particular bit to be $1$ is
\begin{equation}\label{eq1}
    p=1-\left(1-\frac{1}{m}\right)^{nk}\approx \left(1-e^{-\frac{kn}{m}}\right)
\end{equation}
From Equation \eqref{eq1}, the required memory can be calculated by the optimal number of hash functions.

\subsubsection{Scalability}
Scalability is also a key issue in the Bloom Filter. Conventional Bloom Filter is non-scalable. Large number of entries causes more occurrences of false positive. Let the set be \texorpdfstring{$\mathcal{S}$} and the total number of elements in the set be $n$. Elements are $\varepsilon_1,~\varepsilon_2,~\varepsilon_3,~\ldots~\varepsilon_n$. During insertion, the element is mapped to $k$ slots of the Bloom Filter. If the value of $n$ increases, then the number of slots set to 1 increases. This leads to increase in the probability of false positive. Therefore, high scalability is a prominent challenge to achieve.

\subsubsection{Space complexity}
Bloom Filter requires extra space from primary memory to hold the information of the elements. Different kinds of Bloom Filter have different space requirements. For example, CBF has an extra information in each slot i.e. the counter. The counter also requires some memory. Therefore, the CBF \cite{Li_Fan} consumes more memory than other variants. Thus, low space consumption is another challenge to achieve. The space requirement is calculated by \[m=-\frac{n~ln p}{(ln2)^2} \] where $n$ is the number of input items and $p$ is the desired false positive probability. 

\subsubsection{Deletion}
The conventional Bloom Filter does not support deletion. Nowadays, most of the Bloom Filters support deletion operation. However, it is a great challenge to design an efficient deletion algorithm where false negative is also reduced. If a Bloom Filter variant does not support deletion, then the Bloom Filter array sets all slots to 1 over a time period. Thus, the false positive probability leads to 100\%. Therefore, an efficient deletion algorithm for the Bloom Filter is essential. CBF implements deletion operation where deletion does not cause a false negative. However, CBF suffers from high false positive probability.

\subsubsection{Hashing}
Bloom Filter is a hashing data structure that uses a complex hashing technique which is introduced to reduce the false positive probability, for example, Murmur hashing \cite{murmur}, Cuckoo hashing \cite{CH}, SHA-1 \cite{SHA1}, SHA-2 \cite{SHA2}, SHA-3 \cite{SHA3}, MD5, and other cryptographic hashing. However, nested use of such complex hashing scheme introduces computational overhead. The challenge is to have an optimal number of hash functions while maintaining the desired false positive probability. 

\subsubsection{Dependency on number of hash functions}
Bloom Filter depends on the number of hash functions to place an item in various locations of Bloom Filter. Bloom Filter requires an optimal number of hash functions to reduce the false positive probability and improve performance. A large number of hash functions place an item in various places which consumes more bits. Consequently, false positive increases. The Less number of hash functions increases false positive. However, reduction of hash functions improves the performance of Bloom Filter. The optimal number of hash functions $k$ is \[ k= \frac{m}{n} ln2 \] where $m$ is memory size and $n$ is the number of inputs.

\subsubsection{Dependency on input item size}
Bloom Filter consumes memory based on input item sizes. Conventional Bloom Filter does not depend on the input item size. However, fingerprint-based Bloom Filter depends on the input item size. This is another issue in fingerprint-based Bloom Filter.

\subsubsection{Performance}
Bloom Filter is a performance enhancer. Therefore, the performance of Bloom Filter impacts the system. Hierarchical Bloom Filter is able to scale massive input items, however, the performance is not satisfactory. For instance, BloomFlash \cite{BloomFlash}. On the other hand, scaleBF is able to filter the massive amounts of input items \cite{scaleBF}.

\section{Variants of Bloom Filter}
\label{VBF}
Subsequent years saw the development of various variants of Bloom Filters. The main motive to develop various variants despite of good performance of the conventional Bloom Filter is reduction of false positive and false negative issues. In this section, various such variants are discussed in detail. However, only those Bloom Filter variants that are used in the NDN techniques are discussed in this article. Table \ref{BF_table} highlights the advantages and disadvantages of the Bloom Filters.  

\subsection{Compressed Bloom Filter}
Compressed Bloom Filter \cite{compressed} reduces the number of broadcast bits. Moreover, it reduces computations per lookup and false positive probability. Compressed Bloom Filter is used as a proxy. In compressed Bloom Filter, first the Bloom Filter is compressed and transmitted to the destination node. After receiving the compressed Bloom Filter at the destination node, it is decompressed to the original array. Bloom Filter is partitioned into parts and each part is compressed separately. At destination node required parts are decompressed. This reduces the decompression computation time. Compressed Bloom Filter implements a less number of hash functions. Hence, requires less computation per lookup. Compressed Bloom Filters produce less traffic. However, compression and decompression processing are overheads. Moreover, the processing consumes large memory at the endpoint machines.  

\subsection{Stateful Bloom Filter}
Approximate Concurrent State Machines (ACSMs) \cite{Bonomi} helps to compactly present concurrent state machines. Concurrent state machines help to track the state of a huge number of flows simultaneously. In Stateful Bloom Filter (SBF), a slot consists of a cell value and a counter. The cell value saves the flow ID and state of the flow. A state have value [1,v] or $null$. Along with false positive and false negative, SCF returns another error called $don’t~know$ (DK). DK occurs when in a slot the cell value have null, but counter have value 2 or more. This means many flows are hashed to that slot having different states. Initially, the cell value is assigned null and counter as 0. During an insert operation, first flow ID is hashed by $k$ hash functions to obtain $k$ slots. In case counter=0, then cell value is as assigned by a new value and counter is incremented. In case, the cell value is DK or equal to flow ID, the counter is incremented. When cell value $\neq$ flow ID, increment the counter, but cell value is assigned by DK. During modify operation, first $k$ slots are found. In case, the cell value is DK, then ignore. When counter=1, modify the cell value. If counter>1, cell value is assigned to DK. During delete operation, in case counter=1, then cell value=0. If counter is at least 1, then the counter is decremented and cell value is unchanged. During a query, first $k$ slots are found. All slots are checked. In case, all cell values are DK, then return DK as a response. When at least one value is not DK, then return that value. If the slots have different values, then the requested flow is absent. SBF has less space complexity compared to CBF \cite{Li_Fan}. SBF is applicable for application-aware network devices. However, the application has to tolerate some false positive response. Another issue is that if improperly terminated flows are saved in Bloom Filter, then Bloom Filter gives more DK returns. Moreover, if DK error is ignored, then it increases the false negative probability. Also, over a time period, most of the cells are filled with DK values.

\subsection{Stable Bloom Filter}
Stable Bloom Filter (Stable BF)\cite{Deng} is a variant of Bloom Filter. Each element in Stable BF is about $d$ bits and called as cell. Each cell can have maximum value $Max$. Initially, the cell is assigned to a value 0. Stable BF uses $k$ uniform and independent hash functions. During insertion, first Stable BF is checked whether it is a duplicate. Then, randomly $P$ cells are decremented by 1. This step makes space for insertion of new elements. Then, each element is mapped to $k$ cells. Then, the $k$ cells are set to $Max$. The Bloom Filter is called stable because using theorem it is proved, after many iterations the probability that the cells becomes zero is a constant. The array becomes stable when the number of iterations is infinite which is practically impossible. False positive probability is $(1-(\frac{1}{1+\frac{1}{P(1/(Z-1/m))}})^{Max})^Z$ where $Z$=number of cells set to $Max$ and $m$ = array size.  The processing of each element is independent of the array size. Each element processing time is O(1). Stable BF does not have a false negative issue.

\subsection{Attenuated Bloom Filter}
Attenuated Bloom Filter (ABF) \cite{Atten} is a hierarchical Bloom Filter consists of multiple conventional Bloom Filters. The first layer consists of the information about the current node. The second layer consists of information regarding nodes, which are just one hop far away. Therefore, context type information about the nodes $(i-1)$ hops away is found in the $ith$ layer. Saturation of the Bloom Filter is avoided by attenuating bits set that are far away from the sources. Context aggregation is implemented by ABFs. Context discovery using ABFs is a method of combining reactive and proactive discovery mechanisms. Nodes exchanges Bloom Filters to collect context types information and their distribution. When there is a query, the node will first check locally for its existence. If it does not exist, the query is performed on Bloom Filter. After that it is compared to local filters. If a match is found at any layer, a query message consisting of the Bloom Filter received from the stored filter is sent to the neighbor. In case same query is received by the node, the query is dropped. Each node records the path of each query message. If an exact match is found, then a Context-Available reply is sent as a response. If no match is found, it is discarded. The rate of false positive is identical to the conventional Bloom Filters. The performance of ABF highly depends on the appropriate ratio of the query and advertisement rates, and a query range of nodes.

\subsection{Retouched Bloom Filter}
Retouched Bloom Filter (RBF) \cite{Retouched} makes conventional Bloom Filters more flexible. [It reduces selected false positives, but introduces false negatives]. It implements a bit clearing process, in which it resets individually chosen bits in the vector $v$ for 0. This removes some false positives, but it produces false negatives. In case of randomized bit clearing, false positives depend on the size of $U-A$ where $U$ is the universal set and $A$ is the element set. False negatives depend on the size of $A$. $U-A$ is much bigger than $A$, therefore resetting a bit to 0 in $v$ eliminates a large number of false positives compared to the number of false negatives generated. In case of selective clearing, those bits are focused which correspond to elements that trigger false positives. This is done using various algorithms. False positives encountered are recorded in a set. [Elements of the set are false positives that are labeled as troublesome keys]. These are eliminated from the Bloom Filter. RBFs are applied across a wide range of applications. The applications have to identify false positive instances. A false negative is more damaging compared to false positives. Hence, exchanging false negative with false positive is more damaging.  

\subsection{Invertible Bloom Filter}
Goodrich and Mitzenmacher \cite{IBF} proposed an invertible Bloom lookup table (IBLT). The element of IBLT is $key-value$. IBLT is capable of tolerating some natural errors (eg., deletion of an absent element, insertion of keys with different values simultaneously). IBLT along with standard operations (insert, delete and query) also supports listing operation. Listing operation lists all $key-value$s present in IBLT. This operation sometimes returns a partial list with an $list-incomplete$ error. IBLT consist of a lookup table. Each cell in lookup table is a single memory word. Each cell consists of three fields, namely, counter, keySum and valueSum. The counter counts the number of elements hashed to its corresponding cell. keySum stores the total sum of all the $key$s hashed to its corresponding cell. valueSum stores the total sum of all the values hashed to its corresponding cell. Initially, all cells are assigned the value 0. In insert operation of an element $key-value$, first $key$ is hashed by $k$ hash functions to obtain $k$ locations. The element is stored in $k$ locations. If the element is inserted for the first time, then assign counter = 0, keySum=$key$ and  valueSum=$value$. In case of duplicate element, the counter is incremented and keySum = $Previous_{keySum}$+ $key$, valueSum = $Previous_{valueSum}$+$value$. In delete operation, the counter is decremented and keySum = $Previous_{keySum}$- $key$, valueSum = $Previous_{valueSum}$-$value$. During query operation, if the counter is 0, then return $False$. If counter is greater than 0, then return $True$. IBLT handles natural errors by defining an extra cell field and a hash function. IBLT uses quotienting to reduce the space complexity.

\subsection{Mapping Bloom Filter}
Mapping Bloom Filter (MBF) \cite{MaPIT} is an improved Bloom Filter to reduce the space complexity on on-chip memory. MBF consists of a Packet Store and an Index Table. Index Table consists of a Bloom Filter and a Mapping Array (MA). Initially, all bits of Bloom Filter and MA are set to 0. Packet store is accessed using MA. MA gives the offset address of the Packet Store of $n$ bits length. Bloom Filter is partitioned into $n$ parts. Each part maps to one bit of MA. When an element is inserted it is hashed using $k$ hash functions. The part to which the element is hashed, that corresponding bit in MA is set to 1 and other bits remain as 0. Same procedure as convention Bloom Filter is followed to insert element into the Bloom Filter. 

\subsection{Complement Bloom Filter}
Complement Bloom Filter \cite{Lim} is a Bloom Filter variant that determines the trueness of positive responses returned by the Bloom Filter. The Bloom Filter variant uses two Bloom Filters. One is a conventional Bloom Filter and other is the complement of the Bloom Filter. After construction of the Bloom Filter, complement Bloom Filter is constructed using the Bloom Filter. During query operation, the item is checked in both Bloom Filters. In case, both Bloom Filters return $True$, a hash table is checked to verify the false positive. Using two Bloom Filters and a hash table to reduce the false positive probability is a big compromise with memory. Bloom Filter is mostly used to handle the huge number of packets. And, many are duplicate packets. Hence, the majority of the responses of the Bloom Filters are positive. Therefore, the frequency of referring to the hash table increases. 

\subsection{Compression Trie-based Bloom Filter}
Zhang \cite{CT-BF} proposed Compression Trie-based Bloom Filter (CT-BF) for filtering Interest in a content store. It is deployed on on-chip memory for faster searching process. Moreover, the compression trie accommodates a huge number of names in space constraint on-chip Bloom Filter. The names in NDN have multiple segments. These segments are separated by "$_/$". The separator is used in a content store to organize the name trie. A whole name is obtained by the conjunction of name from the root to a leaf node. Moreover, the same node children share the same prefix. In compression trie, only the leaf nodes of the names are inserted into the Bloom Filter. In CT-BF, false positive is based on the depth of the trie. During insert operation, the first segment is added to the first level and the second segment at second level. This process is repeated till all segments are inserted. During delete operation, the last segment is deleted (i.e., leaf node) first, then the second last segment. Again, this process is repeated till all the segments are deleted. Another variant of compression trie is proposed, called Adaptive Compression Trie (ACT). It dynamically adjusts the size of Bloom Filter to increase the filtering efficiency. When more mismatch occurs in any node in compression trie, that space is expanded. The expansion is the process of deleting leaf nodes and inserting its children as leaves. Longer name increases filtering efficiency. This process consumes space, hence, shrinking operation is performed to decrease unwanted space. In shrinking operation, the space where very less request is made is reduced by a new node. The children of the new node are compressed and deleted from compression trie. Construction of ACT-BF is performed in two steps. First, CT-BF is constructed. False positive and filter ratio of the Compression Trie is calculated based on space constraints. Then, CT-BF is reconstructed. When mismatch crosses a threshold value ACT-BF is constructed. This improves the filtering efficiency. When CT-BF reaches the fifth level the false positive probability increases. Hence, the depth of CT-BF is not increased further than fifth level. 

\subsection{Bitmap-mapping Bloom Filter}
Bitmap-mapping Bloom Filter (B-MBF) \cite{B-MaFIB} is a variant of MBF \cite{MaPIT}. B-MBF is proposed by combining bitmap and MBF. The number of slots of MA and the bitmap is exponential. The bitmap is about $N$ dynamic memory spaces. MA value gives the element location in the bitmap. Based on the sequence of insertion of elements, the sequence number is assigned to the element. This sequence number serves as the offset address. Based on the offset address, memory for the element is assigned in off-chip memory. This memory is dynamically allocated. The memory is allocated dynamically to reduce memory consumption. B-MBF includes a CBF \cite{Li_Fan} to permit the delete operation, however, memory consumption increases. Hence, the advantage gained is lost. The false positive probability is same as MBF.

\subsection{Sum-Up Counting Bloom Filter}
Sum-Up Counting Bloom Filter (SCBF) \cite{Hou,sumBF} is a variant of CBF \cite{Li_Fan}. It contains a CBF with $k$ hash functions, Hash-based Lookup Table (LT) and a Sum-up Table (ST). Each ST slot corresponds to a LT slot. The slot value is determined by the corresponding counter in each CBF level. $k$ hash functions of CBF maps the string to $k$ bits. Then the string is encoded. Then, the smallest value slot is chosen from ST. Its index helps in locating the string in LT. ST and LT reduces the false positive probability. In case CBF returns false positive result, ST does not save the string. This results in accurate and fast searching. ST performs mapping between CBF and LT, while LT is responsible for query.

\subsection{Fast Two Dimensional Filter}
Fast Two Dimensional Filter (FTDF) \cite{Shub,Shubbar} is a fingerprint based Bloom Filter. Initially, all cells of FTDF matrix are initialized to zero. Then a quotienting technique is executed. The  hash function returns a string called fingerprint. The fingerprint is separated into two parts, namely, quotient and remainder. The quotient is the most significant bits. And,  the remainder is the least significant bits. The Bloom Filter maps every element into a single bit by finding the fingerprint and using quotient as row index and remainder as column index. FTDF has low memory consumption. It is scalable. Also, it is easy for implementation without dedicated hardware. The cell value 0 indicates the absence of the fingerprint of the item. The cell value 1, indicates the presence of the fingerprint with very high probability.  FTDF delete operation is very fast. For deletion, the content of the position is simply altered from 1 to 0. FTDF is more influenced by hard collision compared to soft collision. A soft collision occurs when two or more elements have different fingerprints, but same quotient fractions. Hard collision is when two or more elements have the same fingerprint.

\begin{longtable}{|p{1.25cm}|p{5cm}|p{5.5cm}|}
    \caption{\textbf{Comparison of Bloom Filters}}\\
    \hline
      \textbf{Name} &\textbf{Advantages} & \textbf{Disadvantages}\\  \hline 
     \endfirsthead
\hline
\centering 
 \textbf{Name} &  \textbf{Advantages} & \textbf{Disadvantages}\\  \hline 
\hline
\endhead

Compress- ed \cite{compressed}, 2002 & \small{$\bullet$ Reduces the number of broadcast bits $\newline$ $\bullet$ Bloom Filter is used as a proxy $\newline$ $\bullet$ Only required part of the compressed Bloom Filter is decompresses which reduces decompression computation time $\newline$ $\bullet$ Reduces network traffic $\newline$ $\bullet$ Reduces number of computations per lookup $\newline$ $\bullet$ Reduces false positive probability} & \small{$\bullet$ Increases processing time due to compression and decompression processes $\newline$ $\bullet$ Large memory consumption at endpoint machines $\newline$ $\bullet$ Performance of compression process depends on the array size} \\ \hline

Stateful \cite{Bonomi}, 2006 & \small{$\bullet$ State of the flow is saved in the Bloom Filter \newline $\bullet$ Space complexity is less compared to conventional CBF \newline $\bullet$ Applicable for application-aware network devices} & \small{$\bullet$ Applicable in system that tolerate some false positive \newline $\bullet$ Apart from false positive and false negative another error called $Don’t~know$ is present \newline $\bullet$ In some applications, ignoring DK error leads to an increase in false negativity \newline $\bullet$ Collision in the slot leads to DK error \newline $\bullet$ Saving improperly terminated flows in Bloom Filter returns more DK error} \\ \hline

Stable \cite{Deng}, 2006 & \small{$\bullet$ With limited storage, it has less false positive probability \newline $\bullet$ False positive probability is dependent on the number of 0s in the array \newline $\bullet$ Processing of each element is independent of array size \newline $\bullet$ Each element processing time is O(1)} & \small{$\bullet$ Decrementing random cells by 1 may set the counter to 0 which leads to increase in false positive probability \newline $\bullet$ The array becomes stable when the number of iterations is infinite which is practically impossible }\\ \hline

Attenuated \cite{Atten}, 2006 & \small{$\bullet$ Context aggregation is easily implemented \newline $\bullet$ Content discovery can be implemented \newline $\bullet$ Context information about nodes are easily found }&\small{ $\bullet$ High false positive and false negative probability due to use of conventional Bloom Filters \newline $\bullet$ Performance highly depends on the appropriate ratio of the query and advertisement rates, and a query range of nodes \newline $\bullet$ Density of network context information sources influences the performance} \\ \hline

Retouched \cite{Retouched}, 2010 & \small{ $\bullet$ Reduces false positive probability }&\small{ $\bullet$ In exchange of false positive false negatives is introduced $\newline$ $\bullet$ Application implementing RBF must  understand the instances of false positive }\\ \hline

Invertible \cite{IBF}, 2011 & \small{$\bullet$ Complete listing of tolerance towards some natural errors \newline $\bullet$ Quotienting reduces the space complexity} & \small{ $\bullet$ Number of duplicate elements are limited to a threshold value \newline $\bullet$ In some cases of listing operation, IBLT returns partial list, and an error \newline $\bullet$ Counter overflow} \\ \hline

Mapping (MBF) \cite{MaPIT}, 2014 & \small{$\bullet$ Consumes less on-chip memory \newline $\bullet$ Increases efficiency of Index table }& \small{ $\bullet$ Uses conventional Bloom Filter which has high false positive and false negative probability \newline $\bullet$ Off-chip memory is wasted} \\ \hline

Comple- ment \cite{Lim}, 2015 & \small{$\bullet$ Determines the trueness of Bloom Filter positives \newline $\bullet$ Reduces false positive probability} & \small{$\bullet$ Compromises with memory \newline $\bullet$ Increase in size of Bloom Filter increases the frequency of hash table reference \newline $\bullet$ Time complexity increases in case of huge number of packets} \\ \hline

Compress- ion Trie-based \cite{CT-BF}, 2017 & \small{$\bullet$ False positive reduces with increase in level of compression \newline $\bullet$ Update operation does not require extra space \newline $\bullet$ Cache hit ratio of content store increases }&\small{ $\bullet$ Increment in level, increases space complexity \newline $\bullet$ Determining the false positive from the Bloom Filter capacity is difficult \newline $\bullet$ Expansion and shrinking process of ACT sometimes negatively affect the query operation \newline $\bullet$ Expansion operation by ACT consumes more memory space \newline $\bullet$ CT-BF level cannot increment further than fifth level due to increase in false positive probability}\\ \hline

Bitmap-mapping \cite{B-MaFIB}, 2018 & \small{$\bullet$ Memory is dynamically allocated }&\small{ $\bullet$ False positive probability is same as MBF. Hence, no improvement \newline $\bullet$ Implements CBF to permit delete operation, but it increases memory consumption }\\ \hline

Sum-up Counting \cite{Hou}, 2018 & \small{$\bullet$ Reduces false positive probability \newline $\bullet$ Enhances data content matching rate \newline $\bullet$ Improves data content searching rate and accuracy \newline $\bullet$ Reduces searching time} & \small{$\bullet$ Consists of CBF, which has higher memory consumption than conventional Bloom Filter} \\ \hline

Fast Two Dimensional Filter \cite{Shub,Shubbar}, 2019  & \small{$\bullet$ Scalable \newline $\bullet$ Easy to implement without dedicated hardware \newline $\bullet$ Fast delete operation \newline $\bullet$ Reduces false positive probability \newline $\bullet$ Not affected by soft collision \newline $\bullet$ Occupies less memory} & \small{ $\bullet$ Influenced by hard collision} \\ \hline

\label{BF_table}
\end{longtable}

\section{Named data networking (NDN)}
\label{NDN}
The NDN is conceptualized from the Information-Centric Networks (ICN) \cite{ICN,ICN1}. Ted Nelson proposed 17 rules in his project, called Project Xanadu in 1979. These 17 rules provide the basic principles of ICN. Project Xanadu was founded in 1960. However, it took a very long time to make Project Xanadu visible. The Wired magazine criticized Project Xanadu and published an article called ``The Curse of Xanadu`` \cite{wired}. Wired magazine called the project as ``the longest-running vaporware story in the history of the computer industry``. Vaporware is mostly used in computer domain which is defined as a product (computer software or hardware) that is announced officially but not actually manufactured or cancelled officially. In 1988, an incomplete implementation was released. However, in 2014, a working version, called OpenXanadu, was released. After the declaration of Project Xanadu, Stanford proposed TRIAD project \cite{TRIAD} in the year 1999. TRIAD is defined as Translating Relaying Internetwork Architecture integrating Active Directories. In this project, the name of an object is used to route to avoid DNS (Domain name system) lookups. TRIAD concept is improved in another project called Data Oriented Network Architecture (DONA) \cite{DONA, ICN1} by UC Berkeley and ICSI in the year 2006. DONA project improves TRIAD by adding security (authenticity) and persistence to the architecture as first-class primitives. In 2009, under CCNx project PARC released an initial open source implementation with content-centric architecture. In 2010, NDN is proposed. NDN is defined as an instance of more general network research direction of ICN \cite{NDN}. 

A new era of Internet expects new things from NDN. A network layer of NDN \cite{NDN1} must be scalable. A single network consists of millions of nodes  and also dynamic in nature due to the addition and removal of mobile devices. These million nodes make millions of requests per second. Therefore, NDN needs to be capable of providing services with less traffic and congestion. NDN \cite{NDN-sec} have to provide high level security. The security includes origin authentication, integrity, and capability of evaluating the importance of routing information. In NDN, every named content is digitally signed by the sender. This helps in storing data in cache for future use. Moreover, it helps in saving bandwidth and efficient utilization of multiple network interfaces \cite{gasti}. NDN has to be resilient \cite{NDN-FS, NDN-FS1}. It constantly maintains efficient packet delivery performance. NDN follows one interest and fetches at most one data packet to maintain the balanced packet flow. Moreover, NDN balances the packet flow in every hop to prevent additional execution of congestion control techniques in between the network path \cite{Jacobson}.  

\subsection{Why NDN architecture is called an hourglass architecture?}
\begin{figure}[!ht]
\centering
\includegraphics[scale=0.35]{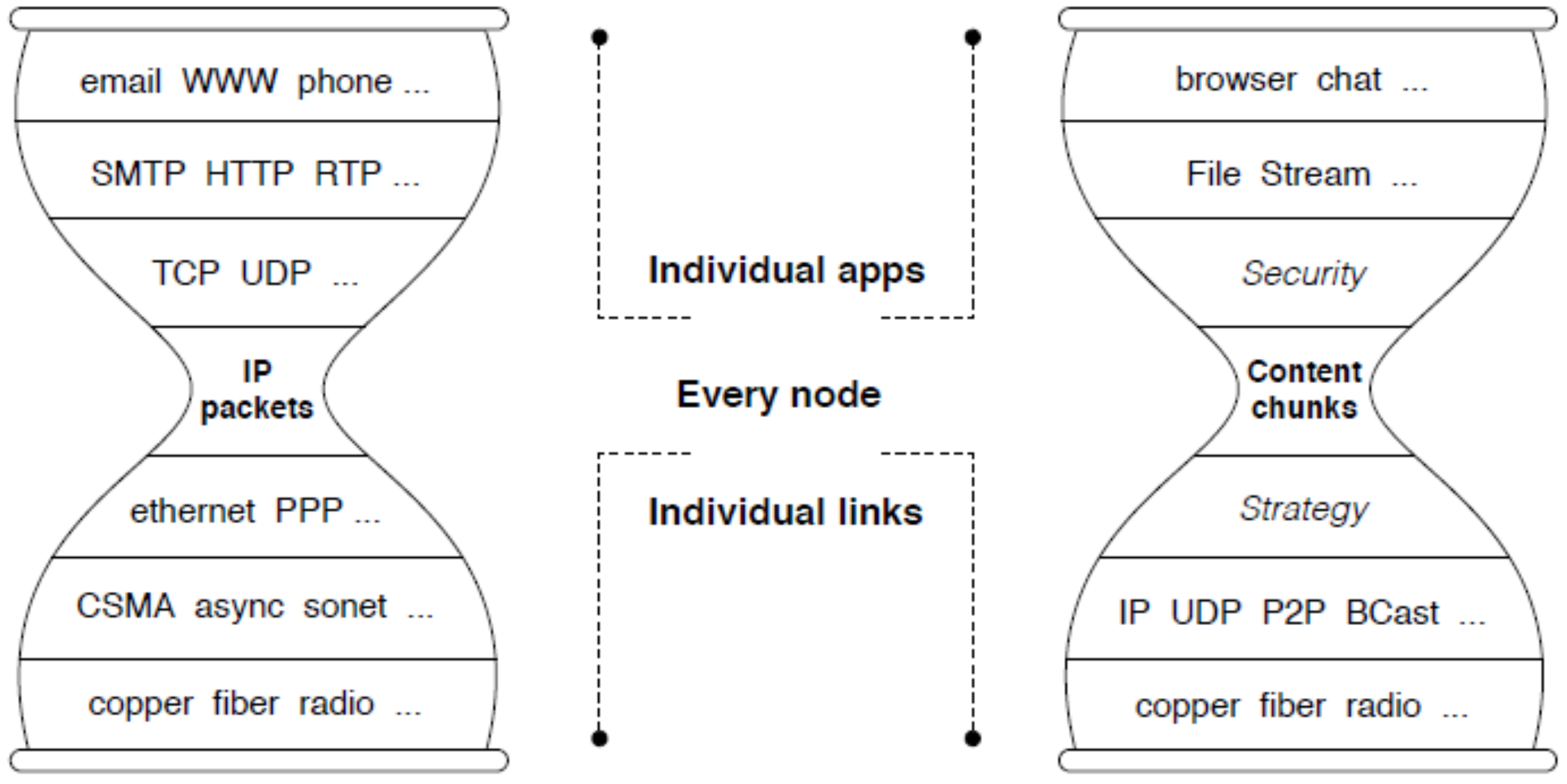}
\caption{\textbf{The main building blocks of NDN [Source \cite{Zhang}].}}
\label{NDN_arch}
\end{figure}

The answer lies with the hourglass architecture of Internet protocol (IP). The layered architecture of IP appears as an hourglass. The architecture of the Internet has seven layers (specifically, application, presentation, session,  transport, network, data-link, and physical layer). The protocols implemented in every layer are represented in Figure~\ref{NDN_arch} (left), and it takes a shape of an hourglass based on a number of protocols. Top and bottom layers have many protocols. Middle layer contains IP protocol, transport layer contains two protocols and network layer contains one protocol. Some modification is performed to the IP hourglass architecture to support NDN. As shown in Figure~\ref{NDN_arch}, the two new layers, i.e., the security and strategy are added to NDN architecture. And, the waist layer is named data/content chunk.

NDN mainly consist of three components, namely, Content Store (CS), Pending Interest Table (PIT) and Forwarding Base Table (FIB). Short elaboration on various components of the NDN is provided in their respective sections. In Table \ref{comp_table}, various features, among CS, PIT and FIB are compared.  

\begin{table}[!ht]
\centering
\caption{\textbf{Comparison of various features among CS, PIT and FIB.} I\MakeTextLowercase{n data structure feature, widely used technique is mentioned.} N/A:\MakeTextLowercase{ not applicable}}
    \begin{tabular}{|p{2cm}|p{2.7cm}|p{2.7cm}|p{2.7cm}|} 
    \hline
    \centering \textbf{Feature} & \centering \textbf{CS} & \centering \textbf{PIT} & \centering \textbf{FIB} \tabularnewline \hline \hline
    \small
Frequency of read operation & High & High & High \\ \hline
Frequency of write operation & High & High & Low \\ \hline
Size of data structure & Different and based on edge and routers & Different and based on edge and routers & Same \\ \hline
Algorithm & Cache Replacement policy & Timeout operation & Forwarding Strategy \\ \hline
Matching algorithm & $\bullet$ Interest packet: All Sub-Name Matching \newline $\bullet$ Data packet: Exact Name Matching & $\bullet$ Interest packet: Exact Name Matching \newline $\bullet$ Data packet: All Name Prefix Matching & $\bullet$Interest packet: Longest Name Prefix Matching \newline$\bullet$ Data packet: N/A \\ \hline
Data structure & Skip List & Bloom Filter & Trie \\ \hline
Issue & Fast name lookup & High memory access frequency & Memory consumption \\ \hline

     \end{tabular}
    \label{comp_table}
\end{table}

\section{Packet}
\label{PKT}
\begin{figure}
    \centering
    \includegraphics[scale=0.35]{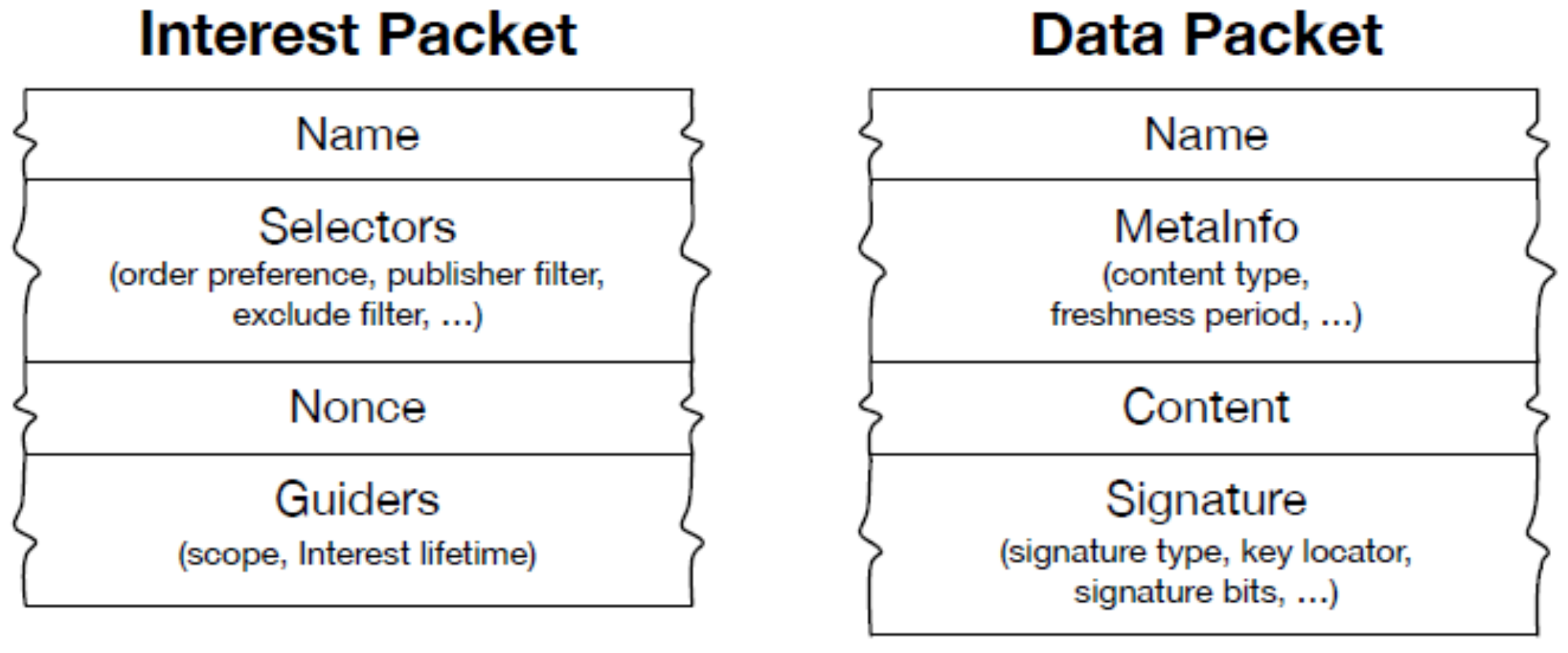}
    \caption{\textbf{Packet Architecture [Source \cite{Zhang}].}}
    \label{packet_arch}
\end{figure}

\begin{figure}
    \centering
    \includegraphics[scale=0.25]{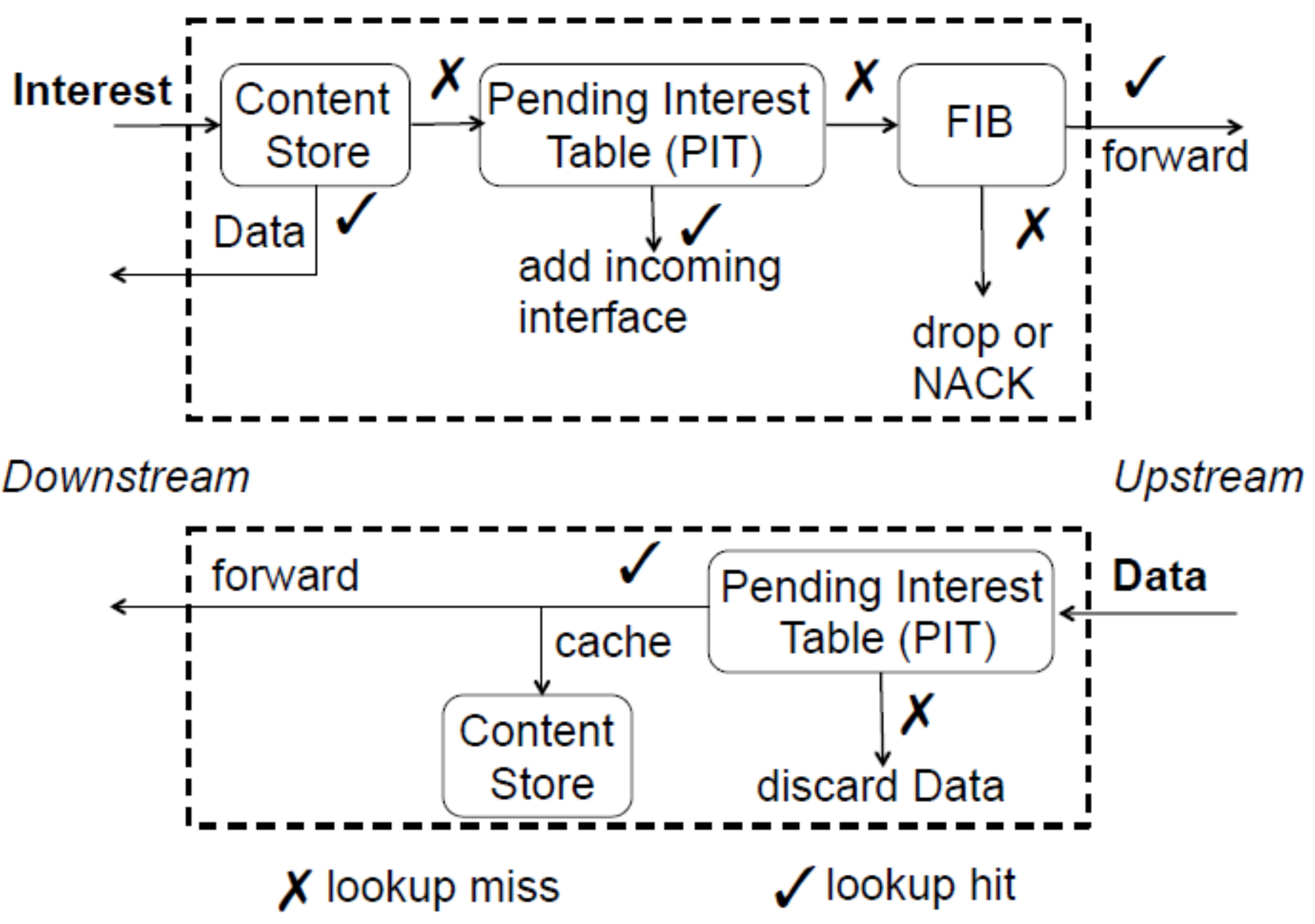}
    \caption{\textbf{Data forwarding and receiving process of NDN [Source \cite{Zhang}].}}
    \label{for}
\end{figure}

NDN communication starts when a node (called a consumer) requests for data by sending packet called Interest packet. Interest packet contains the name (called content name) to identify the requested data. Figure \ref{packet_arch} shows the architecture of forwarding an Interest packet. Content name helps to retrieve content using name and it is independent of the network. First, the content name of the Interest is checked in CS to retrieve data from the cache. CS is a memory buffer that stores the Data packet. More functionalities of CS are elaborated in section \ref{CS}. PIT saves the interest name and the interface from which the interest is received. PIT stores unresolved interest. PIT is discussed in detail in section \ref{PIT}. Interest packet is dispatched to next node after referring FIB. The FIB is a data structure maintained by the router. It stores the list of neighboring nodes and other information. More discussion on FIB is provided in section \ref{FIB}. A node sends a Data packet when the node receives the Interest packet containing the requested data. The sender is called producer. A Data packet contains the name, content (i.e., the requested data) and signature (producer's key). Data packet follows the Interest packet path in reverse to reach the consumer. The packet does not contain any IP address, the routers forward the packet based on the content name. Figure \ref{for} shows the transmission of Interest and Data packet. The dotted box in Figure \ref{for} is a router. A router contains three data structures, namely, CS, FIB and PIT. Table \ref{packet_table} illustrates the advantages and disadvantages of the techniques discussed in this section. 

\subsection{Bloom Filter in Packet}
Mun \textit{et al.} \cite{mun, mun1} proposed a new packet called a summary packet. A summary packet contains a Bloom Filter to represent the summary of the cache present in a node. It helps in increasing the cache efficiency in routers. The Bloom Filter is exchanged with neighboring nodes using summary packet. The summary helps in taking decision regarding caching. The nodes also maintain a new data structure called Summary Store. Summary store saves summary packets. When a node receives a Data packet, it checks with cache summaries of neighbor nodes. If the data exists with neighbor nodes, then the data is not saved in the cache. In case the node receives an Interest packet the summary is checked with Bloom Filter. If Bloom Filter returns $True$, then it is searched in its cache. If cache also returns true, then requested Data packet is sent to the consumer. When the face of matching summary and incoming Interest packet are different, the Interest packet is forwarded to face of a matching summary. However, if both faces of matching summary and incoming Interest packet are same, then it is a case of false positive. In case of no matching summary or false positive, the Interest packet is sent to the next node after referring FIB. In case Data packet is received by a node, first Bloom Filter is searched. If present, the data is ignored. If absent, it is cached in CS and updates its own summary. Summary store in a node maintains $f+1$ Bloom Filters. One Bloom Filter for each face of NDN router. So, $f$ faces require $f$ Bloom Filters. Another Bloom Filter for its own cache. Thus, increase in number of faces increases number of Bloom Filters. Eventually, increases space complexity. Query with Bloom Filters are performed sequentially, leading to a high query time complexity. The proposed method uses SBF \cite{Bonomi} to reduce space and time. SBF supports association queries. Association query recognizes the set to which an element belongs.

A technique called Network Coding (NC)\cite{Bourtsoulatze} deploys Bloom Filter in Interest and Data packets. Bloom Filter stores the IDs of the consumer. When a consumer sends an Interest packet its ID is stored in the Bloom Filter. As the packet gets transmitted towards the producer of the requested data, Bloom Filter is modified based on some rules. After receiving the Interest packet, the Bloom Filter helps to decide (a) aggregates the Interest packet with other Interests stored in PIT, (b) utilizes the Data packets saved in CS, and (c) insert as a new entry into PIT and send to the neighbor node. When the intersection between Bloom Filter of received Interest and existing same content name Interest, is empty, then both Interests are merged. In case the intersection is not empty, then producer receives both Interests. When a Data packet is transmitted, the Bloom Filter of the Interest packet is stored in the Data packet. A node after receiving Data packet, consumes the Interest whose Bloom Filter has a subset of the consumer IDs stored in the Bloom Filter of the Data packet. When the Data packet reaches a node its name is removed from the Bloom Filter. It prevents re-transmission to same consumer. Bloom Filter is also stored in PIT and CS. 

Watano \textit{et al.} \cite{reroute} proposed a re-route algorithm for Interest forwarding based on the degree of similarity of cached contents among each neighboring content routers (CR). Bloom Filter helps in selecting an alternate adequate route with similar contents when original route is lost. This helps to maintain the cache hit ratio. CR cannot forward an Interest, if it cannot reach the next CR hop caused by link disconnection or inoperative state of the next CR hop. This degrades the performance and increases network traffic. The method is based on the similarity of the Bloom Filters among chosen candidate CRs. Each CR consist of a CS, PIT, FIB and a Bloom Filter corresponding to the faces connecting to the CR. CRs can exchange their Bloom Filter value via Interest/Data Packet communications. Similarity is calculated between the initial next CR hop according to its entry in the FIB and the candidate CR. $EXOR$ of the values of the Bloom Filters are considered. Similarity rate is based on the sum of bits calculated as 1. More number of 1s indicates lesser similarity. Since this method uses conventional Bloom Filter, it has a high false positive and false negative probability. The cache hit ratio of this method is much better than the conventional. Although, when the link down time is less, it does not perform very well. Because initially differences in similarities of Bloom Filters is not reflected, hence, appropriate alternate next CR hop cannot be determined.

\subsection{Content Discovery}
Bloom Filter-based Routing (BFR) \cite{BFR,Marandi} is a content discovery technique. It is content oriented, and fully distributed. In this technique, origin server uses Bloom Filter to advertise its content objects. Moreover, it is robust to topology changes. BFR has three phases, namely, representation, advertisement and content retrieval, and FIB population. In the first phase, origin servers construct a Bloom Filter to represent their content objects. Initially, all slots of the Bloom Filter are assigned 0. Then, content object name and name prefixes are mapped to the Bloom Filter. In the second phase, the Bloom Filter is sent to advertise the content objects. A new Interest packet is proposed, called the Content Advertisement Interest (CAI). CAI packet consists of content advertisement Bloom Filter. CAI packets are forwarded to all neighbors. CAI is broadcasted till all nodes receives the packet. 

COntent-driven Bloom filter based Routing Algorithm (COBRA) \cite{COBRA, Marandi} is a content-driven, fully distributed, and Bloom Filter based intra-domain routing algorithm. It maintains routing information without extensive signaling between nodes. Each interface has its own Stable BF \cite{Deng}. When a Data packet is received, it is transmitted to the consumer. Then, the whole name and prefixes are hashed and inserted into the Stable BF. The information on Stable BF is used to rank the interface during each query. When a new Interest is received, the whole name is searched in Stable BFs. The searching process terminates when all interfaces are ranked or new prefix is obtained after removing the last component of the received name. After completion of the search process, all interfaces are assigned to a routing cost. Then, the Interest is forwarded with lowest routing cost. In COBRA, stale information leads to incorrect forwarding. Therefore, Stable BF is updated. Moreover, COBRA deploys an ad-hoc retransmission handling technique to remove stale information. Moreover, the number of used interfaces are incremented as per the requirement. During a retransmission request, Interest is sent both towards the probed before interface and interfaces just after in the global rank. Then, finally the requested whole name is removed from Stable BF. In case of failed link detection, Stable BF associated with the interface is reset. And, the new alternative path is identified. In case of link recovery, directly connected nodes set the Stable BF slots to 1 that are associated with the interface. Then, the recovered link is used as a primary or an additional path.

Pull-based BFR \cite{PB-BFR} algorithm was proposed to advertise only on-demand content objects. This algorithm saves bandwidth, other network nodes requires less memory to save content advertisement information and scalable compared to push-based BFR \cite{BFR}. The algorithm follows two phases (i) obtaining demanded content object and (ii) advertisement. When a consumer sends an Interest. If there is no FIB entry for the content name or no Bloom Filter has content advertisement, then the Interest is not forwarded and kept as pending. The consumer request the server to pull content advertisement information for the demanded content. Moreover, the consumer constructs a Bloom Filter containing the content file name and name prefixes. It also constructs a Content Advertisement Request (CAR) message containing the Bloom Filter. Again, when another consumer sends the same Interest it creates another CAR message. When a router receives both the CAR messages, it unites both CAR messages. This new CAR message is forwarded in a different face. This algorithm is also followed by other routers to create CAR messages. Nodes use sequence numbers in CAR messages to maintain their construction sequence. This process is a Bloom Filter based strategy to inform the server about the demanded content names. All nodes use Bloom Filter of same size and same hash functions for the efficient union of Bloom Filters. Maximum capacity Bloom Filter is not united because all slots of Bloom Filters have value 1. In such case, for any content matching, the Bloom Filter returns a false positive response. Due to false positive response the server advertise the content which is not in demand. Increase in forwarding rate of routing messages increases content advertisement overhead. Similarly, increasing content advertisement refreshes rate increases communication overhead. Memory consumption is less compared to push-based BFR.

\subsection{Wireless NDN}
BlooGo (BLOOm filter based GOssip algorithm) \cite{BLooGo} is a gossip algorithm that transmits the packets using minimum number of transmissions. A gossip algorithm transmits data in a robust manner within a disruptive network. BlooGo uses Bloom Filter to quickly find the neighbor node. In BlooGo, when a node receives the Interest packet, it compares its local Bloom Filter and the Bloom Filter present in the packet. The comparison is performed by $NBF_ {node} ~AND~NBF_ {packet} ~XOR~NBF_ {node} $. NBF contains the list of the nodes within the transmission range of the sender node. $AND$ operation finds the bits common between them. $XOR$ finds the neighbor nodes that the current node can forward the packet in a single transmission. In case Bloom Filter returns a false positive response, then the node may not transmit the packet to its neighbor nodes. When a node is not the producer of the Interest packet, then the node only forward the packet to its neighbor nodes. Breadcrumb technique is implemented to send Data packet back to consumer node. In this technique, the node that received the Interest packet saves the packet id. Data packet has the same id as Interest packet. Advantages of BlooGo are it is completely decentralized, lightweight and uses shortest path to deliver packets. It is easy to hardwire because basic arithmetic operations are performed on Bloom Filter. Moreover, BlooGo is easy to deploy as nodes do not require any knowledge about the network or unique identifier. BlooGo uses conventional Bloom Filter which does not permit deletion. Hence, for every packet new Bloom Filter is constructed. Moreover, dynamic nature of the network causes errors because Bloom Filter is unable to delete failed nodes. 

\subsection{Software Defined Network (SDN)}
A forwarding technique called (NDN-SF) \cite{Kalghoum,Kalghoum1,Kalghoum2} implements Bloom Filter to improve the efficiency of content delivery in SDN. NDN-SF hypothetically partition the network into areas. Each area consists of a single SDN controller, multiple NDN nodes and multiple switches. SDN controller centrally manages the NDN network. SDN controller uses OpenFlow protocol \cite{OpenFlow} to communicate with the switches. Name lookup is enhanced by using CBF based table \cite{Li_Fan}, namely, FIB-Bloom, PIT-Bloom and CS-Bloom. Implementation of Bloom Filter improves prefix searching. During routing step, S-FIB table is constructed/updated. It helps to forward Interest messages to the neighboring node. The controller calculates the most recent path to all available data names in the network. Then, controller broadcast its S-FIB to all switches in the network. During forwarding step, upon receiving an Interest it is searched in CS-Bloom. If present, Data packet is forwarded to requesting interface after referring to CS-Bloom. If absent, the Interest is searched in S-PIT. If absent, then again, it is searched in FIB-Bloom. If present, it is forwarded to the requesting interface after referring to S-FIB. If absent, the Interest is sent to controller to search in other network areas. NDN-SF uses many tables/data structures which leads to increase in memory consumption.

\begin{longtable}{|p{1.25cm}|p{5cm}|p{6cm}|}
    \caption{\textbf{Comparison of Techniques used to improve Packet delivery in NDN}}\\
    \hline
      \textbf{Paper} &\textbf{Advantages} & \textbf{Disadvantages}\\  \hline 
     \endfirsthead
\hline
\centering 
 \textbf{Paper} &  \textbf{Advantages} & \textbf{Disadvantages}\\  \hline 
\hline
\endhead
BlooGo \cite{BLooGo}, 2012 & $\bullet$ Completely decentralized \newline $\bullet$ Appropriate for an application that does not know the producer \newline $\bullet$ Easy to hardware because basic arithmetic operations are performed on Bloom Filter \newline $\bullet$ Lightweight \newline $\bullet$ Easy to deploy as nodes do not require any knowledge regarding the network or unique identifier \newline $\bullet$ Packets delivered using shortest path  & $\bullet$ In case of false positive response the node does not transmit the packet to the neighbor node \newline $\bullet$ Implements Conventional Bloom Filter which gives more false positive \newline $\bullet$ Every new packet contains the newly constructed Bloom Filter \newline $\bullet$ Dynamic nature of the network causes error because Bloom Filter is unable to delete failed nodes \\ \hline
 
COBRA \cite{COBRA}, 2014 & $\bullet$ Fully distributed \newline $\bullet$ Content oriented \newline $\bullet$ Maintains routing information without extensive signaling between nodes \newline $\bullet$ No requirement for any update message exchanges between the nodes & $\bullet$ Increase in number of the interface increases number of Stable BF which increases memory usage \newline $\bullet$ Stale information leads to incorrect forwarding \\ \hline

NDN-SF \cite{Kalghoum,Kalghoum1,Kalghoum2}, 2017 & $\bullet$ Helps in acquiring a global view of the network that enable better control \newline $\bullet$ Implements a non-flooding data publication mechanism to reduce network overloading \newline $\bullet$ Non-flooding data publication mechanism also helps the controller to monitor prefixes in real time \newline $\bullet$ Implementation of Bloom Filter improves prefix searching \newline $\bullet$ Reduces bandwidth usage and delay  &  $\bullet$ Implements many tables/data structure that increases memory consumption \newline $\bullet$ Forwarding to other network requires searching in many data structures \newline $\bullet$ CBF is used in three tables, but it consumes more memory compared to conventional Bloom Filter \\ \hline

BFR \cite{BFR}, 2017 & $\bullet$ Fully distributed \newline $\bullet$ Content oriented \newline $\bullet$ Robust to topology change \newline $\bullet$ Does not store topology information \newline $\bullet$ Resilient to link failure \newline $\bullet$ Delay is less \newline $\bullet$ Takes less congested route & $\bullet$ An Interest reaches a wrong origin server due to the false positive response \newline $\bullet$ Transmit CAI through multiple paths to reach the origin server, which increases traffic congestion \newline $\bullet$ Takes longer route \newline $\bullet$ Efficiency less compared to Pull-based BFR \newline $\bullet$ Total communication overhead depends on the content universe size \newline $\bullet$ Increasing content advertisement refresh rate increases communication overhead \newline $\bullet$ Memory space increases with the increase in content universe  \\ \hline

Mun \textit{et al.} \cite{mun, mun1}, 2017 & $\bullet$ Increases cache hit \newline $\bullet$ Helps to cope with link failure, dynamic network topology, and congestion in some links \newline $\bullet$ Change in the request pattern does not affect content diversity & $\bullet$ Query with Bloom Filters are performed sequentially, which increases time complexity \newline $\bullet$ $f+1$ Bloom Filters are maintained by Summary store, where $f$ = number of faces of NDN router and one for own cache. \\ \hline

Watano \textit{et al.} \cite{reroute}, 2018 & $\bullet$ Cache hit ratio higher than conventional method \newline $\bullet$ Essential when there is an unwanted loss of connection \newline $\bullet$ Space efficient & $\bullet$ Cache hit ratio unsatisfactory for early link down times \newline $\bullet$ Uses conventional Bloom Filter, therefore has a high false positive and false negative probability. \\ \hline
 
Network Coding (NC) \cite{Bourtsoulatze}, 2018 & $\bullet$ Delivers scalable video \newline $\bullet$ Packet delivery rate is based on the popularity of the content \newline $\bullet$ Maximises average video quality \newline $\bullet$ Bloom Filter supports deploying of network coding in NDN \newline $\bullet$ Decision of Bloom Filter helps in deciding the packet movement among the data structure of NDN & $\bullet$ Choosing the network coding coefficients randomly leads to decrease in video quality \newline $\bullet$ Implementing conventional Bloom Filter which gives a high false positive probability. \\ \hline

Pull-based BFR \cite{PB-BFR}, 2019 & $\bullet$ Advertise only on-demand content objects \newline $\bullet$ Bandwidth is saved \newline $\bullet$ Other network nodes require less memory to save content advertisement information \newline $\bullet$ Scalable \newline $\bullet$ Memory consumption is less compared to push-based BFR & $\bullet$ Server is unable to know the demanded content objects in priori \newline $\bullet$ Union of maximum capacity Bloom Filter gives very high false positive response \newline $\bullet$ Due to false positive response the server advertise the content which is not in demand \newline $\bullet$ Increase in forwarding rate of routing messages increases content advertisement overhead \newline $\bullet$ Increasing content advertisement refresh rate increases communication overhead \\ \hline

\label{packet_table}
\end{longtable}
 
\section{Content store}
\label{CS}
After the NDN router receives an Interest packet, first CS searches for a matching data \cite{Li-Xu}. CS is the buffer memory of the router. It caches data to serve future requests. CS consist of index and packet buffer. Index stores the data name. Packet buffer is the cache. CS size is small. Also, CS is of different size as it is based on the size of the edges and the router. CS follows All Sub-Name Matching (ASNM) name matching algorithm. This algorithm searches for any data having the initial name as the requested Interest. After, the Interest packet is forwarded to the corresponding interfaces based on decisions made by PIT. CS decides whether to cache the data. In case of Data packet, Exact Name Matching (ENM) algorithm is followed for name matching. ENM matches the Interest name exactly with the CS entry. If a name mismatch between the data and the Data packet occurs, then data is cached. CS has a high read and write operation frequency. Absence of CS in router does not hinder the routing operations. However, CS is included to store popular Interests. For example, CS greatly helps in live streaming \cite{Fayazbakhsh}. CS requires an efficient cache replacement policy. Widely, skip list data structure is adopted for CS. Linked list in the skip list helps in maintaining the order of data storage. Hence, skip list itself helps in replacement policy. However, the time complexity of query operation in skip list is more. Therefore, Bloom Filter is a better option because the time complexity of all operations of Bloom Filter is $O(1)$. In this section, various technique implementing Bloom Filter in CS is discussed. Table \ref{CS_table} is listing the advantages and disadvantages of the techniques. 

Dai \textit{et al.} \cite{Dai} proposed a technique based on Bloom Filter to store the popularity of content. This technique is deployed in CS to increase the cache hit ratio and reduce the upstream traffic. It uses multiple Bloom Filters where each Bloom Filter is dedicated to a particular range of popularity. Suppose content popularity ranges from 1 to $p$ and $t$ is a content popularity such that $1<t<p$. The first Bloom filter ($Bloom Filter_1$) stores content popularity ranging from 1 to $t$. Second Bloom Filter ($Bloom Filter_2$) from popularity ranging from $t+1$ to $2t$ and so on. While inserting a new element, the content is searched in all Bloom Filters. If the content is new, then its content popularity is 1 and it is stored in $Bloom Filter_1$. Every time an existing content arrives its popularity is increased by 1. When a content has popularity $t+1$, then that content is inserted to $Bloom Filter_2$. During the query, the content is searched in very Bloom Filters. If at least one Bloom Filter returns $True$, then the content is present. Otherwise, the content is absent.

Bloom Filter based Request Node Collaboration (BRCC) \cite{Hou} uses a variant of CBF called sum-up counting Bloom Filter (SCBF). It helps BRCC to optimize the cache performance. BRCC consists of SCBF and caching information table (CIT). During updating, the neighbor nodes exchange their cached data content. Updated caching information is saved in CIT. Collaborative caching helps in reducing data content redundancy. It also improves the data content diversity and hit rate. BRCC does not improve Internet packet's caching, but only improves the data packet's caching. In the improved data packet format, Caching Index (CI) is caching flag. This is initially set to 0. When CI is 1, it means the data content is cached in the data packet. The caching age (CA) records the lifespan of Data packets. Caching Remain Age (CRA) saves remaining lifespan of the Data packet. When a new data packet arrives, it replaces the existing packet with CRA value 0, when the CS caching capacity reaches its limit. If there are no available packets with CRA value 0, the one with lowest CRA value is replaced. And, if there are multiple packets that can be replaced, Least Recently Used (LRU) is used for replacement. NDN router uses SCBF to search the CIT when it performs caching. Collaboration caching helps in reducing data content duplication and makes it more diverse. Therefore, it has a high cache hit ratio. Moreover, First Hop Hit Ratio (FHHR) is enhanced. Because, frequently requested data are stored in the cache for a long time in the NDN router near the consumer. The increment in caching capacity enhances the hit ratio of interest packet's requests and reduces the average hop count. BRCC also has a low Average Route Delay.

Fetching the Nearest Replica (FNR) \cite{FNR} is a technique to deploy content delivery network (CDN)-like \cite{CDN} enhancement in NDN. It fetches data from the nearest replica without checking whether it is the best path from producer to consumer. It determines the popularity of the data using Zipf distribution \cite{zipf}. Based on the request popularity, the technique partition the CS into two parts, namely, Top-N Subset and Heavy-Tailed Subset. Top-N Subset is a subset of size $N$ determined by using the Zipf law. Majority request is satisfied by Top-N Subset. It implements Bloom Filter to reduce the storage and transmission overhead. Heavy-Tailed Subset is a very big size subset that satisfies very less data requests. FNR deploys a Track Server (TS) for each Internet service provider. TS is responsible for the collection and synchronization of replica information. TS merges Top-N Subset of all routers to construct the replica list. Hence, the replica list has all the Bloom Filters. TS is updated periodically. In case of update in Top-N Subset, router notifies TS to retrieve latest Bloom Filter. Then, TS updates its replica list. And, synchronizes with every consumer. In case the router does not update its Bloom Filter for a long period, then TS does the updating. FNR is scalable and lightweight. Using Bloom Filter reduces announcement traffic. However, FNR uses conventional Bloom Filter which gives a high false positive probability. FNR maintains multiple Bloom Filters, and also, stores many data structures which increases memory consumption. Also, FNR deploys a centralized model which may lead to a single point of failure. 

Controller-Based Caching and Forwarding Scheme (CCFS) \cite{Aloulou} is a technique that uses a cooperative caching to improve the forwarding decision. The technique partitions the network into domains. Each domain is controlled by node called a controller. Other nodes in the domain are regular router. The controller makes the forwarding decisions based on the popularity of the content. The controller has a data structure called Cache Information Base (CIB) to store summary of the content along with its neighbor's summary. Each entry of CIB stores the interface ID of the node, type of the neighbor node and a Stable BF \cite{Deng} associated with the interface. Stable BF stores the summary of the node. It helps to check for the summary with less time complexity. CCFS follows cache admission based on content popularity. When the content is highly popular it is stored in the CS. In case the outcoming interface is towards a regular router, then content name is hashed and stored in the Stable BF associated with the interface. Each controller periodically sends Stable BF to one hop neighbor Controllers. Stable BF is sent using a specific Interest packet, and at a low frequency. When a controller receives new Stable BF, then Stable BF of corresponding CIB entry is replaced.  

Scalable Content Routing for Content-Aware Networking (SCAN) \cite{scan, scan1} is a scalable routing scheme to implement content aware networking. In SCAN, a content router (C-router) saves the contents in its attached storage modules. It also guarantees efficient content delivery and retrieval. C-routers adopts Bloom Filter to increase scalability. Bloom Filter compresses the stored content information in a C-router. Each C-router maintains one Bloom Filter for each interface in a Content Router Table (CRT). It first performs an IP routing when it gets a request. Then, it performs scanning in case it can be afforded. For scanning, first the Local Content Table (LCT) is searched. If present, the C-router responds by notifying its existence to host. If absent, Bloom Filter is searched for every outgoing interface. The requested Content ID (CID) is searched in the Bloom Filter of a particular interface. If there are matched interfaces, the scanning request is generated and sent to matched interfaces. For scanning, periodically C-routers exchange CRTs. For example, let $C_1$ and $C_2$ be two C-routers exchanging CRTs. $C_1$ merges Bloom Filters of all interfaces except the interface connected to $C_2$. $C_1$ then sends the merged Bloom Filter to $C_2$. $C_2$ inserts the received information into Bloom Filter of the interface connected to $C_1$. Similarly, $C_2$ forward its own CRT to $C_1$. For reducing the false positive probability, C-router decays all the bits set to 1 of a Bloom Filter probabilistically before exchanging CRT. Therefore, scanning is performed based on a threshold. When the number of matched bits is greater than the threshold, then scanning is performed. Traffic is lowered due to use of cached contents. It does an efficient load balancing. SCAN guarantees reachability of content, regardless of scanning results.

CCNxTomcat\cite{tomcat} is the first practical CCN-enabled web server. It is based on open source $Apache Tomcat$ and the $CCNx library$. It supports CCN-based web applications. CCNxTomcat contains a novel cache called CCNCache. CCNCache is a cache for the Servlet Response. It performs name-lookup for content with high efficiency. For CCNCache, CBF \cite{Li_Fan} is used to check the presence of content packet in the cache. This reduces the number of memory accesses per query. CCNCache implements a composite data structure consisting of hash table, skip list and multiple CBFs. Long CCN names are split into two shorter parts to improve lookup efficiency. The first part contains the Servlet name. The second part contains parameters and values. These parts are stored in hash tables and trie. When there is a huge number of CCN requests, multiple skips lists are constructed to increase the query process. CCNcache performs accurate prefix matching mechanism on all CBFs. When a request is received by CCNcache, it determines the  prefix length. Then CBF is searched to identify whether it is present in the hash table. If present, then CCNcache query the hash table to find the address of the trie tree root node. CCNcache retrieves the requested parameters and values from trie tree. Using these parameters and values skip list is searched. Then CCNcache returns the packet content. CCNxTomcat directly interacts with CCN routers to reduce the cost of HTTP/CCN protocol conversion during development and run-time phases. CCNCache performs faster lookup, although, it occupies additional memory due to multiple CBFs. However, the lookup performance degrades when the number of a name component increases.

 \begin{longtable}{|p{1.25cm}|p{5cm}|p{5.5cm}|}
    \caption{\textbf{Comparison of Techniques used in Content store}}
    \label{CS_table} \\
    \hline 
      \textbf{Paper} &\textbf{Advantages} & \textbf{Disadvantages}\\  \hline 
     \endfirsthead
\hline
\centering 
 \textbf{Paper} &  \textbf{Advantages} & \textbf{Disadvantages}\\  \hline 
\hline
\endhead
SCAN \cite{scan,scan1}, 2011 & \small{$\bullet$ Efficient content delivery \newline $\bullet$ Efficient content reachability \newline $\bullet$ Bloom Filter increases scalability \newline $\bullet$ Efficient load balancing \newline $\bullet$ Cached contents reduces traffic }& \small{$\bullet$ Maintains one Bloom Filter for each interface which increases memory consumption \newline $\bullet$ Due to use of conventional Bloom Filter, false positive and false negative probability is higher} \\ \hline 
   
Dai \textit{et al.} \cite{Dai}, 2014 & \small{$\bullet$ Stores the content popularity \newline $\bullet$ Improves cache hit ratio to reduce upstream traffic} &\small{ $\bullet$ Implements multiple Bloom Filters which increases memory consumption \newline $\bullet$ A content present in higher Bloom Filter also presents in the lower Bloom Filter. Multiple presence of content is wastage of memory \newline $\bullet$ Exact content popularity of a content cannot be determined from Bloom Filter} \\ \hline

CCNx- mcat \cite{tomcat}, 2014 & \small{$\bullet$ First Practical Web Server supporting both HTTP and CCNx \newline $\bullet$ Seamlessly integrates with existing web server kernel and created CCN-enabled web server  \newline $\bullet$ Outperforms conventional methods in terms of throughput \newline $\bullet$ No additional cost of HTTP/CCN conversions \newline $\bullet$ CCNCache performs a very fast look-up }& \small{$\bullet$ CCNCache occupies additonal memory \newline $\bullet$ CCNCache implements CBF which has a higher memory consumption than conventional bloom filters \newline $\bullet$ Look-up performance degrades when number of name components increases} \\ \hline

CCFS \cite{Aloulou}, 2015 & $\bullet$ Improves in-network caching \newline $\bullet$ Stable BF reduces the false positive probability \newline $\bullet$ Stable BF reduces time complexity of searching for summary \newline $\bullet$ Reduces communication load  \newline $\bullet$ Reduces bandwidth usage & $\bullet$ Have high server workload \newline $\bullet$ Each interface has an Stable BF, multiple Stable BFs increase space complexity \newline $\bullet$ Monitoring using the controller may lead to single-point of failure \newline $\bullet$ No backup technique in case of failure of the controller \\ \hline

FNR \cite{FNR}, 2016 & \small{$\bullet$ Provides content delivery network (CDN)-like enhancement in NDN \newline $\bullet$ implements Bloom Filter to reduce storage and transmission overhead \newline $\bullet$ Bloom Filter reduces announcement traffic \newline $\bullet$ Local hashing table prevents repeated data location \newline $\bullet$ Scalable \newline $\bullet$ Lightweight }&\small{ $\bullet$ Implements conventional Bloom Filter, hence have high false positive \newline $\bullet$ Maintains multiple Bloom Filters occupies more memory \newline $\bullet$ Deploys a centralised model which may lead to a single point of failure \newline $\bullet$ CS stores many data structures, increasing memory overhead \newline $\bullet$ CS saves one pointer to each content of Top-N Subset} \\ \hline

BRCC \cite{Hou}, 2018 & \small{$\bullet$ Reduces data content redundancy \newline $\bullet$ Increases data content hit rate, diversity and searching rate and accuracy \newline $\bullet$ High cache hit ratio \newline $\bullet$ High First Hop Hit Ratio \newline $\bullet$ Low Average Route Delay \newline $\bullet$ Enhances data content matching rate \newline $\bullet$ Decreases searching time} & \small{$\bullet$ Employs SCBF, which has a higher memory consumption \newline $\bullet$ Does not improve Interest packet}\\ \hline

\end{longtable}

\section{Pending Interest Table}
\label{PIT}
Pending Interest Table (PIT) \cite{PIT1} is an intermediate router that stores unresolved interest. When a consumer sends an Interest packet, PIT saves the interest name and the interface from which the interest is received. Interest packet is forwarded to nodes till it reaches the producer node. Then a producer sends the Data packet, which follows the path in reverse order to reach the consumer. When the Data packet reaches PIT, it searches for the interface from which the Interest is received. If present, the name is deleted from PIT. Then, the Data packet is forwarded to a consumer. The speed of incoming Interest packet is very high, hence, a large sized data structure is required to store the packets without overflow. The frequency of read and write operation is high. Moreover, all the operations (insert, search, delete) need to be highly efficient to coup with high packet incoming speed. Moreover, different application follows a different mechanism to send its packets. This influence the size and processing speed of PIT. PIT follows an ENM algorithm for matching the content name of the Interest packet. ENM matches the Interest name exactly with the PIT entry. In case of Data packet All Name Prefix Matching (ANPM) algorithm is followed for matching the content name. ANPM algorithm searches for any pending Interest having names similar to prefix of Data name. ANPM helps to achieve Data multicast. PIT deletes entries after a timeout to avoid saturation of the data structure. Timeout also helps against simple flooding attacks \cite{Xie}. Each entry is associated with a timer to determine the timeout. PIT plays an important role in NDN communication \cite{PIT1}. Some important advantages are 
\begin{itemize}
    \item Content concentration: It enables only content concentration in NDN communication. PIT routes the packets (Interest and Data) without including source or destination address.
    \item Security: Lack of source or destination address makes the attacks difficult.
    \item Deduplication: Duplicate data are ignored. 
    \item No Looping: Unique data are stored in the PIT. Hence, the packets does not loop. 
    \item Multipath routing: No looping happens. Hence, Interest packets can send through multiple interfaces. 
    \item Supports Data packet multicast: If same Interests are requested by multiple consumers, then PIT forwards the Data packet to all the consumers
    \item Detects Data packet loss: PIT waits for the data time out, if the Data packet not received it identifies as data loss.
\end{itemize}
In the rest of the section, the various techniques that implement Bloom Filter to enhance PIT which are elaborated. Table \ref{PIT_table} illustrate the advantages and disadvantages of those techniques.

MaPIT \cite{MaPIT, Li} is a technique to achieve an advanced PIT. It implements an improved Bloom Filter called Mapping Bloom Filter (MBF). MaPIT consist of Index Table and CBF. Index Table is constructed on on-chip memory. CBF is constructed in off-chip memory. Packet Store saves the packet information. When an Interest packet arrives, first CS is searched. If absent, Index Table is searched. If also absent from Index Table, then it is stored in the FIB. Its footprint is saved in Bloom Filter and CBF. The packet information is stored in Packet Store using values obtained from MA. In case the Interest exists or Bloom Filter gives a false positive response, then Packet Store is updated and forwarded to FIB. When a Data packet arrives, Index Table's Bloom Filter is searched. If present, then Packet Store is searched. If the data is absent in Packet Store, then data are blocked and the footprint is deleted from the CBF. If data is present, then it is forwarded using the faces saved in Packet Store. Also, the record and footprint are deleted from Packet Store and CBF respectively. Finally, CBF and Bloom Filter are synchronized. MaPIT consumes less on-chip memory. Moreover, it is easy to deploy on faster memory chip. In addition, using MBF reduces the false positive probability. However, its performance is average. 

Fast Two Dimensional Filter Pending Interest Table (FTDF-PIT) \cite{Shub, Shubbar} is a novel PIT implementation which ensures high performance. The PIT employs a variant of FTDF, wherein every FTDF slot consists of a bit vector. The number of bits is equal to the number of faces in the NDN node. The bit vector saves the face number. When an Interest packet is received, first the CS is examined. If there is a match in the CS, the NDN node will return the Data packet. After that, it is mapped to FTDF-PIT. First, hash value appearing in the Interest packet is calculated. Next, quotienting technique is executed. This is performed to retrieve the row and column number. Lastly, inside the FTDF-PIT on the specific position, the corresponding bit in the bit vector is set to 1. This reflects the requested interface. Data packet coming from any interface is cached in CS. Then, content name is hashed to get FTDF-PIT slot position. Data Packet is forwarded to all the interfaces whose bit positions are set to 1. After the requested Data packet is satisfied, the bits in the bit-vector is reset to 0. FTDF-PIT has low false positive probability due to hard collision. Hard collision is solved by using a collision resistant hash function. It also has low construction time because it utilizes one hash function. FTDF-PIT has an efficient memory, hence it is able to store all flooded packets (due to DDoS attack) without consuming large memory. The time complexity of insert and query operation is O(1).

Yu and Pao \cite{hardware} proposes a Hardware Accelerator which is a hardware implementation of PIT for NDN. This design incorporates an off-chip linear chained Hash Table (HT) and an on-chip Bloom Filter. It also contains a name ID table (nidT). nidT stores all distinct name IDs (nid) in the PIT. If content name is found in nidT, then FIB is not referred. This reduces the FIB workload. This technique adopted a software and hardware co-design approach. Hardware accelerator makes the PIT lookup and management independent of software. In the proposed design, on-chip DRAM is used to implement Bloom Filter for the lookup table and external SRAM is used to store the full lookup table. Bloom Filter helps in faster lookup. During lookup, the hardware first query the Bloom Filter. If absent, then it is a PIT-miss. If present, hardware traverses the list of keys mapped to the corresponding bucket. If the input key is present, hardware returns a PIT-hit and deletes the data. If the data structure becomes empty after the deletion, then the Bloom Filter is set to 0. In a Lookup Interest (LI) command, if the key is present, the hardware returns a PIT-hit and the address of the data. If the key is absent, it is inserted in the data structure. The first empty slot is allocated for insertion. Insert and delete operations on nidT and PIT are similar to LD/LI commands. Lower PIT population increases the packet processing rate. The technique improves throughput. A major issue is the requirement of per-packet update in the lookup table.

\begin{table}[!ht]
\centering \footnotesize 
\caption{\small {Comparison of techniques used to improve PIT in NDN}}
   \begin{tabular}{|p{1.25cm}|p{5cm}|p{5.5cm}|}
    \hline
    \centering \textbf{Paper} & \centering \textbf{Advantages} & \centering \textbf{Disadvantages} \tabularnewline \hline \hline
    
 MaPIT \cite{MaPIT}, 2014 & $\bullet$ Reduces on-chip memory consumption \newline $\bullet$ Easy to deploy on faster memory chip \newline $\bullet$ Reduces false positive probability & $\bullet$ Average performance \newline $\bullet$ MBF uses conventional Bloom Filter which has high false positive and false negative issues \\ \hline
    
 Hardware Accelerator \cite{hardware}, 2016 & $\bullet$ Gives an efficient PIT lookup  \newline $\bullet$ Bloom Filter reduces lookup time \newline $\bullet$ Reduces FIB workload \newline $\bullet$ Makes the PIT lookup and management independent of software \newline $\bullet$ Improves throughput & $\bullet$ Requirement of per-packet update in the lookup table \newline $\bullet$ Saturated PIT takes more packet processing time \newline $\bullet$ Conventional Bloom Filter increases the false positive probability \\ \hline

 FTDF-PIT \cite{Shub, Shubbar}, 2019 & $\bullet$ Low memory requirements \newline $\bullet$ Low false positive probability \newline $\bullet$ Quick and efficient deletion \newline $\bullet$ Low building time \newline $\bullet$ Time complexity of insertion, query and retrieving interface number in O(1) \newline $\bullet$ Space and cost-efficient PIT \newline $\bullet$ Provides high security against DDoS attacks \newline $\bullet$ PIT memory independent of length of content name  & $\bullet$ FTDF is influenced by hard collision\\ \hline
   
   \end{tabular}
    \label{PIT_table}
\end{table}
\section{Forward Information Base}
\label{FIB}
Forward Information Base (FIB) \cite{NDN} is a data structure present in a router that stores next hop and other related information of every reachable node. The FIB is referred to forward the Interest packet to the next node. After searching PIT, if the Interest is absent, then the Interest packet is forwarded to FIB. FIB follows Longest Name Prefix Matching (LNPM) for name matching. LNPM searches for longest prefix matching of IP of Interest with the FIB entry. LNPM helps in acquiring most accurate forwarding information. If a matching FIB entry is present, then the Interest packet is sent to the next hop. And, a new entry in PIT is inserted. Sometimes the Interest packet is broadcast to all outgoing interfaces or decision is taken based on the forwarding policy. The frequency of read operations is high, but performs very less write operations. In case of Data packets FIB does not have any role. FIB contains duplicate entry for the same content. In this section, Bloom Filter based FIB techniques is precisely explained. Table \ref{FIB_table} presents the advantages and disadvantages of the techniques. 

So \textit{et al.} \cite{So} proposed a simple technique for NDN software forwarding lookup engine using hash table. A chained hash table is implemented. The hash table has linked list to save collision entries. To avoid unnecessary chain searching for absent data, every hash bucket has a Bloom Filter. When an element is hashed to a bucket, first Bloom Filter is searched. If present, then that data is not stored in the hash table. If absent, then that data is stored in hash table after creating a new link in the linked list. However, this technique implements conventional Bloom Filter that has many issues. Moreover, data cannot be deleted from hash table. Because that data cannot be deleted from the Bloom Filter. Sorted linked list is implemented to improve data searching efficiency.

Bloom Filter-Aided haSh Table (BFAST) \cite{bfast, BFAST1} data structure is proposed for faster FIB indexing. BFAST uses hash table and Bloom Filters. CBF \cite{Li_Fan} balances hash table load. Implementing Bloom Filter reduces search time. BFAST consist of a CBF, a hash table and $k$ auxiliary CBFs. Each slot is associated with a hash table slot to store the elements. Each hash function has an auxiliary CBF (aCBF). In case of collision, chaining is used in the hash table. During insertion, first the element is hashed by $k$ hash functions to obtain $k$ locations of CBF. The element is inserted into CBF. Then, minimum value $min$ among the $k$ location counters is found. The element is inserted into the hash table slot attached to $min$ CBF slot. The aCBF is considered whose $min$ location is returned by the hash function. The location to store the element in aCBF is obtained by performing a modulus operation on its corresponding hash value. During query operations, first CBF is searched. If absent, then element is also absent from hash table. Hence, returns false as a response. If present, then all aCBFs are searched. Rank-Indexed hashing is implemented to reduce memory consumption because the hash table uses lots of pointers. In this technique, the hash table with chaining have multiple buckets in each slot. The elements are stored in an address-contiguous array. Each element is stored in a different array. Only the first element is stored in a single array called First Item Array (FIA). While searching in the hash table $Rank$ operator is used to obtain the location dynamically. Moreover, this operation is possible to execute using hardware instruction which reduces the time complexity of the operation. In this technique each entry of FIB consist of the FIB entity entry, interface ID, signature (content name), and a pointer to the next index entry. This technique is efficient for insertion, deletion and searching in the FIB. But, have very huge space complexity because of using many data structures. 

Bloom-Filter assisted Binary Search (BBS) algorithm \cite{He} improves the efficiency of name lookup. It ensures the correct path searching without additional backtrack costs and less memory consumption. At each branching point a Bloom Filter is deployed to detect the presence of a prefix. Instead of storing virtual name prefixes in hash table, they are stored in Bloom Filters. It reduces memory consumption. BBS adds Bloom Filter to each hash table. BBS checks the presence of prefix by performing three lookups. In the first lookup, if a content is absent in a Bloom Filter, then BBS searches in its right subtree. Second lookup, if content is absent in both Bloom Filter and hash table, then BBS searches in the left subtree. Third lookup, if content is absent in Bloom Filter but present in the hash table. BBS perform two more lookups in Bloom Filter to reduce memory consumption. However, the time complexity increases. When a new name prefix is inserted into the FIB, it is stored in its corresponding hash table. Then, BBS inserts the name prefix into the specified Bloom Filters. Bloom Filter size is dynamic. It changes with the total number of current length-component name prefixes.

MaFIB \cite{Li} is a technique that uses MBF \cite{MaPIT} to enhanced FIB. MaFIB consists of an off-chip and an on-chip memory. On-chip memory is high-speed memory and uses multiple MBFs. Whereas, off-chip memory is low-speed memory and stores multiple CBFs which correspond to on-chip MBFs. Packet store is a static storage used to store packet forwarding information. Each MBF corresponds to one length name prefix which helps in quick query operation. A lookup operation in the packet store is performed using the MA value in MBF. MaFIB supports deletion operation taking the help of timely synchronization between the MBF and CBF. MaFIB has a false positive probability equal to the sum of the MBFs used, and it is lower compared to conventional FIB. The operations on the MaFIB have constant time. MaFIB also has a low on-chip memory consumption.

Lee \textit{et al.} \cite{prematchBF} proposes two efficient forwarding engines to perform quick name lookups. It implements a Bloom Filter and a hashing-based Name Prefix Trie (NPT). The first algorithm accesses the trie node for every positive result of the Bloom Filter. Second algorithm called Name Prefix Trie with Bloom Filter (NPT-BF) queries the Bloom Filter to determine the longest matching length. Trie nodes are processed from the longest possible length. In case no match is found, trie nodes are tracked back. Bloom Filter is stored on-chip and NPT is stored in an off-chip hash table. Every node in an NPT is stored in a hash table (HT). For each Interest packet, hash index is obtained from the content name. The hash index is compared with the hash entry of HT. When a matching entry is found, the search continues. Searching continues till there is no matching entry, or search equals to the input length. The output face of the current matching entry is stored. Query operation is performed on Bloom Filter before accessing HT. It eliminates unwanted HT accesses. In NPT-BF, content name is used to query the Bloom Filter. If present, then HT is accessed. HT gives the output face. Searching continues till Bloom Filter returns false or matching entry is found. When Bloom Filter returns false, HT searches for matching entries. Backtracking is performed until a matching entry is found in the HT. NPT-BF-Chaining increases Bloom Filter size to reduce the false positive probability. The average number of node accesses during implementation of NPT-BF algorithm is higher than that of NPT-BF-Chaining. Moreover, using non-chip Bloom Filter reduces access to an off-chip hash table.

\begin{table}
\centering \small 
\caption{\small {Comparison of techniques used to improve FIB in NDN}}
    \begin{tabular}{|p{1cm}|p{5.5cm}|p{5.5cm}|}
    \hline
    \centering \textbf{Paper} & \centering \textbf{Advantages} & \centering \textbf{Disadvantages} \tabularnewline \hline \hline

So et al. \cite{So}, 2012 & $\bullet$ Hash table reduces data collision \newline $\bullet$ Bloom Filter reduces chain searching \newline $\bullet$ Sorted linked list reduces searching time & $\bullet$ Data is not deleted from the hash table because the data deletion is not permitted in the Bloom Filter \\ \hline

Lee et al \cite{prematchBF}, 2016 & $\bullet$ Non-chip Bloom Filter reduces access to off-chip hash table \newline $\bullet$ Average number of node access is low in NPT-BF-Chaining algorithm \newline $\bullet$ Enables fast look-up & $\bullet$ Conventional Bloom Filter increases false positivity probability \\ \hline

BFAST \cite{bfast}, 2017 & $\bullet$ Increases parallelism \newline $\bullet$ First-Rank-Indexed reduces memory consumption \newline $\bullet$ Have high-throughput and low latency Longest Prefix Match forwarding performance with big FIB entries \newline $\bullet$ Supports incremental update \newline $\bullet$ Bloom Filter reduces searching time & $\bullet$ Implementing lots of data structure, increases space complexity \newline $\bullet$ Many Bloom Filters leads to higher overall false positive and false negative probability \newline $\bullet$ In a CBF, counter overflow may occur \newline $\bullet$ Size of CBF is four times more than conventional Bloom Filter \\ \hline

BBS \cite{He}, 2017 & $\bullet$ No backtrack cost \newline $\bullet$ Storing virtual name prefixes in Bloom Filters reduces memory consumption \newline $\bullet$ Efficient with longer prefix and large database \newline $\bullet$ Bloom Filter size is dynamic and changes with the total number of current length-component name prefixes & $\bullet$ Perform two more lookups in Bloom Filter which increases time complexity \newline $\bullet$ Time complexity of BBS is less compared to Binary search of hash table \\ \hline

MaFIB \cite{Li}, 2017 & $\bullet$ Low on-chip memory consumption \newline $\bullet$ Lower false positive probability compared to conventional FIB \newline $\bullet$ Operations are in constant time & $\bullet$ MBF has high false positive and false-negative probability \newline $\bullet$ CBF for off-chip memory occupies high memory \\ \hline

    \end{tabular}
    \label{FIB_table}
\end{table}

\section{Security}
\label{SEC}
\begin{table}
\centering \footnotesize 
\caption{\small{Comparison of techniques in Security of NDN}}
   \begin{tabular}{|p{1cm}|p{5.3cm}|p{6cm}|}
    \hline
    \centering \textbf{Paper} & \centering \textbf{Advantages} & \centering \textbf{Disadvantages} \tabularnewline \hline \hline

Massawe \textit{et al.} \cite{Massawe}, 2013 & $\bullet$ Provides high security and content privacy \newline $\bullet$ Scalable \newline $\bullet$ Reduces false positive probability by deploying multiple Bloom Filters & $\bullet$ Deploying multiple Bloom Filters occupies more memory \\ \hline

Chen \textit{et al.} \cite{Chen}, 2014 & $\bullet$ Protects access to NDN data contents \newline $\bullet$ Prevents unauthorized consumers to access encrypted data contents \newline $\bullet$ Reduces bandwidth \newline $\bullet$ Prevent download by unauthorized consumers and attempt to decrypt the video & $\bullet$ Maintaining multiple Bloom Filters (one for each site) increases memory consumption \newline $\bullet$ Implements conventional Bloom Filter \newline $\bullet$ No attempt made to reduce the false positive probability \\ \hline

Guo \textit{et al.} \cite{Guo}, 2016 & $\bullet$ Efficient against address spoofing \newline $\bullet$ Occupies less memory \newline $\bullet$ One hash function in Bloom Filter reduces time complexity \newline $\bullet$ Reduces false-locality pollution & $\bullet$ Saving the path information makes the Internet structure exposed to an attacker \newline $\bullet$ Storing path information leads to processing and network overhead \newline $\bullet$ Increases router overhead \newline $\bullet$ Not efficient against address-based attack \\ \hline

CPPM-DAM \cite{zhu}, 2018 & $\bullet$ Provides security against cache privacy snooping \newline $\bullet$ Determines legitimate users \newline $\bullet$ Permutation indicator helps in determining the difference between the packet with high probability & $\bullet$ Permutation indicator is an overhead for new content \newline $\bullet$ Increase in attack increases the round trip time \newline $\bullet$ Less number of hash functions leads to the possibility of guessing the permutation by attackers \newline $\bullet$ Small matrix size increases the false positive probability \newline $\bullet$ Implements conventional Bloom Filter \newline $\bullet$ Permutations are saved because same packet has same permutation \\ \hline

Wu \textit{et al.} \cite{smarthome}, 2019  & $\bullet$ Improves user identity filtering mechanism \newline $\bullet$ Improves user logout mechanism \newline $\bullet$ Reduces total delay \newline $\bullet$ Improves security of the entire system & $\bullet$ Implements CBF, which has a higher memory consumption \newline $\bullet$ High false positive probability \\ \hline

\end{tabular}
    \label{security_table}
\end{table}

NDN architecture is designed to provide some basic security. Every Data packet has a cryptographically signed producer key \cite{NDN-sec}. The signature is present a message and name, or payload. Data packet contains the key to verify the signature. First, consumer verifies the signature, then accept the packet. Intermediate router can verify the signature. But it is not compulsory because the verification increases processing overhead. Content signature helps in providing three basic securities, namely, Data integrity, Origin authentication, and Correctness \cite{NDN-sec}. 
\begin{enumerate}
\item Data integrity: Valid signature ensures intact content.
\item Origin authentication: Signature is verified using the producer’s public key, hence, any node can verify the content signature.
\item Correctness: Due to signature the content is tied to payload. This helps to securely identify whether the content of Data packet is same as the requested data.  
\end{enumerate}

NDN architecture also solves some basic DoS and DDoS attacks \cite{NDN-sec2}. DoS is denial of service and DDoS is the distributed denial of service. The data packet is sent after receiving a request. And, Interests requesting same Data are merged into a single packet. In addition,  timeout record and the number of PIT entries help to analyze and detect attack behaviors. 
However, NDN is vulnerable to many other issues such as \cite{NDN-sec, NDN-sec2}:
\begin{itemize}
\item Privacy: The hierarchically structured name of an Interest packet makes the content vulnerable. This is name privacy. Neighbors uses timing information to get each others content accesses from cache. This is called cache privacy. This information helps to determine the cache hits. The content of the Data packet is not encrypted, hence, they are vulnerable. This is content privacy. The signature leads to identification of the producer and its organization. This information leads to violation of producer's privacy. This is called signature privacy. 
\item Trust: The public key provided to determine the signature may not be trusted. This leads to trust issues in NDN. 
\item Interest flooding: The attacker sends a large number of Interest packets to a specific producer. It flooded the PIT which makes PIT unable to identify the legitimate Interest packets. The data request are of three types, namely, static contents, non-existent contents and dynamically-generated content. Each type gives a different effect. In static contents, the content exists in cache. So, the Interest packets are not forwarded. In case of non-existent contents, the router removes invalid request. However, it occupies PIT memory till timeout. If content is dynamically-generated, then all Interest packets are forwarded to the producer. It wastes the bandwidth and occupies router's PIT.
\item Content/cache poisoning: The attacker sends corrupted data. And, CS caches the corrupted data. When an Interest packet is sent, CS forward the corrupted data in Data packet. 

\end{itemize}

In the rest of the section, Bloom Filter based NDN techniques to solve some security issues are discussed. Table \ref{security_table} illustrates the advantages and disadvantages of the security techniques. 

Massawe \textit{et al.} \cite{Massawe} proposed Scalable and Privacy Preserving Routing Protocol in NDN (SP-NDN) to provide high security and content privacy. The technique is based on Bloom Filter. Each Interest packet has a name and a secret key shared between the consumer and producer. Both packet name and secret key are considered as an element and inserted into the Bloom Filter. During the query, the packet name and secret key are checked with the Bloom Filter. If the packet matches, it is forwarded. Otherwise, the packet is dropped. The technique uses multiple Bloom Filters to reduce the false positive probability. 
Guo \textit{et al.} \cite{Guo} proposed an anti-pollution algorithm against the false-locality pollution. It is based on Bloom Filter. The technique permits the backbone router and traces the path in its CS. Each cached content is associated with a counter and PathTracker. The counter counts the number of times a particular content is hit, since, the last examination against pollution. PathTracker is a data structure to trace the number of unique paths travelled by the Interest messages corresponding to each content. When an Interest packet is received, the cache is searched. The corresponding counter is incremented by 1. And, PathTracker stores the travelled path information. When a cache content reaches a threshold value, then the content is either a popular content or a victim of an attack. Saving the path information makes the Internet structure exposed to an attacker. To prevent this, in case of a cache miss the path information is deleted. Moreover, storing path information also leads to processing and network overhead. The technique implemented the PathTracker using probabilistic counting with stochastic averaging and Bloom Filter. The Bloom Filter uses only one hash function. The content name is hashed by the hash function, then the slot is set to 1. The technique is resilient against address spoofing. However, it's not efficient in case of address-based approach. 

Wu \textit{et al.} \cite{smarthome}  proposed an access control algorithm to provide an effective security solution for Smart Homes. The access control algorithm uses Cipher Block Chaining (CBC) symmetric encryption algorithm, Identity Based Encryption (IBE), Proxy Re-encryption (PR) and CBF \cite{Li_Fan}. This algorithm  effectively utilizes the cache. Cache is used to process user requests for data to cancel user privileges, improve user logout, and reduce delays. In a smart home, the home administrator is called a home manager that performs functions to grant user privileges. The access control algorithm has four stages. First stage, initialization uses CBC and IBE. Second stage is a user application permission stage. It requires the user signs the Interest packet using a private key and send to the home manager. After successful authentication, the home manager stores user information in the user registry. Transform key is generated using PR and is sent to the user along with registration confirmation Data packet. A third stage is user access data stage. The algorithm permits the users to access data in two ways, namely, caching and not caching. The intermediate routers do not cache the data or it caches the data. The data keys are converted to cipher text which is required by authorized users. Fourth stage, user logout stage occurs in two situations. One is active cancellation and the other is passive cancellation. In passive cancellation, the home manager delete users from the registry based on time. The deleted users are broadcast to the intermediate routers. The intermediate router updates the CBFs. CBFs are also used to filter ineffective Interest. This algorithm improves both user identity filtering and user logout mechanism. 

Chen \textit{et al.} \cite{Chen} proposed an access control mechanism based on Bloom Filter. The mechanism implements Bloom Filter to pre-filter Interest from consumers to avoid unauthorized access. In this technique, for each site a Bloom Filter is constructed to identify the authorized users. Each site constructs a Bloom Filter using digests of all public keys present in active user tables. Then, the Bloom Filter is sent to routers. Bloom Filter is piggybacked with Data packet to the routers. The router uses Bloom Filter to pre-filter Interests from users not in the active user tables. Continuously, the users get added to the network. Hence, Bloom Filter is periodically sent to the router for the update. In case of mobile users, a virtual subscription is given to the user. When a new user moves to a different network and sends an Interest, if it is rejected, then the Interest is a resend. This Interest triggers update of Bloom Filter in producer site. The method is using conventional Bloom Filters. Moreover, it is not using any technique to reduce the false positive probability.

Cache privacy protection mechanism (CPPM-DAM) \cite{zhu} is a Bloom Filter based security mechanism. It determines legitimate users and prevents cache privacy snooping. Bloom Filter has $k$ independent hash functions and $D$ $k$-dimensional matrices, where $D$ refers to the number of classes for content popularity. And, each $k$-dimensional matrix stores the elements belong to the corresponding class. From $n!$ permutation of hash functions, $n$ number of hash functions are randomly chosen. This permutation indicator is included in the Interest packet by adding as a new string field to the content name of the packet. However, the same packet should have the same permutation. After receiving an Interest packet, the content name and the permutation indicator are extracted from the packet. Using the permutation indicator the name is mapped to $n$ addresses. Based on the addresses the matrix cell is verified whether the addresses are 1. If all are 1, then the content is present in CS. Then the data is sent to the consumer. If any one cell is 0, then the content is new and the addresses are set to 1. Permutation indicator helps in determining the difference between the packet with high probability. However, permutation indicator is an overhead in case of new content. Increase in attack increases the round trip time. Moreover, less number of hash functions leads to the possibility of guessing the permutation by attackers. Matrix size depends on the popularity. The less number of hash functions leads to the possibility of guessing the permutation by attackers.

\section{Simulator}
\label{sim}
Currently, the only and most popular simulator for NDN is ndnSIM. It is an open source NS-3 based simulator. In this section, the available three ndnSIM versions are discussed. In addition, Table \ref{simu} highlights the advantages and disadvantages of Bloom Filter based ndnSIM. Table \ref{simu} illustrates some features and URL of the Bloom Filter based ndnSIM.

ndnSIM \cite{ndnSIM1} is an open source NS-3 based simulator. The first version of ndnSIM was proposed in the year 2012. It represents the NDN network in a modular fashion. It has an independent protocol stack. It is possible to install the stack on a simulated network node. It includes various helper classes and traffic generator applications. This helps ndnSIM to provide a simplified simulation scenario. ndnSIM provides maximum extensibility and flexibility by making the design independent of the underlying transport. This is achieved by abstracting the interlayer interactions. A Face abstraction is implemented to provide communication between core protocol. For CS implementation, an interface is provided to implement various indexing and element query designs, cache replacement policies and other features. Similarly, for PIT ndnSIM provides a template-based realizations for limiting number of PIT entries and various replacement policies. PIT realizations are based on a trie data structure with hash-based indexing on Data names. Likewise, realization of FIB is based on a trie data structure with hash-based indexing on Data names. ndnSIM provides various forwarding strategies without the need for core component modification. It provides a forwarding strategy abstraction responsible for handling Interest and Data packets like an event. This simulation has some accuracy limitations in the simulation results. 

ndnSIM 2.0 \cite{ndnSIM2} is the extension version of ndnSIM \cite{ndnSIM1} to include the latest advancements of NDN. It was proposed in the year 2015. It uses NDN C++ library with eXperimental eXtensions (ndn-cxxlibrary) and deploys NDN Forwarding Daemon (NFD) to perform experiment with real code. Similar to the previous version,  NDN protocol stack is the core component of the ndnSIM 2.0. NFD performs packet processing. Both ndnSIM simulations and real NFD use forwarding plane extensions. CS implementation uses NFD codebase. However, it is not flexible regarding cache replacement policies. The version has added the feature to port old ndnSIM 1.0 to extend flexibility. Similarly, PIT, FIB and forwarding strategy abstraction use NFD. Basic applications of ndnSIM 2.0 are similar to the previous version. However, due to the use of ndn-cxx library some minor changes are performed. The disadvantages of ndnSIM 2.0 are it is slower and occupies more memory compared to previous version. 

ndnSIM \cite{ndnSIM} is the version that is the first open source tool to model Bloom Filter in NDN. It is considered as the most complete one because of its accurate representation of all the facets of the NDN architecture. It allows simulation of various Bloom Filters for different purposes, for example: checking membership of locally cached components, or name lookup in forwarding strategies. ndnSIM also implements NDN stack. This stack consists of all basic data structures characterizing an NDN node, such as the Interest and Data packets, CS, PIT and FIB and network entities. It also provides helper classes, applications, and traffic generators that help to facilitate the construction of a complete simulation scenario. The module of this version is completely written in C++. Bloom Filters also help in further specializing their implementations and integration of the tool within the whole module. Bloom Filter is implemented using $Dynamic\_bitset$ data structure. $Dynamic\_bitset$ helps in an easy and direct access to its bits, that allows efficient insert, delete and query operations. $Dynamic\_bitset$ permits defining the Bloom Filter size during run time. Moreover, it permits resizing of Bloom Filter based on user requirements. $BloomFilterBase$ is the basic class that defines all parameters and functionalities that are common in different Bloom Filters. $ForwardingStrategy$, $PitImpl$ and $StackHelper$ are properly extended to integrate Bloom Filters. The $ForwardingStrategy$ class manages the transmission and reception of Interest and Data packets. $PitImpl$ class manages the lookup and construction of PIT entries. The module enables integration of Bloom Filter without noticeable impact on the memory usage. The simulation time is strictly influenced by the number of events that are handled during the simulation. Implementing Bloom Filter enhances the forwarding speed and scalability of routing protocols. The presented extension is consistent, accurate and flexible.

\begin{table}[!ht]
    \centering
    \caption{\small{Available simulator for NDN}}
    \begin{tabular}{|p{1cm}|p{1.7cm}|p{1.5cm}|p{6cm}|}
  \hline
    \textbf{Name} & \textbf{Language} & \textbf{Licence type}& \textbf{Link}\\ \hline
    ndnSIM \cite{ndnSIM} & NS-3, $C++$ & Open Source & \url{https://ndnsim.net/current/index.html}\\ \hline
        
    \end{tabular}
    \label{simu}
\end{table}

    


\section{Miscellaneous}
\label{OTH}
\begin{longtable}{|p{1.25cm}|p{5cm}|p{5.5cm}|}
    \caption{\textbf{Comparison of the other components of NDN}}
    \label{other_table} \\
    \hline 
      \centering \textbf{Paper} & \textbf{Advantages} & \textbf{Disadvantages}\\  \hline 
     \endfirsthead
\hline
\centering  \textbf{Paper} &  \textbf{Advantages} & \textbf{Disadvantages}\\  \hline 
\hline
\endhead
Name- Filter \cite{NameFilter}, 2013 & \footnotesize{$\bullet$ Single hash function in first stage reduces time complexity \newline $\bullet$ Merged Bloom Filter operations are performed in parallel, reducing time complexity }&\footnotesize{ $\bullet$ Both the stages have multiple Bloom Filters consuming large memory \newline $\bullet$ CBF consumes more memory compared to conventional Bloom Filter} \\ \hline

NLAPB \cite{Quan}, 2014 & \footnotesize{$\bullet$ Trie is highly flexible and scalable when the memory is dynamically allocated \newline $\bullet$ Constructing different CBF for each specific length B-prefix prevents asymmetric behaviors among different CBFs \newline $\bullet$ Reducing number of entries in CBF, reduces the false positive probability \newline $\bullet$ Adaptive lookup technique increases the lookup speed} & \footnotesize{$\bullet$ Due to variable lengths of the T-suffix, Bloom Filter is not used for searching \newline $\bullet$ Multiple CBFs are maintained for each specific B-prefix length which increases memory consumption \newline $\bullet$ Increase in prefix length degrades the lookup speed} \\ \hline

BADONA \cite{badona, badona1}, 2014 & \footnotesize{$\bullet$ Efficient membership \newline $\bullet$ Bloom Filter improves route selection \newline $\bullet$ Allows filter updates \newline $\bullet$ Generates less traffic \newline $\bullet$ Updates deployment information without increasing data delivery latency \newline $\bullet$ Retouched Bloom Filter reduces false positive probability} & \footnotesize{ $\bullet$ Maintaining multiple Bloom Filters increases memory consumption \newline $\bullet$ Maintains many copies of the same data \newline $\bullet$ CBF has a higher memory consumption than conventional Bloom Filters}  \\ \hline

PURSUIT \cite{topology}, 2014 & \footnotesize{$\bullet$ Bloom Filter reduces memory consumption \newline $\bullet$ Scalable \newline $\bullet$ Stateless multicast forwarding \newline $\bullet$ zFormation technique improves security} & \footnotesize{$\bullet$ conventional Bloom Filters gives high false positive probability \newline $\bullet$ Large fill factor of FID makes the technique vulnerable to brute-force attack \newline $\bullet$ When the topology manager requires highly optimized multicast trees, using an improved tree routing algorithm increases computation cost}\\ \hline

iSync \cite{iSync}, 2015 & \footnotesize{$\bullet$ Permits different reconciliation process to efficiently skip no update datasets \newline $\bullet$ No requirement for prior context \newline $\bullet$ Out-of-date collections is found within one data exchange} &\footnotesize{ $\bullet$ Maintaining multiple local IBFs increases memory consumption \newline $\bullet$ Packet size is bigger compared to CCNx \newline $\bullet$ Consumes more local computing resources during IBF decoding process} \\ \hline

Chandana \textit{et al.} \cite{Chandana}, 2015 & $\bullet$ Reduces search time \newline $\bullet$ Occupies less memory \newline $\bullet$ Provides high access rate in memory & $\bullet$ Packet already contains the destination name which is different from the principle of NDN \newline $\bullet$ CBF occupies more memory compared to conventional \\ \hline

NDN-NIC \cite{NDN-NIC}, 2016 & \footnotesize{$\bullet$ Reduced CPU usage \newline $\bullet$ Less space complexity \newline $\bullet$ Consumes less memory \newline $\bullet$ Hybrid Bloom Filter allows deletion \newline $\bullet$ Supports variety of name matching rules as per NDN semantics \newline $\bullet$ Basic CS and Active CS reduce the number of names in Bloom Filters to improve the filtering efficiency \newline $\bullet$ Direct Mapping update technique is efficient with fewer entries of the table \newline $\bullet$ Permit processing of partially updated Bloom Filter without increasing false negative issue }& \footnotesize{ $\bullet$ In case of false positive response, packets are checked with matching table that increases CPU overhead \newline $\bullet$ Requires frequent Bloom Filter updates to synchronize with NFD tables \newline
Direct Mapping update technique:
\newline $\bullet$ Performance degrades with increase in the number of table entries \newline $\bullet$ Majority computation overhead is due to hash function computation \newline $\bullet$ Saving hash values increase memory consumption \newline $\bullet$ Receives huge number of Bloom Filter updates \newline
Basic CS update technique:
\newline $\bullet$ When a producer quits, FIB entries are deleted, but the producer data is valid in cache leading to huge overhead \newline $\bullet$ Passively waits for reuse of FIB entries \newline
Active CS update technique: 
\newline $\bullet$ Complex than Direct mapping and Basic CS \newline $\bullet$ Maintains additional information to perform Transformation, Aggregation, or Reversion which increases space complexity \newline $\bullet$ BF-FIB usage increases} \\ \hline

PSync \cite{namesync}, 2017 & \footnotesize{$\bullet$ Efficiently handle both full and partial data synchronization \newline $\bullet$ Efficiently synchronizes sub namespace \newline $\bullet$ Bloom Filter allows efficient computation of set differences \newline $\bullet$ Scalable \newline $\bullet$ Robustness against producer failures} & \footnotesize{$\bullet$ When the subset of a data stream is large, Bloom Filter is compressed but increases false positive probability} \\ \hline

\end{longtable}

In this article, we classified the Bloom Filter based NDN techniques into five major components of NDN (i.e. packet, CS, PIT, FIB and security). However, Bloom Filter is also used in many other topics of NDN. In this section, we tried to discuss those techniques. Table \ref{other_table} illustrates those techniques advantages and disadvantages. 

\subsection{Namespace}
iSync \cite{iSync} is an NDN synchronization protocol. It provides an efficient data reconciliation using two-layer invertible Bloom Filter (IBF) \cite{IBF}. iSync uses two types of IBFs, namely, global and local. Global IBF contains all. Every node maintains their own local IBF. A collection in iSync is a set of content items having a common prefix. iSync has a repository and a sync agent. When a new content name is received, it is matched with the existing local declared collections. Sync Agent notifies the new content name to iSync repository and indexes the inserted content name. Then, the sync agent updates a digest that contains all the content names of the collection. It also sends periodic broadcasts of local digests to remote nodes and receives the remote node digests. Local and remote digests are compared. If they are different, reconciling process is initiated. The digest difference is found by comparing remote IBF tables with local and global IBF tables. iSync uses two-layer IBF, namely, Digest sync IBF and collection sync IBFs. Digest sync IBF maintains status records of the whole repository. A collection sync IBF is maintained for each sync collection. Collection sync IBF record the inserts and deletes of sync collection. When a new content name is added to collection, the content name is hashed and inserted to collection sync IBF. And, when a collection digests changes, then repository IBF is updated. iSync finds out-of-date collections within the single data exchange. When a new content name is detected, the number of updates is checked whether it exceeds the maximum value. If it exceeds, then current IBF is made local IBF and saved as a backup table. The number of updates is set to 0. To reduce memory consumption, the number of IBFs is limited. With expire of iSync timer the current IBF is made of global IBF. Storing the global and local IBFs helps in tracing the difference between IBFs. To reduce false positive results, iSync follows a blacklist scheme. However, consumes more local computing resources during IBF decoding process. iSync maintains multiple local IBFs which increases memory consumption. iSync send bigger packets compared to CCNx. 

A novel Name Lookup engine with Adaptive Prefix Bloom Filter (NLAPB) \cite{Quan} is proposed to enhance the searching efficiency of content name. NLAPB is a hybrid lookup engine that implements CBF \cite{Li_Fan} and trie. It is based on the idea of partitioning the NDN name into B-prefix followed by T-suffix. A lengthy NDN name is partitioned into a fixed sized B-prefix and a variable length T-suffix. Partition of NDN name reduces the entries to CBF, which reduces the false positive probability. When the memory of trie is dynamically allocated, it is highly flexible and scalable. After constructing the T-suffix trie it is binded to B-prefixes using hash table. The hash table stores the root of the trie to assess the T-suffixes. For each length of the B-prefix a corresponding CBF is constructed. During insertion, first the B-prefix is inserted to CBF. Then, update the hash table connecting B-prefixes and T-suffixes. Finally, the suffix of the name is inserted into the trie. During query operation, the B-prefix is searched in the CBF. If found, the hash table is used to access the T-suffix trie. The trie is traversed to match the data name suffix. If found, the face(s) are forwarded.

Chandana \textit{et al.} \cite{Chandana} proposed a new technique for efficient name lookup that have implemented NLAPB Adaptive. The technique divides the content name into two parts, namely,  prefix and suffix. The prefix is stored in the Bloom Filter, hence called B-prefix. The suffix is stored in a trie, hence called a T - suffix. The technique has two stages. The first stage is B-prefix is inserted into a CBF \cite{Li_Fan}. So, CBF is called Adaptive Prefix Bloom Filter (APBF). Second stage is T-suffix which uses a hash table to find the location. It uses root information of trie structure to search the longest matching prefix. For each character the technique assigns a Bloom Filter. A content name is first searched in the Bloom Filter. Bloom Filter returns the longest prefix matching (LPM). T-suffix and LPM is used to search in the hash table to obtain remaining T-suffix. The technique has three modules, namely, Packet listener, lookup destination and index. It also has two databases (a) index and destination name, and (b) next node. Each node has a packet listener. Each packet contains the destination information. The destination name is divided into B-prefix and T-suffix. Next node is determined from the index by matching the B-prefix with the T-suffix. A  route table is also partitioned into two parts. The first part contains the Bloom Filter information. The second part contains the mapping T-suffix information.

NameFilter \cite{NameFilter} is a name lookup technique based on two-stage Bloom Filter. The first stage has multiple Bloom Filters. Each Bloom Filter has a different length. First the length of the name prefix is found. Then, the prefix is inserted in respective Bloom Filter based on length. The Bloom Filter is called One Memory Access Bloom Filter. Because, the Bloom Filter uses a single hash function. Single hash function reduces time complexity. The first hash value is determined by using DJB hash function \cite{DJB}. The next hash function result is obtained from the previous hash result. After $k-1$ loops a single hash result is obtained. Returned hash [0] is utilized to determine the addresses in Bloom Filters. Rest bits of the word are hashed from the respective address determined by the $AND$ operation. When all $k-1$ bits are 1, then prefix is present. Second stage searches for prefix within a small group of Bloom Filters taking the output of the first stage as input. In this stage, the prefixes are grouped based on their associated next-hop port(s). Each group is stored in a Bloom Filter. Searching the Bloom Filters gives the destination port of the respective longest prefix. The number of Bloom Filters used is equal to the number of the router's ports. Size of Bloom Filters are same. Same hash functions are used in all the Bloom Filters. All the Bloom Filters are combined and mapped to a single bit string. This bit string is stored in a merged Bloom Filter. Moving from the most significant bit, the $N^{th}$ bit stores the respective hash result of $N^{th}$ Bloom Filter. The unused remaining bits are assigned 0. Merged Bloom Filter operations are performed in parallel. Hence, it reduces time complexity. NameFilter first uses CBF \cite{Li_Fan}, then it gets converted to boolean Bloom Filter. And, in the second stage it is mapped to the merged Bloom Filter. All operations (insertion, deletion and update) are done incrementally.

PSync \cite{namesync} efficiently addresses different types of data synchronization, full data synchronization as well partial data synchronization. 
PSync uses IBF \cite{IBF} to represent the latest data names. It takes advantage of the subtraction operation on IBF to identify the list of new data names that have been generated. Using this data list and the subscription information of consumers, a producer notifies a consumer if new data matching the subscription are generated. PSync uses Bloom Filters to express their subscription list in their Sync Interests. The subset of data streams is hashed into a Bloom Filter. If the subset is large, then Bloom Filter is compressed to make the subscription list small. But it increases the false positive probability. IBF contains the latest data. IBF only contains one data name from each data stream. IBF instantly finds the differences between two nodes in different synchronization states. PSync is scalable to serve a large number of customers and handle a large number of subscriptions. It is robustness against producer failures.

\subsection{Network Interface Card}
NDN-NIC \cite{NDN-NIC} is network interface card proposed to filter unrequited NDN packets. The purpose is achieved using less CPU, memory and energy resources. NDN-NIC occupies less on-chip memory. NDN-NIC implements a hybrid Bloom Filter of conventional Bloom Filter and CBF \cite{Li_Fan}. CBF is maintained in software and conventional Bloom Filter is maintained at a hardware level. NDN saves rulesets to filter the packets based on the content. But they occupy lots of memory on network cards. NDN-NIC stores these rulesets using three Bloom Filters. Three Bloom Filters are (a) Bloom Filter Forward Information Base (BF-FIB), (b) Bloom Filter Pending Interest Table (BF-PIT), and (c) Bloom Filter Content Store (BF-CS). BF-FIB and BF-PIT saves FIB or PIT entries respectively. BF-CS saves prefixes of CS entry. NDN-NIC has two components, namely, NDN-NIC hardware and NDN-NIC driver. NDN-NIC hardware is the network interface card which filters the incoming packets. NDN-NIC driver instructs the NDN Forwarding Daemon (NFD) what to filter. NDN-NIC driver consists of three Bloom Filters and an update technique. The changes made in table entries need to be reflected in the Bloom Filter. Hence, three Bloom Filter update techniques are proposed, namely, direct mapping, basic CS, and active CS. The update algorithm updates the CBF, then the updates are saved in the Bloom Filter in the hardware. If CBF $counter \neq 0$, then corresponding bits of CBF are set to 1. If CBF counter = 0, then corresponding bits of Bloom Filter are set to 0. This allows to reflect delete updates also in the conventional Bloom Filter. First, all 0-to-1 updates are performed, then all 1-to-0 updates are performed. This helps the hardware to process incoming packets on partially updated Bloom Filter while avoiding false negative issue. In direct mapping technique, each entry is updated in the Bloom Filters. This technique is efficient in less numbered entries of the table. When the number of entries of the table increases the performance of Bloom Filter degrades. Majority computation overhead is due to hash function computation. To lesser the overhead, the hash values are computed once and saved in the name tree nodes. However, this increases memory consumption. In addition, this technique receives a huge amount of Bloom Filter updates dues to saving all changes of the tables. Basic CS updates those names in the Bloom Filter that are not present in the FIB. It reduces the false positive probability. Efficiency of Basic CS is more in case of data received from local producers. BF-CS only stores prefixes smaller than FIB entry. Computational overhead is less compared to direct mapping. NDN-NIC driver checks every name in the name tree node and insert into BF-CS. When a name already present in FIB is encountered the checking is terminated. When a producer quits, FIB entries are deleted, but the producer data is valid in the cache. This leads to huge overhead. To solve the issue basic CS re-enters the CS entries to BF-CS. Active CS is more efficient than Basic CS but more complex. It reduces BF-CS access by inserting appropriate prefixes into BF-FIB without referring corresponding FIB entries. Active CS performs two operations, namely, Transformation and Aggregation. Transformation performs replacement of a single BF-CS name with a BF-FIB prefix. Aggregation performs aggregation of many BF-FIB prefixes to a shorter and common prefix in the BF-FIB. Transformation reduces the access of BF-CS but increases the access of the equal BF-FIB. Whereas, aggregation reduces the access of the BF-FIB. Hence, overall use of BF-FIB increases. In the active CS technique, for every insertion of CS entry, the name tree is updated. Names that are not matching in BF-CS and BF-FIB are updated in BF-CS.


\subsection{Storage}
Bloom Filter Aggregated Data Oriented Networking Architecture (BADONA) \cite{badona,badona1} is a Bloom Filter based approach to solve the storage issue of DONA \cite{DONA}. Addition of Possibility and route selection improve the resolution process of DONA. DONA applies a hierarchical approach to a data oriented network. Route-by-name paradigm is used for name resolution. In BADONA, each router constructs a Bloom Filter for all registrations received. Then, it combines these data structures from the routers and forwards the resulted filter. Bloom Filter is helpful since it enables merge operations. BADONA also uses a clearing mechanism using Retouched Bloom Filters \cite{Retouched}. Location information is inserted like the hop count in a Bloom Filter. CBF \cite{Li_Fan} is used with two vectors. CBF is used since it enables delete operation. The principle vector is used to include data registration and respond to membership queries. If the cell is not used and is set to 0, it means the element is not inserted. If all the cells are used and have the same value, the element is present in the filter and the real hop count is less than or equal to the value. If all the cells are used, but have different values, then the element is only present in the Bloom Filter. The lowest value gives the approximate hop count. For adding new data registered in the filter, the nearest router to the data provider saves it in the registration table and Bloom Filter. The router hashes new registration with every hash function and checks for related generated position. When the data provider forwards deletion requests, the nearest router deletes its entry in the table and updates the filter. It then forwards this information to higher level routers. In this approach, membership queries are easily solved using Bloom Filters. Filter updating is also allowed. Traffic is limited because Bloom Filters are exchanged only between routers. However, there is a possibility of having many duplicate copies of the same data. Another advantage is it has the ability to update deployment information without increasing data delivery delay.

\section{Trending topics in NDN}
\label{trend}
In this section, some emerging topics are discussed. A short description to understand the new technology and precise elaboration of the Bloom Filter based technique proposed to improve the technology. Table \ref{trending_table} present the list of advantages and disadvantages of the techniques discussed in the section. 

\subsection{5G}
With the growth of dependence on mobile devices, the wireless world is now moving to 5G (fifth generation). 5G is capable to provide increased wireless data rates, coverage, bandwidth, and connectivity. Moreover, the power consumption and round trip latency is less. 5G satisfies eight major requirements \cite{Agiwal}; (i) in real networks,  provides 1-10 Gbps data rates, (ii) round trip latency takes 1 ms, (iii) Provide high bandwidth in unit area, (iv) capable to handle large number of connected devices, (v) always available, (vi) complete coverage, (vii) low power consumption, and (viii) Long battery life. However, with the 5G new security requirement is also essential. New security solutions and architecture is required due to the new services and technologies in 5G \cite{Ahmad}. 5G in NDN is also being explored.

B-MaFIB \cite{B-MaFIB} is proposed to enhance the FIB performance in NDN to support heterogeneous networks for 5G. It uses B-MBF as an index. B-MaFIB allocate memory dynamically to reduce on-chip memory consumption. It consists of two storages. First is SRAM on-chip memory. It deploys multiple MBFs. Second is DRAM off-chip memory. It deploys  CBFs, multiple bitmaps, and named packet stores. B-MBF is searched to provide a corresponding offset address. CBF synchronizes with Bloom Filter of on-chip memory. Each bitmap is one packet store. It stores the forwarding information. Moreover, offset addresses of packet store are the sequence numbers stored in the bitmap. Entries mapped to the same location in the bitmap are chained to the corresponding packet store entry. This resolves the hash conflicts. After receiving an Interest, first its presence is checked in B-MBFs. If the longest matching name prefix is present, the corresponding offset address stored in the bitmap is obtained using MA value. Offset address is used to obtain the base address of packet store. This helps to retrieve the forwarding information. The time complexity of B-MaFIB is $O(1)$. However, the usage of multiple B-MBFs increases the total false positive probability. 

\subsection{IoT}
IoT played a big role in making the devices smarter. It provides services that enable the devices to hear, see, think, and communicate with each other. Moreover, people are connected to each other even when they are moving. It provides data storage and data processing round the clock. It also ensures availability, security and many more functionalities \cite{Al-Fuqaha}. Another emerging technology is fog computing \cite{Mukherjee}. It supports delay-sensitive service requests from user using less power and low traffic. Fog nodes based on the request decides whether to process the request in its node or send to cloud server. Fog computing achieves higher performance, and bandwidth with lower power consumption. Also, Edge computing is an emerging technology for mission-critical applications. It has high performing capabilities for providing real-time data analysis, high scalability, low operational cost, reduced latency and improved quality of service \cite{El-Sayed}. Moreover, with the emergence of 5G, mobile edge computing is also emerging to support 5G \cite{Mao}. 

Bloom Filter Based Hierarchical Hybrid Mobility Management (BFH\textsuperscript{2}M\textsuperscript{2}) \cite{iot} consists of three layers: fog, cloud and edge. This method implements various Bloom Filters: CBF \cite{Li_Fan}, Mapping Bloom Filter information base (MaBFIB) and Attenuated Bloom Filter (ABF) \cite{Atten}. Each layer uses Bloom Filters for storing location information because BFH\textsuperscript{2}M\textsuperscript{2} uses different location schemes. Bloom Filter is used for the lookup operation to locate the node in the indirect location scheme. The direct location scheme uses CBF for location update operation. CBF is utilized to support the insertion or deletion of pattern. MBF manages querying and mapping the set elements in the memory. Using MBF also lowers on-chip memory consumption. Fog routers maintain MaBFIB that minimizes memory space and accelerates the lookup process. In case of hot devices, upon entering geographically into local edge networks, the Edge router sends location information using the Bloom Filter to the upstream Fog router. Bloom Filter is recognized using the device ID and local locator information (LLI). And, aggregated Bloom Filters are stored in the CBF of MaBFIB for in-network routing computation. Fog router then sends the CBF of ABF to upstream cloud. For cold devices, each router aggregates their routing information towards Fog router periodically by performing bitwise OR operation in its Bloom Filters. Then, it eventually sends the aggregated Bloom Filters to the upstream cloud gateway.

\begin{table}[!ht]
\centering \small
\caption{\textbf{Techniques in trending topics in NDN}}
    \begin{tabular}{|p{1.25cm}|p{5cm}|p{5.5cm}|}
    \hline
    \centering \textbf{Paper} & \centering \textbf{Advantages} & \centering \textbf{Disadvantages} \tabularnewline \hline \hline
    
B-MaFIB \cite{B-MaFIB}, 2018 & $\bullet$ Supports heterogeneous network for 5G \newline $\bullet$ Dynamically allocate memory to reduce off-chip memory consumption \newline $\bullet$ Stable forwarding performance even when the number of names stored is high \newline $\bullet$ Reduces searching time & $\bullet$ Multiple B-MBFs increases the total false positive probability \newline $\bullet$ Conventional Bloom Filter in on-chip memory, have high false positive which affects the performance of B-MaFIB \\ \hline

BFH\textsuperscript{2}M\textsuperscript{2} \cite{iot}, 2019 & $\bullet$ MaBFIB minimizes memory space \newline $\bullet$ MaBFIB accelerates the lookup process \newline $\bullet$ Edge nodes minimizes delay, bandwidth, and cross-domain traffic \newline $\bullet$ 
Edge nodes also make better context-awareness and location-awareness decisions & $\bullet$ In MBF, off-chip memory is wasted \newline $\bullet$ UCBF increases memory consumption \\ \hline 

 \end{tabular}
    \label{trending_table}
\end{table}

\section{Discussion} 
\label{dis}
\begin{longtable}{|p{2.5cm}|p{0.7 cm}|p{2cm}|p{2cm}|p{0.4cm}|p{0.4cm}|p{2cm}|}
    \caption{\textbf{Comparison of NDN techniques (FP: F\MakeTextLowercase{alse positive and} FN: F\MakeTextLowercase{alse Negative}} *N\MakeTextLowercase{ot mentioned that} CBF \MakeTextLowercase{is implemented}. A\MakeTextLowercase{lso not mentioned any counting mechanism})} 
    \label{overall_table} \\ 
    \hline 
     \centering \textbf{Paper} & \centering \textbf{Year} & \centering \textbf{Bloom Filter} & \centering \textbf{Method to reduce False Positive} & \centering \textbf{FP} & \centering \textbf{FN} & \centering \textbf{Bloom Filter Usage}  
     \endfirsthead
\hline \hline
\centering \textbf{Paper} & \centering \textbf{Year} &\centering \textbf{Bloom Filter} & \centering \textbf{Method to reduce False Positive} & \centering \textbf{FP} & \centering \textbf{FN} & \centering \textbf{Bloom Filter Usage} 
\endhead \hline 
SCAN \cite{scan,scan1} & 2011 & Conventional & Information Decaying & \checkmark & \checkmark & Content Router\\ \hline
BlooGo \cite{BLooGo} & 2012 & Conventional & None & \checkmark & X & Packet  \\ \hline
So \textit{et al.} \cite{So} & 2012 & Conventional & None & \checkmark & X & Forwarding lookup engine \\ \hline
Massawe \textit{et al.} \cite{Massawe} & 2013 & Conventional & Multiple Bloom Filters & X & X & Packet \\ \hline
NameFilter \cite{NameFilter} & 2013 & CBF & None & X & \checkmark & Lookup Engine \\ \hline
MaPIT \cite{MaPIT} & 2014 & Mapping \cite{MaPIT} & MBF & \checkmark & \checkmark & PIT \\ \hline
NLAPB \cite{Quan} & 2014 & CBF & Partition of NDN name & X & \checkmark & Lookup Engine \\ \hline
Dai \textit{et al.} \cite{Dai} & 2014 & Conventional* & None & \checkmark & X & CS \\ \hline
COBRA \cite{COBRA} & 2014 & Stable \cite{Deng} & None & \checkmark & X & Interface \\ \hline
BADONA \cite{badona, badona1} & 2014 & CBF and Retouched \cite{Retouched} & Introduces false negative & X & \checkmark & Storage space reduction\\ \hline
Chen \textit{et al.} \cite{Chen} & 2014 & Conventional & None & \checkmark & X & CS \\ \hline
PURSUIT \cite{topology} & 2014 & Conventional & None & \checkmark & X & Forwarding \\ \hline
CCNxTomcat \cite{tomcat} & 2014 & CBF & None & X & \checkmark & Cache and Web Server \\ \hline
iSync \cite{iSync} & 2015 & Invertible \cite{IBF} & Blacklist scheme & X & X & Reconciliation process \\ \hline
CCFS \cite{Aloulou} & 2015 & Stable \cite{Deng} & Stable BF has less false positive probability & \checkmark & X & CS \\ \hline
NDN-NIC \cite{NDN-NIC} & 2016 & Hybrid of CBF and Conventional & Absent content name is inserted into Bloom Filter & \checkmark & \checkmark & Network interface card \\  \hline
Guo \textit{et al.} \cite{Guo} & 2016 & Conventional & None & \checkmark & X & Cache \\ \hline
FNR \cite{FNR} & 2016 & Conventional & None & \checkmark & X & CS \\ \hline
Lee \textit{et al.} \cite{prematchBF} & 2016 & Conventional & Increases Bloom Filter size & X & X &FIB \\ \hline
Hardware Accelerator \cite{hardware} & 2016 & Conventional & Increase number of hash functions & X & \checkmark & PIT \\ \hline
Mun \textit{et al.} \cite{mun, mun1} & 2017 & Stateful \cite{Bonomi} & None & \checkmark & X & Packet \\ \hline
BFAST \cite{bfast} & 2017 & CBF & None & X & \checkmark & FIB \\ \hline
CT-BF \cite{CT-BF} & 2017 & Conventional & None & \checkmark & X & CS \\ \hline
BFR \cite{BFR} & 2017 & Conventional & None & \checkmark & X & Packet \\ \hline
BBS \cite{He} & 2017 & Conventional & None & \checkmark & X & Lookup Engine \\\hline
MaFIB \cite{Li} & 2017 & Mapping \cite{MaPIT} and CBF & None & X & \checkmark & FIB \\ \hline
PSync \cite{namesync} & 2017 & Invertible \cite{IBF} & None & \checkmark & X & Data Synchronization \\ \hline
Watano \textit{et al.} \cite{reroute} & 2018 & Conventional & None & \checkmark & X & Router \\ \hline
BRCC \cite{Hou} & 2018 & Sum-up Counting \cite{Hou} & Sum-up table and hash-based Look-up Table & \checkmark &  \checkmark & Cache \\ \hline
Bourtsoulatze \textit{et al.} \cite{Bourtsoulatze} & 2018 & Conventional & None & \checkmark & X & Packet \\ \hline
B-MaFIB \cite{B-MaFIB} & 2018 & CBF & None & \checkmark & \checkmark & FIB \\ \hline
Pull-based BFR \cite{PB-BFR} & 2019 & Conventional &  None & \checkmark & X & Packet \\ \hline
FTDF-PIT \cite{Shub, Shubbar} & 2019 & Fast Two Dimensional \cite{Shub,Shubbar} & Fingerprint & X & X & PIT \\ \hline
BFH\textsuperscript{2}M\textsuperscript{2} \cite{iot} & 2019 & CBF, Mapping \cite{MaPIT} and Attenuated \cite{Atten} & None & X & \checkmark & FIB \\ \hline
Wu \textit{et al.} \cite{smarthome} & 2019 & CBF & None & \checkmark & \checkmark & Packet \\ \hline

\end{longtable}

\subsection{Bloom Filter}
As from our past experience, the performance of the NDN depends on Bloom Filter. It depends on false positive, false negative, true positive and true negative. False positives adds an extra lookup, albeit the lookup cost is negligible. If NDN is implemented using high accuracy Bloom Filter, the performance will also increase. Moreover, there are false negatives in Bloom Filter other that CBF variants. However, the trade-off is that CBF exhibits a higher false positive rate than a non-CBF while there is no false negative in CBF. The key advantage of using Bloom Filter is true negative. The true negative prevents unnecessary lookup/update/write in slow storage media. Therefore, Bloom Filter should be used only when there is a huge probability of true negative. Otherwise, it adds extra complexity to the system. For instance, a true positive results lookup operation in Bloom Filter along with other operations in storage media. 

\subsubsection{Implementation of Bloom Filter}
The performance of a Bloom Filter depends on the number of arithmetic operations performed to insert, delete, lookup, or update an item. We must avoid costly operator while implementing a Bloom Filter. For instance, $2^n$ can be rewritten as $2<<n$. The bitwise operators are the fastest operators and these operators are used to implement Bloom Filter. The modulus operator (\%) is the costliest operator which must be avoided to achieve higher performance of Bloom Filter. However, it is very difficult to avoid modulus operator, since, Bloom Filter inherits the hashing property. Also, the operators, $\times$, and $\div$ are avoided in the implementation of Bloom Filter. However, it is observed that the implementation of a Bloom Filter in C/C++ programming language is faster than Java and Python programming languages.

\subsubsection{Number of hash functions}
As we experienced in implementing Bloom Filter, the false positive, false negative, true positive and true negative of Bloom Filter depend on the number of hash functions. In Bloom Filter, it is the most important to choose the optimal number of hash functions. The large number of hash functions in Bloom Filter causes a huge false positive, because, most of the slots are filled with `1' in the Bloom Array with a few input items. Also, less number of hash functions in Bloom Filter causes same issue, because, lower number of hash functions cause higher collision probability. Therefore, the number of the hash functions must be taken care of while implementing NDN system. 

\subsubsection{Separation of the hash functions}
As we know that Bloom Filter uses a tiny amount of main memory to store huge information about data, since, it only stores the membership information of the data. Typically, memory consumption of Bloom Filter ranges from a few KB to MB. The false positive is a big issue of Bloom Filter. However, the false positive probability can be reduced by separating the hash functions which introduce multiple Bloom Filters. In this method, a few amount of main memory is sacrificed to achieve a higher accuracy. This approach has been implemented in HFil \cite{HFil} and PassDB \cite{PassDB}, and drastically improve the accuracy. However, the performance remains same.

\subsection{NDN}
NDN architecture consists of many data structures for the smooth forwarding of packets. These data structures need to handle Interest/Data packets at a frequency of millions per second. Mostly, those packets are duplicate packets. Hence, the application of Bloom Filter in NDN is greatly explored. In this article, it can be concluded that major components of NDN are exploring the use of Bloom Filter. However, one major concern is the underestimation of issues of Bloom Filter. As shown in the Table \ref{overall_table}, except a few, almost all techniques are implemented using conventional Bloom Filters. But, conventional Bloom Filter has a high false negative probability with deletion support. Therefore, the conventional Bloom Filter does not permit delete operation. Bloom Filter stores content names. Interests are requested at a frequency of millions per second. All unique content name is inserted into Bloom Filter. And, when saturated, Bloom Filter gives more false positive responses. Moreover, the Bloom Filter have to keep storing old packets due to restriction on performing delete operation. In addition, when the techniques are using the conventional Bloom Filter no extra mechanism or method is implemented to reduce the false negative probability. 

To permit the delete operation, CBF is utilized. However, CBF has four times more memory consumption compared to conventional Bloom Filter. Moreover, many NDN techniques construct a single Bloom Filter for each interface/faces (eg. Chen \textit{et al.} \cite{Chen}, Mun \textit{et al.} \cite{mun, mun1}) or maintains multiple Bloom Filters (eg. BADONA \cite{badona}, NameFilter \cite{NameFilter}, FNR \cite{FNR}). For example, BlooGo \cite{BLooGo} constructs Bloom Filter for each packet. Hence, the technique has to maintain many Bloom Filters, and, have to perform construction and deletion of Bloom Filters very frequently. But, NDN router has a limited memory. It has to maintain three data structures, namely CS, PIT and FIB. Hence, maintaining multiple Bloom Filters occupies a large amount of memory. Therefore, Bloom Filter needs to occupy less memory while stores millions of unique contents.  

Bloom Filter's simplicity improves its flexibility which enables the deployment of Bloom Filter in any component of NDN. Table \ref{overall_table} proves Bloom Filter is implemented nearly in all components of NDN. Mostly, Bloom Filter is implemented in the three data structures (CS, PIT and FIB) of NDN. However, only deploying Bloom Filter is not the optimal solution. Along with Bloom Filter, its issues need to be solved. Along with new NDN techniques, there is a requirement for new variants of Bloom Filter. These variants can help NDN to achieve the optimal performance. Numerous variants of Bloom Filters are introduced and incorporating these modern Bloom Filter which can enhance the performance of NDN dramatically. For instance, rDBF \cite{rDBF}.
  
\subsubsection{Content Store}
CS is the cache that stores data for future requests. It also has a high read and write operation frequency. Therefore, deploying Bloom Filter before the buffer will improve efficiency. Before searching in CS the Bloom Filter is searched. Bloom Filter returns the response in O(1). In the case of $False$, searching in CS is not required and the data are directly written. And, in the case of $True$, data does not need to be cached. Hence, eliminating write operation. CS size is based on the size of the edges and the router. And, Bloom Filter occupies less memory. Therefore, deploying Bloom Filter is a tiny overhead. CS requires an efficient cache replacement policy. Linked list in the skip list helps in maintaining the order of data storage. Hence, skip lists are widely adopted for CS. However, skip list time complexity of query operation is more. Query operation is a highly executed operation for making caching decisions. Therefore, using a skip list is costly. Again, Bloom Filter is more efficient than the skip list. CS is implemented to store popular Interests. Therefore, a Bloom filter variant capable of storing huge input and occupying tiny memory is appropriate for CS. 

\subsubsection{Pending Interest Table}
In PIT, the incoming Interest packet speed is very high. Bloom Filter insertion time complexity is O(1), hence, it is able to cope with the high speed insertion of packets. PIT requires a large sized data structure to store the packets without overflow. And, Bloom Filter occupies less memory. Therefore, Bloom Filter is not much of an overhead to PIT. All the operations (insert, search and delete) performed in PIT need to be highly efficient to coup with the higher packet incoming speed. As discussed in section \ref{BF}, the time complexity of all Bloom Filter operations is O(1). Another important requirement of PIT is deletion of entries to avoid saturation of the data structure. Therefore, PIT requires a Bloom Filter that permits deletion operation along with reducing Bloom Filter issues (i.e., false positive and false negative). Some examples are Deletable Bloom Filter \cite{DBF} and Bloofi \cite{Adina}.

\subsection{rDBF: A Bloom Filter solution for CS, PIT and FIB}
rDBF \cite{rDBF} is a novel $r$-dimensional Bloom Filter. It is a very fast Bloom Filter. It has less false positives and is free from false negatives. It has high scalability and high adaptability. A small sized rDBF also exhibits very good performance. The input element is mapped to a single bit in rDBF. Hence, it consumes very tiny memory and stores a huge number of elements while maintaining a low false positive probability. Therefore, deploying rDBF in CS is ideal. Again, as per the requirements of PIT, rDBF does not have to be very big sized to store lots of packets. Moreover, all operations of rDBF are very efficient. In addition, rDBF supports delete operation without increasing false negative probability. Similarly, rDBF is ideal for PIT. As mentioned earlier, a small sized rDBF also exhibits very good performance. Likewise, rDBF is an ideal choice for FIB. rDBF has high accuracy compared to other Bloom Filters. Henceforth, rDBF is an ideal Bloom Filter which is capable of satisfying all the requirements of NDN. The derivatives of rDBF, acBF \cite{acBF} and HFil \cite{HFil}, can also be deployed in NDN for higher accuracy while maintaining same performance.

\section{Conclusion}
\label{con}
With a billion nodes and million requests per second, NDN is becoming extremely popular to meet current requirements of upcoming Internet users. Along with NDN, researchers are exploring Bloom Filter for increasing the performance of the NDN. NDN consists of many data structures for enhancing the movement of packets to serve the users. As discussed in this article, all main components of NDN are implementing Bloom Filter. Even the latest emerging technology of NDN, such as 5G and IoT is utilizing Bloom Filter in their implementation. However, Bloom Filter has two main issues, namely, false positive and false negative. It can be concluded from the last table presented in the article, researchers are focusing on an efficient technique but ignoring the issues of Bloom Filter. Luo \textit{et al.} \cite{Luo} also demonstrates the techniques to optimize the Bloom Filter which is important for NDN design and development. Therefore, along with new NDN techniques, there is a requirement for optimization of Bloom Filter with less false positive and false negative probability.

\bibliographystyle{abbrv}
\bibliography{mybibfile}

\begin{thebibliography}{100}

\bibitem{badona}
A.~Abidi, S.~Mettali~Gammar, F.~Kamoun, W.~Dabbous, and T.~Turletti.
\newblock Towards a new internetworking architecture: A new deployment approach
  for information centric networks.
\newblock In M.~Chatterjee, J.-n. Cao, K.~Kothapalli, and S.~Rajsbaum, editors,
  {\em Distributed Computing and Networking}, pages 519--524, Berlin,
  Heidelberg, 2014. Springer Berlin Heidelberg.

\bibitem{Aboodi1}
A.~{Aboodi}, T.~{Wan}, and G.~{Sodhy}.
\newblock Survey on the incorporation of ndn/ccn in iot.
\newblock {\em IEEE Access}, 7:71827--71858, 2019.

\bibitem{zipf}
L.~A. Adamic and B.~A. Huberman.
\newblock Zipf's law and the internet.
\newblock {\em Glottometrics}, 3(1):143--150, 2002.

\bibitem{NDN3}
A.~Afanasyev, J.~Burke, T.~Refaei, L.~Wang, B.~Zhang, and L.~Zhang.
\newblock A brief introduction to named data networking.
\newblock In {\em MILCOM 2018-2018 IEEE Military Communications Conference
  (MILCOM)}, pages 1--6. IEEE, 2018.

\bibitem{ndnSIM1}
A.~Afanasyev, I.~Moiseenko, L.~Zhang, et~al.
\newblock ndnsim: Ndn simulator for ns-3.
\newblock {\em University of California, Los Angeles, Tech. Rep}, 4, 2012.

\bibitem{Agiwal}
M.~{Agiwal}, A.~{Roy}, and N.~{Saxena}.
\newblock Next generation 5g wireless networks: A comprehensive survey.
\newblock {\em IEEE Communications Surveys Tutorials}, 18(3):1617--1655,
  thirdquarter 2016.

\bibitem{Ahed}
K.~{Ahed}, M.~{Benamar}, and R.~{El Ouazzani}.
\newblock Content delivery in named data networking based internet of things.
\newblock In {\em 2019 15th International Wireless Communications Mobile
  Computing Conference (IWCMC)}, pages 1397--1402, June 2019.

\bibitem{ICN}
B.~Ahlgren, C.~Dannewitz, C.~Imbrenda, D.~Kutscher, and B.~Ohlman.
\newblock A survey of information-centric networking.
\newblock {\em IEEE Communications Magazine}, 50(7):26--36, 2012.

\bibitem{Ahmad}
I.~{Ahmad}, S.~{Shahabuddin}, T.~{Kumar}, J.~{Okwuibe}, A.~{Gurtov}, and
  M.~{Ylianttila}.
\newblock Security for 5g and beyond.
\newblock {\em IEEE Communications Surveys Tutorials}, pages 1--1, 2019.

\bibitem{Al-Fuqaha}
A.~{Al-Fuqaha}, M.~{Guizani}, M.~{Mohammadi}, M.~{Aledhari}, and M.~{Ayyash}.
\newblock Internet of things: A survey on enabling technologies, protocols, and
  applications.
\newblock {\em IEEE Communications Surveys Tutorials}, 17(4):2347--2376,
  Fourthquarter 2015.

\bibitem{Aloulou}
N.~{Aloulou}, M.~{Ayari}, M.~F. {Zhani}, and L.~{Saidane}.
\newblock A popularity-driven controller-based routing and cooperative caching
  for named data networks.
\newblock In {\em 2015 6th International Conference on the Network of the
  Future (NOF)}, pages 1--5, Sep. 2015.

\bibitem{NDN-FS}
N.~{Aloulou}, M.~{Ayari}, M.~F. {Zhani}, L.~{Saidane}, and G.~{Pujolle}.
\newblock Taxonomy and comparative study of ndn forwarding strategies.
\newblock In {\em 2017 Sixth International Conference on Communications and
  Networking (ComNet)}, pages 1--8, March 2017.

\bibitem{topology}
B.~A. Alzahrani, M.~J. Reed, J.~Riihijärvi, and V.~G. Vassilakis.
\newblock Scalability of information centric networking using mediated topology
  management.
\newblock {\em Journal of Network and Computer Applications}, 50:126 -- 133,
  2015.

\bibitem{Marandi}
A.{Marandi}, T.{Braun}, K.{Salamatian}, and N.{Thomos}.
\newblock A comparative analysis of bloom filter-based routing protocols for
  information-centric networks.
\newblock In {\em 2018 IEEE Symposium on Computers and Communications (ISCC)},
  pages 00255--00261, June 2018.

\bibitem{BLooGo}
F.~Angius, M.~Gerla, and G.~Pau.
\newblock Bloogo: Bloom filter based gossip algorithm for wireless ndn.
\newblock In {\em Proceedings of the 1st ACM Workshop on Emerging Name-Oriented
  Mobile Networking Design - Architecture, Algorithms, and Applications}, NoM
  '12, pages 25--30, New York, NY, USA, 2012. ACM.

\bibitem{murmur}
A.~Appleby.
\newblock Murmur hashing.
\newblock Retrieved from https://sites.google.com/site/murmurhash/.

\bibitem{Syambas}
W.~T. {Ariefianto} and N.~R. {Syambas}.
\newblock Routing in ndn network: A survey and future perspectives.
\newblock In {\em 2017 11th International Conference on Telecommunication
  Systems Services and Applications (TSSA)}, pages 1--6, Oct 2017.

\bibitem{iot}
I.-H. Bae.
\newblock Design and evaluation of a bloom filter based hierarchical hybrid
  mobility management scheme for internet of things.
\newblock In K.~J. Kim and H.~Kim, editors, {\em Mobile and Wireless Technology
  2018}, pages 3--15, Singapore, 2019. Springer Singapore.

\bibitem{DJB}
D.~J. Bernstein.
\newblock Djb hash.
\newblock Accessed on 26 August 2019 from
  \url{http://www.partow.net/programming/hashfunctions/#DJBHashFunction}.

\bibitem{Bloom}
B.~H. Bloom.
\newblock Space/time trade-offs in hash coding with allowable errors.
\newblock {\em Comm. of the ACM}, 13(7):422--426, 1970.

\bibitem{Bonomi}
F.~Bonomi, M.~Mitzenmacher, R.~Panigrah, S.~Singh, and G.~Varghese.
\newblock Beyond bloom filters: from approximate membership checks to
  approximate state machines.
\newblock {\em ACM SIGCOMM Computer Communication Review}, 36(4):315--326,
  2006.

\bibitem{Bourtsoulatze}
E.~{Bourtsoulatze}, N.~{Thomos}, J.~{Saltarin}, and T.~{Braun}.
\newblock Content-aware delivery of scalable video in network coding enabled
  named data networks.
\newblock {\em IEEE Transactions on Multimedia}, 20(6):1561--1575, June 2018.

\bibitem{Broder}
A.~Broder and M.~Mitzenmacher.
\newblock Network applications of bloom filters: A survey.
\newblock {\em Internet Mathematics}, 1(4):485--509, 2004.

\bibitem{FNR}
J.~{Cao}, D.~{Pei}, X.~{Zhang}, B.~{Zhang}, and Y.~{Zhao}.
\newblock Fetching popular data from the nearest replica in ndn.
\newblock In {\em 2016 25th International Conference on Computer Communication
  and Networks (ICCCN)}, pages 1--9, Aug 2016.

\bibitem{Chandana}
K.~L. {Chandana}, S.~B. {Patil}, N.~L. {Taranath}, and P.~{Patil}.
\newblock Efficient lookup for nlapb in named data networking.
\newblock In {\em 2015 International Conference on Applied and Theoretical
  Computing and Communication Technology (iCATccT)}, pages 27--32, Oct 2015.

\bibitem{NDN-sec}
T.~{Chatterjee}, S.~{Ruj}, and S.~D. {Bit}.
\newblock Security issues in named data networks.
\newblock {\em Computer}, 51(1):66--75, January 2018.

\bibitem{Chen3}
Q.~{Chen}, R.~{Xie}, F.~R. {Yu}, J.~{Liu}, T.~{Huang}, and Y.~{Liu}.
\newblock Transport control strategies in named data networking: A survey.
\newblock {\em IEEE Communications Surveys Tutorials}, 18(3):2052--2083,
  thirdquarter 2016.

\bibitem{Chen2}
S.~Chen and F.~Mizero.
\newblock A survey on security in named data networking.
\newblock {\em CoRR}, abs/1512.04127, 2015.

\bibitem{Chen}
T.~{Chen}, K.~{Lei}, and K.~{Xu}.
\newblock An encryption and probability based access control model for named
  data networking.
\newblock In {\em 2014 IEEE 33rd International Performance Computing and
  Communications Conference (IPCCC)}, pages 1--8, Dec 2014.

\bibitem{CDN}
J.~{Choi}, J.~{Han}, E.~{Cho}, T.~{Kwon}, and Y.~{Choi}.
\newblock A survey on content-oriented networking for efficient content
  delivery.
\newblock {\em IEEE Communications Magazine}, 49(3):121--127, March 2011.

\bibitem{Adina}
A.~Crainiceanu and D.~Lemire.
\newblock Bloofi: Multidimensional bloom filters.
\newblock {\em Information Systems}, 54(Supplement C):311 -- 324, 2015.

\bibitem{PIT1}
H.~Dai, B.~Liu, Y.~Chen, and Y.~Wang.
\newblock On pending interest table in named data networking.
\newblock In {\em Proceedings of the Eighth ACM/IEEE Symposium on Architectures
  for Networking and Communications Systems}, ANCS '12, pages 211--222, New
  York, NY, USA, 2012. ACM.

\bibitem{BFAST1}
H.~{Dai}, J.~{Lu}, Y.~{Wang}, and B.~{Liu}.
\newblock Bfast: Unified and scalable index for ndn forwarding architecture.
\newblock In {\em 2015 IEEE Conference on Computer Communications (INFOCOM)},
  pages 2290--2298, April 2015.

\bibitem{bfast}
H.~{Dai}, J.~{Lu}, Y.~{Wang}, T.~{Pan}, and B.~{Liu}.
\newblock Bfast: High-speed and memory-efficient approach for ndn forwarding
  engine.
\newblock {\em IEEE/ACM Transactions on Networking}, 25(2):1235--1248, April
  2017.

\bibitem{Dai}
H.~{Dai}, Y.~{Wang}, H.~{Wu}, J.~{Lu}, and B.~{Liu}.
\newblock Towards line-speed and accurate on-line popularity monitoring on ndn
  routers.
\newblock In {\em 2014 IEEE 22nd International Symposium of Quality of Service
  (IWQoS)}, pages 178--187, May 2014.

\bibitem{BloomFlash}
B.~Debnath, S.~Sengupta, J.~Li, D.~J. Lilja, and D.~H.~C. Du.
\newblock Bloomflash: Bloom filter on flash-based storage.
\newblock In {\em 2011 31st International Conference on Distributed Computing
  Systems}, pages 635--644, 2011.

\bibitem{Deng}
F.~Deng and D.~Rafiei.
\newblock Approximately detecting duplicates for streaming data using stable
  bloom filters.
\newblock In {\em Proceedings of the 2006 ACM SIGMOD International Conference
  on Management of Data}, SIGMOD '06, pages 25--36, New York, NY, USA, 2006.
  ACM.

\bibitem{NDN-sec2}
W.~{Ding}, Z.~{Yan}, and R.~H. {Deng}.
\newblock A survey on future internet security architectures.
\newblock {\em IEEE Access}, 4:4374--4393, 2016.

\bibitem{Retouched}
B.~Donnet, B.~Baynat, and T.~Friedman.
\newblock Improving retouched bloom filter for trading off selected false
  positives against false negatives.
\newblock {\em Comput. Netw.}, 54(18):3373--3387, Dec. 2010.

\bibitem{SHA3}
M.~J. Dworkin.
\newblock Sha-3 standard: Permutation-based hash and extendable-output
  functions.
\newblock Technical report, National Institute of Standards and Technology,
  2015.

\bibitem{SHA1}
D.~Eastlake and P.~Jones.
\newblock Us secure hash algorithm 1 (sha1), 2001.

\bibitem{Bakkouchi}
A.~{EL-Bakkouchi}, A.~{Bouayad}, and M.~E. {Bekkali}.
\newblock A hop-by-hop congestion control mechanisms in ndn networks – a
  survey.
\newblock In {\em 2019 7th Mediterranean Congress of Telecommunications (CMT)},
  pages 1--4, Oct 2019.

\bibitem{El-Sayed}
H.~{El-Sayed}, S.~{Sankar}, M.~{Prasad}, D.~{Puthal}, A.~{Gupta}, M.~{Mohanty},
  and C.~{Lin}.
\newblock Edge of things: The big picture on the integration of edge, iot and
  the cloud in a distributed computing environment.
\newblock {\em IEEE Access}, 6:1706--1717, 2018.

\bibitem{Cuckoo}
B.~Fan, D.~G. Andersen, M.~Kaminsky, and M.~D. Mitzenmacher.
\newblock Cuckoo filter: Practically better than bloom.
\newblock In {\em Proceedings of the 10th ACM Intl. Conf. on Emerging
  Networking Experiments and Technologies}, CoNEXT '14, pages 75--88, Sydney,
  Australia, 2014.

\bibitem{countingBF}
L.~Fan, P.~Cao, J.~Almeida, and A.~Z. Broder.
\newblock Summary cache: A scalable wide-area web cache sharing protocol.
\newblock {\em IEEE/ACM Trans. Netw.}, 8(3):281--293, June 2000.

\bibitem{Fang}
C.~{Fang}, H.~{Yao}, Z.~{Wang}, W.~{Wu}, X.~{Jin}, and F.~R. {Yu}.
\newblock A survey of mobile information-centric networking: Research issues
  and challenges.
\newblock {\em IEEE Communications Surveys Tutorials}, 20(3):2353--2371,
  thirdquarter 2018.

\bibitem{ICN2}
C.~{Fang}, H.~{Yao}, Z.~{Wang}, W.~{Wu}, X.~{Jin}, and F.~R. {Yu}.
\newblock A survey of mobile information-centric networking: Research issues
  and challenges.
\newblock {\em IEEE Communications Surveys Tutorials}, 20(3):2353--2371,
  thirdquarter 2018.

\bibitem{Fayazbakhsh}
S.~K. Fayazbakhsh, Y.~Lin, A.~Tootoonchian, A.~Ghodsi, T.~Koponen, B.~Maggs,
  K.~Ng, V.~Sekar, and S.~Shenker.
\newblock Less pain, most of the gain: Incrementally deployable icn.
\newblock In {\em Proceedings of the ACM SIGCOMM 2013 Conference on SIGCOMM},
  SIGCOMM '13, pages 147--158, New York, NY, USA, 2013. ACM.

\bibitem{Feng1}
B.~Feng, H.~Zhou, and Q.~Xu.
\newblock Mobility support in named data networking: a survey.
\newblock {\em EURASIP Journal on Wireless Communications and Networking},
  2016(1):220, 2016.

\bibitem{iSync}
W.~Fu, H.~B. Abraham, and P.~Crowley.
\newblock Synchronizing namespaces with invertible bloom filters.
\newblock In {\em 2015 ACM/IEEE Symposium on Architectures for Networking and
  Communications Systems (ANCS)}, pages 123--134. IEEE, 2015.

\bibitem{gasti}
P.~Gasti, G.~Tsudik, E.~Uzun, and L.~Zhang.
\newblock Dos and ddos in named data networking.
\newblock In {\em 2013 22nd International Conference on Computer Communication
  and Networks (ICCCN)}, pages 1--7. IEEE, 2013.

\bibitem{GERAVAND1}
S.~Geravand and M.~Ahmadi.
\newblock Bloom filter applications in network security: A state-of-the-art
  survey.
\newblock {\em Computer Networks}, 57(18):4047 -- 4064, 2013.

\bibitem{IBF}
M.~T. Goodrich and M.~Mitzenmacher.
\newblock Invertible bloom lookup tables.
\newblock In {\em 2011 49th Annual Allerton Conference on Communication,
  Control, and Computing (Allerton)}, pages 792--799. IEEE, 2011.

\bibitem{GreenICN}
GreenICN.
\newblock Architecture and applications of green information centric
  networking.
\newblock Accessed on 26th June 2019 from \url{http://www.greenicn.org/}.

\bibitem{Guo}
H.~{Guo}, X.~{Wang}, K.~{Chang}, and Y.~{Tian}.
\newblock Exploiting path diversity for thwarting pollution attacks in named
  data networking.
\newblock {\em IEEE Transactions on Information Forensics and Security},
  11(9):2077--2090, Sep. 2016.

\bibitem{Gupta}
D.~{Gupta} and S.~{Batra}.
\newblock A short survey on bloom filter and its variants.
\newblock In {\em 2017 International Conference on Computing, Communication and
  Automation (ICCCA)}, pages 1086--1092, May 2017.

\bibitem{He}
D.~{He}, D.~{Zhang}, K.~{Xu}, K.~{Huang}, and Y.~{Li}.
\newblock A fast and memory-efficient approach to ndn name lookup.
\newblock {\em China Communications}, 14(10):61--69, Oct 2017.

\bibitem{Hou}
R.~Hou, L.~Zhang, T.~Wu, T.~Mao, and J.~Luo.
\newblock Bloom-filter-based request node collaboration caching for named data
  networking.
\newblock {\em Cluster Computing}, Mar 2018.

\bibitem{hussaini}
M.~Hussaini, S.~A. Nor, and A.~Ahmad.
\newblock Producer mobility support schemes for named data networking: A
  survey.
\newblock {\em International Journal of Electrical and Computer Engineering},
  8(6):5432, 2018.

\bibitem{IRTF}
IRTF.
\newblock Internet research task force.
\newblock Accessed on 26th June 2019 from \url{https://irtf.org/}.

\bibitem{Abyss}
S.~D. Jackman, B.~P. Vandervalk, H.~Mohamadi, J.~Chu, S.~Yeo, S.~A. Hammond,
  G.~Jahesh, H.~Khan, L.~Coombe, R.~L. Warren, et~al.
\newblock {ABySS 2.0}: resource-efficient assembly of large genomes using a
  bloom filter.
\newblock {\em Genome research}, 27(5):768--777, 2017.

\bibitem{Jacobson}
V.~Jacobson, D.~K. Smetters, J.~D. Thornton, M.~F. Plass, N.~H. Briggs, and
  R.~L. Braynard.
\newblock Networking named content.
\newblock In {\em Proceedings of the 5th international conference on Emerging
  networking experiments and technologies}, pages 1--12. ACM, 2009.

\bibitem{Kalghoum}
A.~{Kalghoum}, S.~M. {Gammar}, and L.~A. {Saidane}.
\newblock Towards a novel forwarding strategy for named data networking based
  on sdn and bloom filter.
\newblock In {\em 2017 IEEE/ACS 14th International Conference on Computer
  Systems and Applications (AICCSA)}, pages 1198--1204, Oct 2017.

\bibitem{Kalghoum1}
A.~{Kalghoum} and L.~A. {Saidane}.
\newblock Fcr-ns: a novel caching and forwarding strategy for named data
  networking based on software defined networking.
\newblock {\em Cluster Computing}, 22(3):981--994, Sep 2019.

\bibitem{Kalghoum2}
A.~Kalghoum and L.~A. Saidane.
\newblock Fcr-ns: a novel caching and forwarding strategy for named data
  networking based on software defined networking.
\newblock {\em Cluster Computing}, pages 1--14, 2019.

\bibitem{Khelifi}
H.~{Khelifi}, S.~{Luo}, B.~{Nour}, H.~{Moungla}, Y.~{Faheem}, R.~{Hussain}, and
  A.~{Ksentini}.
\newblock Named data networking in vehicular ad hoc networks: State-of-the-art
  and challenges.
\newblock {\em IEEE Communications Surveys Tutorials}, pages 1--1, 2019.

\bibitem{DONA}
T.~Koponen, M.~Chawla, B.-G. Chun, A.~Ermolinskiy, K.~H. Kim, S.~Shenker, and
  I.~Stoica.
\newblock A data-oriented (and beyond) network architecture.
\newblock {\em ACM SIGCOMM Computer Communication Review}, 37(4):181--192,
  2007.

\bibitem{kumar1}
N.~Kumar, A.~K. Singh, A.~Aleem, and S.~Srivastava.
\newblock Security attacks in named data networking: A review and research
  directions.
\newblock {\em Journal of Computer Science and Technology}, 34(6):1319--1350,
  2019.

\bibitem{OpenFlow}
A.~{Lara}, A.~{Kolasani}, and B.~{Ramamurthy}.
\newblock Network innovation using openflow: A survey.
\newblock {\em IEEE Communications Surveys Tutorials}, 16(1):493--512, First
  2014.

\bibitem{prematchBF}
J.~Lee, M.~Shim, and H.~Lim.
\newblock Name prefix matching using bloom filter pre-searching for content
  centric network.
\newblock {\em Journal of Network and Computer Applications}, 65:36 -- 47,
  2016.

\bibitem{scan}
M.~Lee, K.~Cho, K.~Park, T.~Kwon, and Y.~Choi.
\newblock Scan: Scalable content routing for content-aware networking.
\newblock In {\em 2011 IEEE International Conference on Communications (ICC)},
  pages 1--5. IEEE, 2011.

\bibitem{scan1}
M.~Lee, J.~Song, K.~Cho, S.~Pack, J.~Kangasharju, Y.~Choi, et~al.
\newblock Content discovery for information-centric networking.
\newblock {\em Computer Networks}, 83:1--14, 2015.

\bibitem{Li}
Z.~Li, K.~Liu, D.~Liu, H.~Shi, and Y.~Chen.
\newblock Hybrid wireless networks with fib-based named data networking.
\newblock {\em EURASIP Journal on Wireless Communications and Networking},
  2017(1):54, Mar 2017.

\bibitem{MaPIT}
Z.~{Li}, K.~{Liu}, Y.~{Zhao}, and Y.~{Ma}.
\newblock Mapit: An enhanced pending interest table for ndn with mapping bloom
  filter.
\newblock {\em IEEE Communications Letters}, 18(11):1915--1918, Nov 2014.

\bibitem{B-MaFIB}
Z.~{Li}, Y.~{Xu}, K.~{Liu}, X.~{Wang}, and D.~{Liu}.
\newblock 5g with b-mafib based named data networking.
\newblock {\em IEEE Access}, 6:30501--30507, 2018.

\bibitem{Li-Xu}
Z.~{Li}, Y.~{Xu}, B.~{Zhang}, L.~{Yan}, and K.~{Liu}.
\newblock Packet forwarding in named data networking requirements and survey of
  solutions.
\newblock {\em IEEE Communications Surveys Tutorials}, 21(2):1950--1987,
  Secondquarter 2019.

\bibitem{Li_Fan}
{Li Fan}, {Pei Cao}, J.~{Almeida}, and A.~Z. {Broder}.
\newblock Summary cache: a scalable wide-area web cache sharing protocol.
\newblock {\em IEEE/ACM Transactions on Networking}, 8(3):281--293, June 2000.

\bibitem{Lim}
H.~{Lim}, J.~{Lee}, and C.~{Yim}.
\newblock Complement bloom filter for identifying true positiveness of a bloom
  filter.
\newblock {\em IEEE Communications Letters}, 19(11):1905--1908, Nov 2015.

\bibitem{Atten}
F.~Liu and G.~Heijenk.
\newblock Context discovery using attenuated bloom filters in ad-hoc networks.
\newblock In T.~Braun, G.~Carle, S.~Fahmy, and Y.~Koucheryavy, editors, {\em
  Proceedings 4th International Conference on Wired/Wireless Internet
  Communications, WWIC 2006}, Lecture Notes in Computer Science, pages 13--25.
  Springer, 5 2006.

\bibitem{surv1}
L.~{Luo}, D.~{Guo}, R.~T.~B. {Ma}, O.~{Rottenstreich}, and X.~{Luo}.
\newblock Optimizing bloom filter: Challenges, solutions, and comparisons.
\newblock {\em IEEE Communications Surveys Tutorials}, 21(2):1912--1949,
  Secondquarter 2019.

\bibitem{Luo}
L.~{Luo}, D.~{Guo}, R.~T.~B. {Ma}, O.~{Rottenstreich}, and X.~{Luo}.
\newblock Optimizing bloom filter: Challenges, solutions, and comparisons.
\newblock {\em IEEE Communications Surveys Tutorials}, 21(2):1912--1949,
  Secondquarter 2019.

\bibitem{Mao}
Y.~{Mao}, C.~{You}, J.~{Zhang}, K.~{Huang}, and K.~B. {Letaief}.
\newblock A survey on mobile edge computing: The communication perspective.
\newblock {\em IEEE Communications Surveys Tutorials}, 19(4):2322--2358,
  Fourthquarter 2017.

\bibitem{BFR}
A.~{Marandi}, T.~{Braun}, K.~{Salamatian}, and N.~{Thomos}.
\newblock Bfr: A bloom filter-based routing approach for information-centric
  networks.
\newblock In {\em 2017 IFIP Networking Conference (IFIP Networking) and
  Workshops}, pages 1--9, June 2017.

\bibitem{PB-BFR}
A.~{Marandi}, T.~{Braun}, K.~{Salamatian}, and N.~{Thomos}.
\newblock Pull-based bloom filter-based routing for information-centric
  networks.
\newblock In {\em 2019 16th IEEE Annual Consumer Communications Networking
  Conference (CCNC)}, pages 1--6, Jan 2019.

\bibitem{Massawe}
E.~A. {Massawe}, S.~{Du}, and H.~{Zhu}.
\newblock A scalable and privacy-preserving named data networking architecture
  based on bloom filters.
\newblock In {\em 2013 IEEE 33rd International Conference on Distributed
  Computing Systems Workshops}, pages 22--26, July 2013.

\bibitem{ndnSIM2}
S.~Mastorakis, A.~Afanasyev, I.~Moiseenko, and L.~Zhang.
\newblock ndnsim 2.0: A new version of the ndn simulator for ns-3.
\newblock {\em NDN, Technical Report NDN-0028}, 2015.

\bibitem{Xie}
{Mengjun Xie}, I.~{Widjaja}, and {Haining Wang}.
\newblock Enhancing cache robustness for content-centric networking.
\newblock In {\em 2012 Proceedings IEEE INFOCOM}, pages 2426--2434, March 2012.

\bibitem{compressed}
M.~Mitzenmacher.
\newblock Compressed bloom filters.
\newblock {\em IEEE/ACM Trans. Netw.}, 10(5):604--612, Oct. 2002.

\bibitem{Mukherjee}
M.~{Mukherjee}, L.~{Shu}, and D.~{Wang}.
\newblock Survey of fog computing: Fundamental, network applications, and
  research challenges.
\newblock {\em IEEE Communications Surveys Tutorials}, 20(3):1826--1857,
  thirdquarter 2018.

\bibitem{mun1}
J.~H. {Mun} and H.~{Lim}.
\newblock Cache sharing using a bloom filter in named data networking.
\newblock In {\em 2016 ACM/IEEE Symposium on Architectures for Networking and
  Communications Systems (ANCS)}, pages 127--128, March 2016.

\bibitem{mun}
J.~H. Mun and H.~Lim.
\newblock Cache sharing using bloom filters in named data networking.
\newblock {\em Journal of Network and Computer Applications}, 90:74--82, 2017.

\bibitem{Nayakbio}
S.~{Nayak} and R.~{Patgiri}.
\newblock A review on role of bloom filter on dna assembly.
\newblock {\em IEEE Access}, 7:66939--66954, 2019.

\bibitem{access}
S.~{Nayak} and R.~{Patgiri}.
\newblock A review on role of bloom filter on dna assembly.
\newblock {\em IEEE Access}, 7:66939--66954, 2019.

\bibitem{isBF}
I.~Nikolaevskiy, A.~Lukyanenko, T.~Polishchuk, V.~Polishchuk, and A.~Gurtov.
\newblock isbf: Scalable in-packet bloom filter based multicast.
\newblock {\em Computer Communications}, 70(Supplement C):79 -- 85, 2015.

\bibitem{NSF}
NSF.
\newblock National science foundation.
\newblock Accessed on 26th June 2019 from
  \url{https://www.nsf.gov/publications/pub\_summ.jsp?ods\_key=nsf13538}.

\bibitem{CH}
R.~Pagh and F.~F. Rodler.
\newblock Cuckoo hashing.
\newblock {\em Journal of Algorithms}, 51(2):122 -- 144, 2004.

\bibitem{HFil}
R.~{Patgiri}.
\newblock Hfil: A high accuracy bloom filter.
\newblock In {\em 2019 IEEE 21st International Conference on High Performance
  Computing and Communications; IEEE 17th International Conference on Smart
  City; IEEE 5th International Conference on Data Science and Systems
  (HPCC/SmartCity/DSS)}, pages 2169--2174, Aug 2019.

\bibitem{DDoS}
R.~{Patgiri}, S.~{Nayak}, and S.~K. {Borgohain}.
\newblock Preventing ddos using bloom filter: A survey.
\newblock {\em arXiv preprint arXiv:1810.06689}, 2018.

\bibitem{scaleBF}
R.~Patgiri, S.~Nayak, and S.~K. Borgohain.
\newblock scalebf: A high scalable membership filter using 3d bloom filter.
\newblock {\em International Journal of Advanced Computer Science and
  Applications}, 9(12), 2018.

\bibitem{acBF}
R.~Patgiri, S.~Nayak, and S.~K. Borgohain.
\newblock {acBF}: A high accuracy membership filter using {rDBF}.
\newblock {\em Proceedings of {ADCOM}}, sep 2019.

\bibitem{patgiri1}
R.~Patgiri, S.~Nayak, and S.~K. Borgohain.
\newblock Hunting the pertinency of bloom filter in computer networking and
  beyond: A survey.
\newblock {\em Journal of Computer Networks and Communications}, 2019, 2019.

\bibitem{PassDB}
R.~{Patgiri}, S.~{Nayak}, and S.~K. {Borgohain}.
\newblock Passdb: A password database using 3d bloom filter.
\newblock In {\em 2019 IEEE 21st International Conference on High Performance
  Computing and Communications; IEEE 17th International Conference on Smart
  City; IEEE 5th International Conference on Data Science and Systems
  (HPCC/SmartCity/DSS)}, pages 1147--1154, Aug 2019.

\bibitem{rDBF}
R.~Patgiri, S.~Nayak, and S.~K. Borgohain.
\newblock {rDBF}: A r-dimensional bloom filter for massive scale membership
  query.
\newblock {\em Journal of Network and Computer Applications},
  136(June):100--113, 2019.

\bibitem{BD}
R.~{Patgiri}, S.~{Nayak}, and S.~K. {Borgohain}.
\newblock Role of bloom filter in big data research: {A} survey.
\newblock {\em CoRR}, abs/1903.06565, 2019.

\bibitem{Shed}
R.~Patgiri, S.~Nayak, and S.~K. Borgohain.
\newblock Shed more light on bloom filter's variants.
\newblock {\em arXiv preprint arXiv:1903.12525}, 2019.

\bibitem{SHA2}
W.~Penard and T.~van Werkhoven.
\newblock On the secure hash algorithm family.
\newblock {\em Cryptography in Context}, pages 1--18, 2008.

\bibitem{badona1}
I.~Publishers.
\newblock Storage management in dona content routers.
\newblock {\em Int. J. Internet Protoc. Technol.}, 11(4):205--218, Jan. 2018.

\bibitem{tomcat}
X.~Qiao, G.~Nan, W.~Tan, L.~Guo, J.~Chen, W.~Quan, and Y.~Tu.
\newblock Ccnxtomcat: An extended web server for content-centric networking.
\newblock {\em Computer Networks}, 75:276--296, 2014.

\bibitem{Quan}
W.~{Quan}, C.~{Xu}, J.~{Guan}, H.~{Zhang}, and L.~A. {Grieco}.
\newblock Scalable name lookup with adaptive prefix bloom filter for named data
  networking.
\newblock {\em IEEE Communications Letters}, 18(1):102--105, January 2014.

\bibitem{Rahel}
S.~{Rahel}, A.~{Jamali}, and S.~E. {Kafhali}.
\newblock Energy-efficient on caching in named data networking: A survey.
\newblock In {\em 2017 3rd International Conference of Cloud Computing
  Technologies and Applications (CloudTech)}, pages 1--8, Oct 2017.

\bibitem{Rath}
C.~Rathgeb, F.~Breitinger, C.~Busch, and H.~Baier.
\newblock On application of bloom filters to iris biometrics.
\newblock {\em IET Biometrics}, 3(4):207--218, 2014.

\bibitem{ren2016}
Y.~Ren, J.~Li, S.~Shi, L.~Li, G.~Wang, and B.~Zhang.
\newblock Congestion control in named data networking--a survey.
\newblock {\em Computer Communications}, 86:1--11, 2016.

\bibitem{DBF}
C.~E. Rothenberg, C.~A.~B. Macapuna, F.~L. Verdi, and M.~F. Magalhaes.
\newblock The deletable bloom filter: a new member of the bloom family.
\newblock {\em IEEE Communications Letters}, 14(6):557--559, 2010.

\bibitem{Biomet}
D.~Sadhya and S.~K. Singh.
\newblock Providing robust security measures to bloom filter based biometric
  template protection schemes.
\newblock {\em Computers \& Security}, 67(Supplement C):59 -- 72, 2017.

\bibitem{Sangeetha}
R.~{Sangeetha} and N.~{Ramasubramanian}.
\newblock A survey of hardware signature implementations in multi-core systems.
\newblock In {\em 2015 3rd International Conference on Signal Processing,
  Communication and Networking (ICSCN)}, pages 1--5, March 2015.

\bibitem{Sasaki}
K.~Sasaki and A.~Nakao.
\newblock Packet cache network function for peer-to-peer traffic management
  with bloom-filter based flow classification.
\newblock In {\em 2016 18th Asia-Pacific Network Operations and Management
  Symposium (APNOMS)}, pages 1--6, 2016.

\bibitem{NDN}
D.~Saxena, V.~Raychoudhury, N.~Suri, C.~Becker, and J.~Cao.
\newblock Named data networking: a survey.
\newblock {\em Computer Science Review}, 19:15--55, 2016.

\bibitem{NDN-NIC}
J.~Shi, T.~Liang, H.~Wu, B.~Liu, and B.~Zhang.
\newblock Ndn-nic: Name-based filtering on network interface card.
\newblock In {\em Proceedings of the 3rd ACM Conference on Information-Centric
  Networking}, ACM-ICN '16, pages 40--49, New York, NY, USA, 2016. ACM.

\bibitem{Shubbar}
R.~{Shubbar} and M.~{Ahmadi}.
\newblock Efficient name matching based on a fast two-dimensional filter in
  named data networking.
\newblock {\em International Journal of Parallel, Emergent and Distributed
  Systems}, 34(2):203--221, 2019.

\bibitem{Shub}
R.~Shubbar and M.~Ahmadi.
\newblock A filter-based design of pending interest table in named data
  networking.
\newblock {\em Journal of Network and Systems Management}, Mar 2019.

\bibitem{Singh}
A.~Singh, S.~Garg, S.~Batra, N.~Kumar, and J.~J. Rodrigues.
\newblock Bloom filter based optimization scheme for massive data handling in
  iot environment.
\newblock {\em Future Generation Computer Systems}, 2017.

\bibitem{So}
W.~So, A.~Narayanan, D.~Oran, and Y.~Wang.
\newblock Toward fast ndn software forwarding lookup engine based on hash
  tables.
\newblock In {\em Proceedings of the Eighth ACM/IEEE Symposium on Architectures
  for Networking and Communications Systems}, ANCS '12, pages 85--86, New York,
  NY, USA, 2012. ACM.

\bibitem{Soniya}
M.~M.~S. {Soniya} and K.~{Kumar}.
\newblock A survey on named data networking.
\newblock In {\em 2015 2nd International Conference on Electronics and
  Communication Systems (ICECS)}, pages 1515--1519, Feb 2015.

\bibitem{TRIAD}
stanford.
\newblock Triad project.
\newblock Accessed on 28th June 2019 from
  \url{http://gregorio.stanford.edu/triad/nbrp.html}.

\bibitem{Talpur}
A.~Talpur, T.~Newe, F.~K. Shaikh, A.~A. Sheikh, E.~Felemban, and A.~Khelil.
\newblock Bloom filter based data collection algorithm for wireless sensor
  networks.
\newblock In {\em 2017 Intl. Conf. on Information Networking (ICOIN)}, pages
  354--359, 2017.

\bibitem{NDN-FS1}
A.~{Tariq}, R.~A. {Rehman}, and B.~{Kim}.
\newblock Forwarding strategies in ndn based wireless networks: A survey.
\newblock {\em IEEE Communications Surveys Tutorials}, pages 1--1, 2019.

\bibitem{Tarkoma}
S.~{Tarkoma}, C.~E. {Rothenberg}, and E.~{Lagerspetz}.
\newblock Theory and practice of bloom filters for distributed systems.
\newblock {\em IEEE Communications Surveys Tutorials}, 14(1):131--155, First
  2012.

\bibitem{COBRA}
M.~{Tortelli}, L.~A. {Grieco}, G.~{Boggia}, and K.~{Pentikousisy}.
\newblock Cobra: Lean intra-domain routing in ndn.
\newblock In {\em 2014 IEEE 11th Consumer Communications and Networking
  Conference (CCNC)}, pages 839--844, Jan 2014.

\bibitem{ndnSIM}
M.~Tortelli, G.~Piro, L.~Grieco, and G.~Boggia.
\newblock On simulating bloom filters in the ndnsim open source simulator.
\newblock {\em Simulation Modelling Practice and Theory}, 52:149 -- 163, 2015.

\bibitem{bitmap1}
L.~Torvalds.
\newblock Linux kernel source tree.
\newblock Accessed on 10/07/2019 from
  \url{https://github.com/torvalds/linux/blob/master/lib/bitmap.c}.

\bibitem{NameFilter}
Y.~{Wang}, T.~{Pan}, Z.~{Mi}, H.~{Dai}, X.~{Guo}, T.~{Zhang}, B.~{Liu}, and
  Q.~{Dong}.
\newblock Namefilter: Achieving fast name lookup with low memory cost via
  applying two-stage bloom filters.
\newblock In {\em 2013 Proceedings IEEE INFOCOM}, pages 95--99, April 2013.

\bibitem{reroute}
H.~Watano and T.~Shigeyasu.
\newblock Interest re-route control according to degree of similarity on cached
  contents using bloom filter on ndn.
\newblock In L.~Barolli, F.~Xhafa, and J.~Conesa, editors, {\em Advances on
  Broad-Band Wireless Computing, Communication and Applications}, pages
  230--240, Cham, 2018. Springer International Publishing.

\bibitem{wired}
Wired.
\newblock The curse of xanadu.
\newblock Accessed on 27th June 2019 from
  \url{https://www.wired.com/1995/06/xanadu/}.

\bibitem{smarthome}
R.~Wu, B.~Cui, and R.~Li.
\newblock Research on access control of smart home in ndn (short paper).
\newblock In H.~Gao, X.~Wang, Y.~Yin, and M.~Iqbal, editors, {\em Collaborative
  Computing: Networking, Applications and Worksharing}, pages 560--570, Cham,
  2019. Springer International Publishing.

\bibitem{sumBF}
T.~Wu, L.~Zhang, J.~Lei, R.~Hou, and Z.~Song.
\newblock Sum-up counting bloom filter-based name lookup method for named data
  networking.
\newblock {\em Recent Advances in Electrical \& Electronic Engineering
  (Formerly Recent Patents on Electrical \& Electronic Engineering)},
  11(2):176--180, 2018.

\bibitem{Xiao2}
P.~Xiao, Z.~Li, H.~Qi, W.~Qu, and H.~Yu.
\newblock An efficient ddos detection with bloom filter in sdn.
\newblock In {\em 2016 IEEE Trustcom/BigDataSE/ISPA}, pages 1--6, 2016.

\bibitem{Xiong}
S.~Xiong, Y.~Yao, S.~Li, Q.~Cao, T.~He, H.~Qi, L.~Tolbert, and Y.~Liu.
\newblock {kBF}: Towards approximate and bloom filter based key-value storage
  for cloud computing systems.
\newblock {\em IEEE Transactions on Cloud Computing}, 5(1):85--98, 2017.

\bibitem{ICN1}
G.~Xylomenos, C.~N. Ververidis, V.~A. Siris, N.~Fotiou, C.~Tsilopoulos,
  X.~Vasilakos, K.~V. Katsaros, and G.~C. Polyzos.
\newblock A survey of information-centric networking research.
\newblock {\em IEEE communications surveys \& tutorials}, 16(2):1024--1049,
  2013.

\bibitem{yovita}
L.~V. Yovita and N.~R. Syambas.
\newblock Caching on named data network: a survey and future research.
\newblock {\em International Journal of Electrical \& Computer Engineering
  (2088-8708)}, 8, 2018.

\bibitem{hardware}
W.~Yu and D.~Pao.
\newblock Hardware accelerator to speed up packet processing in ndn router.
\newblock {\em Computer Communications}, 91-92:109 -- 119, 2016.

\bibitem{NDN2}
L.~Zhang, A.~Afanasyev, J.~Burke, V.~Jacobson, P.~Crowley, C.~Papadopoulos,
  L.~Wang, B.~Zhang, et~al.
\newblock Named data networking.
\newblock {\em ACM SIGCOMM Computer Communication Review}, 44(3):66--73, 2014.

\bibitem{NDN1}
L.~Zhang, D.~Estrin, J.~Burke, V.~Jacobson, J.~D. Thornton, D.~K. Smetters,
  B.~Zhang, G.~Tsudik, D.~Massey, C.~Papadopoulos, et~al.
\newblock Named data networking (ndn) project.
\newblock {\em Relat{\'o}rio T{\'e}cnico NDN-0001, Xerox Palo Alto Research
  Center-PARC}, 157:158, 2010.

\bibitem{Zhang}
L.~Zhang, D.~Estrin, J.~Burke, V.~Jacobson, J.~D. Thornton, D.~K. Smetters,
  B.~Zhang, G.~Tsudik, D.~Massey, C.~Papadopoulos, et~al.
\newblock Named data networking (ndn) project.
\newblock {\em Relat{\'o}rio T{\'e}cnico NDN-0001, Xerox Palo Alto Research
  Center-PARC}, 157:158, 2010.

\bibitem{namesync}
M.~Zhang, V.~Lehman, and L.~Wang.
\newblock Scalable name-based data synchronization for named data networking.
\newblock In {\em IEEE INFOCOM 2017-IEEE Conference on Computer
  Communications}, pages 1--9. IEEE, 2017.

\bibitem{CT-BF}
R.~{Zhang}, J.~{Liu}, T.~{Huang}, T.~{Pan}, and L.~{Wu}.
\newblock Adaptive compression trie based bloom filter: Request filter for ndn
  content store.
\newblock {\em IEEE Access}, 5:23647--23656, 2017.

\bibitem{Zhang1}
Y.~{Zhang}, A.~{Afanasyev}, J.~{Burke}, and L.~{Zhang}.
\newblock A survey of mobility support in named data networking.
\newblock In {\em 2016 IEEE Conference on Computer Communications Workshops
  (INFOCOM WKSHPS)}, pages 83--88, April 2016.

\bibitem{zhu}
Y.~Zhu, H.~Kang, and R.~Huang.
\newblock A cache privacy protection mechanism based on dynamic address mapping
  in named data networking.
\newblock {\em KSII Transactions on Internet \& Information Systems}, 12(12),
  2018.

\end{thebibliography}

\end{document}